	\documentclass[twoside,fleqn]{ActaStyle}
       \usepackage[dvips]{graphicx}
       \usepackage{times,cite}

\usepackage[british, slovak]{babel}
\usepackage[T1]{fontenc}
\usepackage[cp1250]{inputenc}
\usepackage{setspace,amssymb,amsmath,amsthm,graphicx,graphics}
\usepackage{indentfirst}
\usepackage{subfig,psfrag}
\usepackage{lastpage}
\usepackage{rotating}
\usepackage{titlesec}
\usepackage{amssymb}
\usepackage{amscd}
\usepackage{mathrsfs}
\usepackage{amsmath}
\usepackage{hyperref}
\addto\captionsslovak{
  \renewcommand{\contentsname}
    {Table of content}}

\newcommand{\beqa}{\begin{eqnarray}}
\newcommand{\eeqa}{\end{eqnarray}}
\newcommand{\nn}{\nonumber}

\numberwithin{equation}{section} 
	\def\authorheading{Gáliková, Ková\v{c}ik, Pre\v{s}najder}
	\def\shorttitle{Quantum Mechanics in Noncommutative space}
	
\begin{document}

	\pagerange{1}{108}   

	\title{QUANTUM MECHANICS IN NONCOMMUTATIVE SPACE}

	\author{Veronika Gáliková$^*$, Samuel Ková\v{c}ik\email{kovacik@fmph.uniba.sk}$^*$, Peter 	Pre\v{s}najder\email{presnajder@fmph.uniba.sk}$^*$ }
              {$^*$Faculty of Mathematics, Physics and Informatics,\\
Comenius University Bratislava, Mlynská dolina\\
Bratislava, 842 48, Slovakia}

	\abstract{\textbf{Abstract}}
This paper provides an examination of how are prediction of standard quantum mechanic (QM) affected by introducing a non-commutative (NC) structure into the configuration space of the considered system (electron in the Coulomb potential in the present case). The parameter controlling the extent of the modification is denoted as $\lambda$. The coordinates in the NC space are realized via creation and annihilation operators acting in an auxiliary Fock space, this one being chosen in such a way that the rotational invariance of the system remains intact also in NCQM. Analog of the Schr\"{o}dinger equation for hydrogen atom is found and analytically solved, both for bound and scattering states. The exact formulas for NC corrections are given. None of the NC predictions contradicts experimentally verified QM results, since in the correspondence limit $\lambda \rightarrow 0$ both QM and NCQM coincide. Highly surprising feature of the NC version is the existence of bound states for repulsive potential at ultra-high energies. However, these disappear from the Hilbert space in the mentioned limit. The whole problem is solved also using a Pauli method. Besides rotational invariance, the dynamical symmetry related to the conservation of NC analog of Laplace-Runge-Lenz vector is being used and the results obtained this way are in full agreement with those given by ''Schr\"{o}dinger-like'' approach.\\
The presented NC deformation of QM  preserves all those mysterious
properties of the Coulomb system that made it a
distinguished cornerstone of the modern physics.

\tableofcontents

	\section{Introduction}
This paper focuses on the way in which introducing certain non-trivial small-scale structure into the geometry of the configuration space can modify the predictions of the ''standard'' quantum mechanic (QM from now on).\\
\\
 The geometry of a space with ''non-trivial structure on small scales'' \, expressed via nontrivial commutation relations for coordinates that restrict their simultaneous knowledge is said to be  ''noncommutative''. There are many models suiting this description, distinguished by the form of the mentioned commutators of coordinates. The feature they have in common is that  the notion of a single point has to be given up. We will often use the shorthand ''NC'' to denote anything related to such a space; for example  we will write ''NCQM'' \, instead of ''quantum mechanics in a space with noncommutative coordinates''.

\subsection{The history and motivation behind NC geometry}

The idea of NC coordinates is quite old, it was first suggested by Heisenberg in order to get rid of the appearing infinities. This task was overtaken by the process of renormalization and NC coordinates vanished into oblivion for several decades. Despite the initial success, the method of renormalization had reached its limits and additional tools able to cure the theory were sought. This is why NC models became popular again during the eighties of the previous century in connection with the following two fundamental issues of modern theoretical physics: 1. existence of ultraviolet (UV) divergences in quantum field theory, and 2. building a framework for  quantization of gravity.\\
\\
The basic ideas of noncommutative geometry were developed in \cite{a1} and \cite{a2} with the spectral triplet as the main technical tool, and in the form of matrix geometry in \cite{b1} and \cite{b2}.\\
\\
They were followed by number of papers, and sometimes their goals were anything but humble - the ambitions involved the aspects mentioned above: the removal of UV- divergences from quantum field theories, and maybe even building a base for quantum gravity.\\
\\
Within the spectral triples formalism the Standard Model has been formulated in {\it almost commutative spaces}, and even the gravity was included into the models, \cite{a3}, \cite{a4}, \cite{a5}. However, the considered minimal noncommutativity did not lead to the desired absence of UV divergences, but it led to additional restriction among Standard Model parameters.\\
\\
The investigations within matrix noncommutative geometry started with field theoretical models in one of the simplest {locally Euclidean} NC spaces - the fuzzy sphere $S^2_F$. The space  $S^2_F$  was first introduced in \cite{berezin} as the QM phase space emerging from the quantization of the standard sphere $S^2$ considered as a particular classical phase space, or in the quantum description of two-dimensional membranes  \cite{hoppe}. In \cite{madore1} and \cite{madore2} the fuzzy sphere $S^2_F$ was described as a simple model of matrix geometry that was used in \cite{grossemadore} for the formulation of NC (or fuzzy) version of the 2-dimensional field-theoretical model - the Euclidean QED on $S^2$ formulated in \cite{Jayewardena}.\\
\\
The construction of matrix fuzzy spaces was extended in \cite{GP0} to compact Lie group orbits, the main tool were Perelomov's generalized coherent states introduced in \cite{Perelomov1}, \cite{Perelomov2}, or alternatively, using various Lie group representations representations in terms of annihilation and creation operators, \cite{GP1}, \cite{GP2}, \cite{GP3}. Such fuzzy Euclidean QFT models contained finite number of degrees of freedom, like in QM, and consequently, the UV divergencies were automatically absent. However, the generalization to relativistic QFT remained unclear.\\
\\
So far we have mentioned why are people interested in NC theories (e.g. NC removes infinities) and how have we been trying to define them. However, there is another point of view. NC theories are not (only) a lust, but a must.\\
\\
If we try to merge quantum mechanics with general relativity, we inevitably reach the conclusion, that two extremely close points of space cannot be distinguished (for two points separated by Plancks length a black hole would be created). Many theories capable of unifying gravity with (quantum) field theories, string theories being a prominent example, hold NC configuration space as a certain energetic limit. In addition, the latest research in M-theory shows, that the degrees of freedom in nature could be represented by matrices and therefore proper understanding of NC coordinates is a necessity.\\
\\
In 1995  two papers \cite{doplicher1} and \cite{doplicher2} appeared where it was shown that inclusion of gravity into QFT generates non-observability of space points that can be explained by  noncommutativite space-time coordinates satisfying  relations
$$[x_\mu, x_\nu]\,=\,i\theta_{\mu\nu}\,,$$
where $\theta_{\mu\nu}$ are  constant parameters commuting with coordinates. However, the Lorentz invariance is violated here. Later fuzzy QFT emerged as effective low-energy models of particular string theories, \cite{schomerus}, \cite{witten}. This incited an enormous activity in the field, but the problems with the violation of relativistic invariance persisted.\\
\\
Many different approaches to NC theories have been tried so far, differing not only in the NC relations, but also in the way they are realized and models they are studying (covering both quantum theories and gravity). Many successes were reached, however they are only partial and we still lack THE noncommutative quantum field theory which we seek and which would hopefully allow a peaceful merger of gravity and quantum theory.
\\
Learning the lesson from those attempts, we prefer to start small. Instead of trying to formulate new consistent fundamental models in fuzzy spaces we decided to test the ideas of NC geometry in a much simpler area: to test the consequences of the NC space in QM, i.e. within NCQM, since while QFT is quite problematic on its own, QM is quite well understood. Although the  primary ambition of NCQM has never been the ''improvement'' of  the precision of standard QM, one has to begin somewhere; and it is at least safer to introduce the concept of NC space while dealing with QM problems, to test it there and learn from toy models before handling any of the QFT issues.

\subsection{Outline}
This outline is meant to help to decide about the relative importance of particular sections - the reader is going to be informed about what  can be skipped and missed at what price. 
\\
\\
This current chapter 1. (''Introduction'') - is focused mostly on the motives, gives certain historical background and can be skipped without losing the main ''scientific message''. This part is not designed to please readers interested solely in dense academic reports. (Therefore such a reader is invited to turn their attention to the next parts, which are less or more marked by the appearance of the first numbered equations.) However, science is not only about technical details of solving equations,  and we decided to include  here a fraction of the other aspects as well. There is something captivating in exploring of how physics works,  and searching for certain balance regarding the extent to which one has to take models seriously. If some mathematical approach seems to work, it is often tempting to attribute more ''validity'' to it than is its due; on the other hand, an encounter with some troublesome features following  from restrictions of a  model may result in its hasty rejection, when one gets rid of both the problem and, unfortunately, of a possibly instructive experience as well.\\
The point is, the  model we have been working on for the last years is obviously restricted to a particular class of problems, it is a toy designed for exploring features of certain mathematical ideas, particularly those which may have something to say about physical space and its quantum mechanical aspects.  However,  if anything, we learned that there is nothing disreputable about toy models, as long as we try to modify them little by little according to how the Nature works and not the other way round. It seems the best we can do is to (smartly!) guess, look how much we have erred, and try to hit closer the next time. It would be cocky to expect hitting the nail on the first attempt, maybe it is more than enough  to expect we will ever hit it. (Some people even  doubt there is a nail to be hit; but it is probably hard to get anything done if one pays too much respect to such ideas.) Anyways, the process seems fascinating - that is, if one avoids starting on an overly confident level (in which case it tends to get accordingly painful). As to the ''guessing method'', it seems to be quite an effective one, and it is captivating - and we believe also instructive - to follow the reasoning of those who were particularly good at it  - even if their approaches are not compatible with each other up to these days. In fact all theories in the present science have problems, but it does not stop us from building on those which also have some redeeming virtues. In the strictest sense every part of contemporary physics is a toy theory, even though relativity or quantum mechanics are probably specimens of the most sophisticated toys we ever got to play with.\\
So these ideas are discussed in the introductory section (I.). It can be skipped without any grave consequences. However, as already has been mentioned, it is recommended to start paying attention when the first numbered equations appear.\\
\\
In  Chapter 2.  the noncommutative space is introduced. This is one of the parts that should be read if the following ones are to give any sense at all.  The aim is to get the reader acquainted with the NC space, to ''show them around''. A parameter that measures the extent to which the space is ''blurred'' is introduced,  important features of QM problems and the  requirements which are being put on their NC analogs are being searched for.\\
 \\
In Chapter 3. there is a brief overview of the known results of QM. Not because there would be any danger of its being new for anyone, but it is simpler to write the relevant relations down here, assign them  numbers and refer to them when the need arises  than to constantly recalling them fron an outer source. Since one of the goals is comparison of noncommutative quantum mechanics with the standard one, including some notoriously known facts should not be regarded as completely out of place. \\
\\
Chapters 4., 5. and 6. constitute the very core of this story, so skipping them approximately equals to not reading the paper.
Here are a few words about what to expect: One of the first recognized successes of QM has been the explanation of the discrete energy spectrum  labeling the bound states of electron in hydrogen. In QM, this was done in a twofold way, using both differential (Schr\"{o}dinger) and algebraical (Pauli) approach. Both methods agreed with each other and the observed values and  QM began to be taken seriously. Analogs of both approaches to the hydrogen atom problem are found also in the NC case. Moreover, the predictions are analytically worked out and to our pleasure found to be compatible with QM results and experimental data as well. There are of course NC corrections present and  presented - however, their magnitude is dependent on the one parameter that accounts for the measure of blurring the space - and this parameter is not fixed within our model. Important thing is, that in the commutative limit (which corresponds to setting the mentioned parameter equal to zero) all the corrections disappear and the former quantum mechanical results are restored.  \\
As to the investigation of scattering processes, one simply has to  flip the sign of energy in the previous calculations. Due to the potential approaching zero at the infinity, the energetic spectrum  is continuous in this case.  One curious feature of the Rutherford formula is  the fact that quantum and classical mechanics agree on the angular dependence. The NC version of this problem does not seem to be willing to break the tradition  - there are some modifications, but all of them vanish in the ''commutative limit'', just like it was the case with the bound states. However, there are certain points of interest worth dwelling on a little. One of them is appearance of the energetic cut-off, which is not completely unexpected, since it is a common feature of all theories concocted with a less or more hidden ambition to deal with energy divergences. Another, and probably  more surprising point is a kind of ''mirror symmetry'' with respect to the energy equal to the half of the maximal possible value. This symmetry goes as far as providing a set of bound states at the ultra-high energies.  More about this in the corresponding sections.\\
One chapter (number 5) is devoted to examining somewhat more general (not focused solely on the Coulomb problem) features of NCQM, namely the velocity operator, together with many interesting aspects it brings about - e.g. the analogs of Heisenberg uncertainty relations are derived, kinematic symmetry is studied, etc. The proper formula for velocity operator is essential for NC generalization of Pauli's algebraic approach.\\
\\
The last chapter - Conclusions - is meant to provide a concise overview of results and discussion thereof, to suggest problems which may be possibly addressed in the future, in short.\\
\\
There are several appendices included at the very end. People differ in their favorite way of approaching a paper of this kind,  some prefer having all the details derived and written down, others consider it to stand in the way of a concise account of the physics behind.  Trying to take the best of both worlds,  we decided to include relatively detailed calculations, but to store them mostly in the separate sections, so as to not overload the main part. These appendices are being mentioned abundantly throughout the whole paper, but we tried to make it  somewhat readable also for someone who prefers to refrain from taking too close look at the derivations.  It seemed inappropriate to leave more of the computations out completely, since they represent a good deal of the process. Undoubtedly an  idea counts more than the related algebra. However, this algebra is necessary to somehow bring ideas ''down to earth'', not to mention that it is also a tool to convict the ''bad ones'' coming in an appealing disguise. Although those  calculations which are provided here may not be the most efficient and ingenious ones,  at least those which led nowhere or to even worse places are omitted. There were numerous specimens of that kind accompanying the whole process. Some of them may have been as useful as a mistake can be - as a learning tool for the present authors, but these authors doubt that potential  reader would thoroughly enjoy the complete account of that ''learning process''. It is being said that physicists and detectives have a lot of in common, but presumably it is not the way they write their reports.  Whilst it may add to the appeal of a Sherlock Holmes story to watch him go astray for a while,  we suspect Sir Arthur would have lost some readers if hypergeometric functions had been used as a description of  his hero following the wrong hint.

\subsection{Why do we ask these questions}
This section aims to give some reasons why we have decided to ask the questions presented in this paper and why it seemed reasonable to choose hydrogen atom as a respondent to fill up our questionnaire.\\
\\
Exploring how Nature works is in a way like going to the theater.  \emph{All the world's a stage}, according to the play titled \emph{As You Like It}. Hopefully the reader will at least not mind it if we make certain use of this parable in the present section. Of course every metaphor has its limits; however, finding and pointing them out may teach us comparably much as do the features which our simile manages to catch faithfully.\\
\\
There are certain peculiarities about the drama that Nature performs. For example the exposition we are given is far from being exhaustive - the   information about setting, list of characters and their back-stories,  prior plot events, ...  is incomplete. We have the task  to fill it in, notwithstanding our being a few billions of years late for the opening scene.  A scientific paper can be viewed as a  kind of report providing some piece of information, or at least speculation, about the theater, the play, its cast and script. It is appealing to  think of it as one shared report mankind is writing, and everyone is welcomed to add a few lines. A small paragraph written by a playful mind, left just as a reminder of some curious,  amusing idea, may inspire someone else to add a breakthrough chapter a few decades later. \\
\\
This seems to be a good time for being such a reporter. There have already been people extremely skilled in gaining information, so we have worthy models to follow and learn from; on the other hand, there are so many questions unanswered, even unasked, that there is no danger of our ending without job. \\
\\
So at the beginning, let us try to sum up how did our forerunners proceed with asking questions and  what did they make out of the play so far.  People start really learning when they were given the opportunity to ''stand and stare'' besides caring for the mere survival. This ''watching the scene'' resulted in the basic notions of what is the stage of the theater, which characters are most common, what is the basic rhythm of changing the acts.  As to the stage, it was not paid enough attention. When watching such radiant actors as stars or atoms playing their part, one is prone to give only a little thought to the stage which hosts their performance.  Of course some assortment of tools to describe which actor is where and when  is needed for every play,  and people soon became quite efficient in handling such tools. Ren\'e Descartes brought this efficiency right to the top as far as the spatial part was concerned. What he did for space, Einstein and Minkowski  did for space-time, and physics seemed to have at least the question of the stage for its spectacles solved. With some intriguing facts to be admitted and our confidence in own intuition considerably shaken, but still.  \\
\\
However, the feeling about having the issues regarding this aspect of the drama all cleared up did not last long, and once again people had to realize this is one strange theater we are dealing with, since the stage not only provides a background with some contra-intuitive features - it seems to take an active role, too. This part of the report was given name ''general relativity''. So we had to admit we have this on extra actor to be interviewed when writing about our observations of the world theater. Curious, but so much more interesting.  There are certain issues with making an interview with the stage, since it is prone to provide the more twisted storylines the more co-actors are around (and the more they ''matter''), and slightly flat ones when left without company.  \\
\\
A few years later another surprise came, an astounding one. The stage even cannot be left without company of co-actors and still be interviewed, because there is no way of asking a question and not take a role of an actor, too. This part of the report was given name ''quantum theory''.\\
\\
 Those two  chapters in the report are  telling us that the world theater provides no clear distinction between stage, cast and spectators.   Anyway, it is quite amusing to find the idea of ''\emph{a stage where every man must play a part}''  already in \emph{Merchant of Venice}.   One almost wonders whether the division of books in library departments really is such a trivial matter....\\
\\
The answers to the questions which remained unresolved even  after the arrival of relativity and quantum theory  seemed to be obtainable from examining acting of the characters in new scenes, possibly in waiting for acts requiring more players or different kind of them. So the more parts of the play were watched, many times a need came to call for repeated performance, so that we could try some newly designed  opera glasses - in the case of this theatre both telescope and microscope are used, together with any other equipment to look into the things  more distant or more hidden. Indeed, some actors appeared that were unknown up to then- often not because they would have been absent, but because the spotlights had never been turned on them before. The list of newly discovered cast grows longer  every year, adding particles appearing in the high energy collisions under the  Franco-Swiss border or quasars revealed in a galaxy (far) far away where ''no  man could have looked before''.  Undoubtedly  every new item on the list is worth a celebration, but we are  still missing some clue. This unsettling fact manifests itself in various ways  -  for example, all the efforts to bring the two perhaps  most exciting  chapters of our report under one title have been futile.  Consistency  has always been considered a worthy cause; however, pursuing it resulted in  suspicious infinities having flooded our calculations.\\
Revision is necessary, and such a process is usually  only as good as are the question we  ask. So let us try to review what questions are being asked now.
\\
\\
 Having seen many scenes - having analyzed  many experiments,  people naturally start wondering, which actors are the key ones. Some  appear far oftener than others, some are difficult to glimpse, but their lines may be of vital importance all the same. So which character is common for all the acts and scenes? The stage of course (we know now we have to count it among the actors). After all, it was here  whenever, wherever and whatever happened. This may lead to deep thoughts as well as to intricate blind alleys, but one certainly cannot help thinking whether the stage should not be reexamined once again. We have heard this song before...and it was an instructive one at the time. But notwithstanding that breakthrough from century ago maybe we still underestimated  how much ''different'' the stage issues can get without us noticing.\\
\\
Having presented the status of the report so far, let us proceed to the few lines we would like to add to it. In comparison to the main chapters mentioned above and written by much skilled investigators, these few  lines can be nothing more than just a footnote written in  small font. Luckily the shared report has place for such items too. Now back to the promise from the beginning of this section - why did we choose to ask our questions and why to have the hydrogen atom as the main respondent. Having read the abstract, the reader is acquainted with the topic of this paper - to investigate how can some nontrivial structure of space have influence on quantum-mechanical properties of the hydrogen atom. At this phase it should be clear that the bottom line is questioning the nature of space and exploring possible modification that could be done in our ideas about the stage. Hydrogen is here in the role of some guide that would hopefully be willing to make a tour of the stage for us. Let us explain the selection of this guide. Hydrogen meets criteria of relevance, simplicity and beauty....which makes it a logical first choice.\\
\\
 As to the relevance, hydrogen  is one of the longest-serving actors in the theatre, takes part in so many acts that it is usually hard to find one where it would be absent. It remembers times from about three or four hundred thousands of years after the first scene.  One may argue it missed the curtain rise anyway, so it cannot have all the answers. However, common sense keeps us far from expecting all the answers just now, we do not even have all the questions yet, and will be glad enough if those we do have prove themselves to be somewhat meaningful.  Moreover, let us remind ourselves of what we are about to do - if we want to explore the stage which we suspect may considerably differ from the image we have in our minds, we need a reliable guide we are at least to some extent familiar with, one that is willing to give interviews, is relatively easy to understand, available under almost any conditions, not inclined to appear only now and then when by us unpredictable chances are kind to us, and, (not trying to be disrespectful to any field of physics interest), one not prone to have escaped into some alien dimensions or another realm of hardly reproducible circumstances.  Besides, while it is true that there is much more unknown than known to us in this universal theater, hydrogen accounts for the staggering majority of what we do ''know''. (Due to  the numerous surprises the Nature provides, the last word of the preceding sentence should not appear without apostrophes.) \\
 Beauty is often spelled ''symmetry'' in physics - and indeed, in this case the spherical one  is quite striking, due to the exclusive radial dependence of the Coulomb potential, and the specific form thereof provides for an additional ''hidden'' symmetry which makes it even more appealing.\\
 Regarding simplicity, honesty is anything but out-of-place when writing a contribution to the ''report on the theater\,''. \,Of course this is the simplest atom around and probably the only one which allows us to hope for analytical solution for a long time coming. There is nothing wrong with this...after all, with the puzzling mathematical issues with infinities scattered all over the place which challenge today's writers of the report, there seems to be an obvious need to carry out a revision of  basics of the basics. So we hereby declare that if there were simpler atom than hydrogen, this paper would be dealing with it. And yes, it even is just a toy model (in the sense that it does not include all the features we think the stage has). However, what provides more possibilities to learn than a good game?

\section{NC space}

In the introduction it has been suggested that there may be something going on about the ''stage '' of the universal theatre - the space-time. The standard way of handling it is the manifold view - however is this twisted, it is locally isomorphic to Minkowski flat space, with the whole Poincar\'{e} group worth of symmetry. This model given by general relativity  is the most efficient one we have - so far no other has managed to capture  so many features of what we see in Nature.  Indeed, we would hardly dream of giving it up  if it were not for the fact that it fails to accommodate requirements of quantum theory  - which is another quite successful  enterprise,  although  in rather different aspects.  So we are looking for some alternatives, using various models. Let us say something about what kind of modifications do we have in mind first, and then  let us look at the problem of various models.\\
\\
As to the modifications: All the relativistic machinery  works efficiently up until we found ourselves in the realm of  tiny intervals of space-time.  So if there is a room for some deformation of space-time (meaning if there is some  model other than a manifold locally isomorphic to Minkowski space) it should happen on these small scales. There is, however, a fact that the standard quantum theory does not consider such matters and still succeeds in its predictions. Well, if the mentioned deformations are subtle enough, it may all fit together in such a way that they do not alter the predictions of standard QM much  - if that were the  case, we simply may have overlooked them.  \\
\\
The basic idea is to try out whether such deformations are worth examining at all.  Whether it is even possible to find such that they  would both incorporate some non-trivial structure into the space-time and at the same time  would be compatible with  those experimentally well confirmed QM predictions. Another wish is to make them capture as much of the symmetries given by Poincar\'{e} group as possible. Needless to say, for the time being it is just a wish to have all this requirements met. However, as far as the testing of  the idea of non-trivial space-time structure and compatibility with standard QM is concerned, there is still a chance. Even if we do not manage to capture all of the nice features Poincar\'{e} symmetry has, we may still be able to catch enough of them to sufficiently describe some simple system, solve for it, and compare with the corresponding QM prediction.\\
\\
Now to ''choose a model'' means to choose which aspects of the observed world do we incorporate into our approach.  Of course it is strongly recommended to do this  with regard to the choice of the ''simple system we are about to describe''. People are working on various approaches, each of them has undeniable shortcomings (therefore it is called a model), but there seem to be some sense in doing this anyway - one just has to be careful to not restrict their attention exclusively to their point of view. It is acceptable to have one's  preferred  approach to play with, instructive to  find out what other kind of methods do the others have, and nice  to arrange for some exchange stays in those various toy universes. Hopefully we manage to not forget there is a bigger one outside. \\
\\
As to the usefulness of various sensible, although incomplete models that are being examined  and appear  to be incompatible sometimes, there is an ancient parable from the Middle East which seems to apply here.  This may not be the original version,  the point is  preserved nevertheless. It tells a story about a group of people who  were  examining an elephant in a dark room by touching  it. One of them felt the tusk and claimed the creature is solid and similar to a stake, another one grabbed the ear and concluded the animal  is soft and resembles burdock leaves; the man sitting on the elephant's back  insisted it is comparable  to a big pumpkin, and the man holding the leg was adamant about comparison of the animal to a tree trunk. A strong disagreement about what an elephant is like arose among the people. The resolution differs in various versions of the story; the participants  of the research end up wiser or sorer according to what attitude they adopt.\\
\\
 Maybe one day we will be able to see how the Nature really is;  in the absence of the evidence to the contrary  let us hope and till then try to touch at least  some particular aspects. Although not all at once - the elephant is too  big for us  - but  we can step by step, model by model examine its features. Important thing is to not condemn various approaches just because we are yet unable to make them seamlessly fit together. Models are to be learned from,  not to be believed. \\
 This collecting of partial pieces of information is quite tempting by itself, but maybe there is something even better in the store for us. Perhaps one day, while stumbling in the dark, checking for a yet unexamined area we manage to hit a light switch, or to find out the walls have windows in them.  Maybe even in that case we will have to wait for the Sun to rise and shine, but it would be nice to have at least the shutters unfastened by then.

\subsection{ What is meant by deformed geometry}
Now it is the time to talk about models of space(time) with a ''non-trivial structure '' more specifically. First of all, since we will be testing how do predictions of QM change under such modification, and since QM is most successful when dealing with stationary problems, the  models  oriented this way can  afford  to forget about time, so that alterations which are being made are related to the spatial issues only. Certainly a considerable flaw as to relativistic aspects, but this one is present in the standard QM, too. We are not giving up on tackling also space-time, but let us try first whether we can reproduce good features of stationary QM. To be honest, if we are thinking about  starting to  work on deforming 3+1 dimensional manifold,  we should better collect all the useful hints we can get regarding how to proceed. One substantial source of hints is considering a simpler case first. \\
\\
As to the expression ''non-trivial structure of space'', it usually refers to the absence of local isomorphism to ${\bf R}^3$. Perhaps this is not a particularly fitting expression, because it may suggest that continuous space structure is  ''trivial''. Well, imagining a  continuous space on the small scales, with its distinct points infinitesimally close to one another, is  difficult comparably to imagining a space somewhat ''blurred'' with no points to visualise whatsoever. The calculus on continuous manifolds is, however, considerably more developed these days. This is to be expected - the point in such a space can be described by n-tuples of classical numbers. We do not need to be very general here, so let us talk about 3 dimensional case. These three classical numbers tell us what are the three coordinates of the point. If the configuration space can be represented as one isomorphic to ${\bf R}^3$, it means there is no limitation as to the simultaneous knowledge of all three coordinates.  On the other hand, if we are considering a ''fuzzy'' configuration space, it means there is such a fundamental   restriction.

\subsubsection{How QM deforms phase space }
\noindent Deforming some space so that the notion of its point is to be abandoned is not a new thing, we have encountered it when moving from classical to quantum mechanics. Recall there is an elegant hamiltonian formalism which is strongly connected to the phase space of a system. Its coordinates are  all the positions $q_i$ and momenta $p_i$ of whatever constitutes the system.  States of the system can be represented by points of this phase space, physical observables as functions defined on it. After letting the Heisenberg uncertainty principle into the play, the restriction on the simultaneous knowledge of position and momentum  (mathematically this ''blurring'' is expressed via non-vanishing commutator  of the former coordinates $q_i$, $p_i$ ) makes the notion of points in that space ill-defined. This implies that if we do not want to lose the notion of a physical state well, we have to reconsider and redefine it. This way the whole new  probability-based  approach operating with wave functions came into physics. It is perhaps appropriate to emphasize some aspect here, namely that the form of the commutator $[q_i,p_i]$  had a paramount impact on the whole QM. The Hilbert space of states represented by wave functions, operators given as hermitian operators acting on those etc.  are what they are to a great extent because that commutator is what it is. As to the extent to which QM modifies classical mechanics, it is given by the magnitude of the constant appearing there, $\hbar$, which gives  a measure of how much has  the phase space been blurred by letting Heisenberg uncertainty principle in. According to the correspondence principle, the limit  $\hbar \rightarrow 0$ is expected to reduce the prediction to those of classical mechanics.\\

\subsubsection{How NC model deforms configuration space}
\noindent Something similar is happening when we are deforming the configuration space, whose coordinates represent positions of whatever constitutes the system. Some uncertainty principle is conceivable here too, expressed via non-zero commutator of the coordinates and the concept of continuum of points has to abandoned. \\
\\
So blurring our ${\bf R}^3$ space means to prescribe some nonzero commutation relation $[x_i , x_j]$ for its coordinates. These cannot remain c-numbers anymore; we have to find some mathematical objects to be the new representatives. In quantum mechanics, the former phase-space coordinates became operators on some Hilbert space of wave-functions. So in our case the NC coordinates will be operators acting somewhere; the answer to the question where exactly and how distinguishes between particular models of NC spaces. The choices of a model and a system we wish to examine are to be made with respect to each other, so that we stay out of unnecessary trouble when doing the actual calculations.

\subsection{Our NC model }
Finally  to our case. The considered system is hydrogen; the Coulomb problem configuration space  is ${\bf R}^3_0\,=\,
{\bf R}^3\setminus \{0\}$, the exact solution of QM is known, let us try to do no worse in NCQM. \\
Now to how our model should be like: We wish to preserve those nice features which made the problem solvable in standard QM. The Schr\"{o}dinger equation is exactly solvable in QM thanks to the fact that it was separable in spherical coordinates, which was in turn possible due to the rotational symmetry of the whole model  - the system studied and the space it was placed in. As to the system, hydrogen is simply hydrogen, its rotational invariance is taken as granted - we just have to make sure that our NC space has that symmetry too. Since we are about to mention it rather frequently, let us introduce its symbol  here: ${\bf R}^3_\lambda $ stands for ''rotationally invariant NC space,  an analog of the Coulomb problem configuration space ${\bf R}^3_0$.'' That role of the little  $\lambda$-accessory will be cleared up shortly. It is our NC parameter having the dimension of length, its magnitude expresses to what extent is the NC space ''fuzzy''. There is one happy coincidence in this notation. The $0$-subscript in ${\bf R}^3_0$ is meant to indicate that the origin has to be excluded from the configuration space for the Coulomb problem - however,  since this $0$ appears at the same position as $\lambda$-subscript in  ${\bf R}^3_\lambda $, it is possible to interpret it also this way - that the two spaces are the more alike the more $\lambda$ approaches $0$, and  coincide for $\lambda=0$.

 \subsubsection{NC coordinates and their realization}
\noindent What follows are commutation relations for the ''coordinates'' in ${\bf R}^3_\lambda $ (apostrophes account for the fact that they are not c-numbers, but we will refrain from  stressing this every time we come to mention them).
\be\label{com} [x_i,x_j]\ =\ 2i\,\lambda\,\varepsilon_{ijk}\,x_k\,,
\ee
So the first  numbered equation has appeared in this paper, and since these relations have impact on almost everything that will follow,  we  hope to have now the attention also of the reader who has chosen to skip the introductory speeches after skimming the  ''Outline''-section. So we hereby welcome this reader; and let us move on to discuss what is going on in (\ref{com}). \\
\\
As the notation suggests, $x_i$ are analogs of the former Cartesian coordinates. The factor 2 is here just for convenience, to avoid other factors 2 which would appear elsewhere had we not put it here.  \\
\\
$\lambda$ has been introduced a few lines above. It  maintains the proper dimension and takes care of the correspondence principle as well. Zero $\lambda$ would mean that coordinates \textit{do} commute; small $\lambda$ suggests that space is a little blurred, to the extent small enough to avoid any experimental evidence so far. The magnitude of $\lambda$ is not fixed within our model; it is related to the smallest distance relevant in the given NC space. It may well be even at the Planck's scale ($\approx 10^{-35}$ m); if it is so, then the NC corrections are well hidden beyond the scale of ordinary QM experiments. To have the correspondence principle satisfied, we hope the limit $\lambda \rightarrow 0$ to reduce the predictions of NCQM to those of QM. \\
\\
As to the the commutator (\ref{com}), it reminds us of $so(3)$ algebra - the generators of rotations in ${\bf R}^3$ obey the relations of the same form. Obviously, since the related symmetry of space is just that which we want to have in our model in the first place.\\
\subsubsection{Auxiliary Fock space ${\cal F}$ }
\noindent Now that we know the key relations our NC  coordinates have to obey, we have to find some way to bring them from the realm of platonic forms ''down to earth'' and find some concrete realization. This  will account for how our calculations in NC case will look like. \\
To examine certain symmetry, we have to find the relevant group, and its representation can be obtained as linear operators acting in some suitable space. Many times it is useful to choose a Fock space. Such spaces come with a set of creation and annihilation (c/a) operators which have many appealing properties when it comes to doing calculations; moreover, any relevant linear operator can be expressed in their terms. The choice of particular kind of Fock space is given by the symmetry we want to explore. In our case, the symmetry is given  by the  rotational invariance of our  ${\bf R}^3_\lambda $, so  Fock space suitable for our purposes is that one whose c/a operators can be used to construct $x_i$ so that (\ref{com}) holds. It turns out that we need two pairs of c/a operators $a_\alpha, \, a_\alpha^\dagger$, $\alpha=1,2$, which satisfy the standard relations
\be\label{acom} [a_\alpha,a^\dagger_\beta]\,=\,\delta_{\alpha\beta },\ \
[a_\alpha,a_\beta]\,=\,[a^\dagger_\alpha, a^\dagger_\beta]\,=\,0\, .\ee
The Fock space ${\cal F}$ related to them is  spanned by normalized
vectors
\be\label{funct} |n_1,n_2\rangle\ =\ \frac{(a^\dagger_1)^{n_1}\,(a^\dagger_2)^{n_2}}{
\sqrt{n_1!\,n_2!}}\ |0\rangle\,. \ee
Here $|0\rangle\,\equiv\,|0,0\rangle$ denotes the normalized
vacuum state: $a_1\,|0\rangle\ =\ a_2\,|0\rangle\ =\ 0$. We shall use the
notation ${\cal F}_n\, =\, \{ |n_1,n_2\rangle\,|\ n_1+n_2 = n\}$.
\\
Those  $\,a^+_\alpha, \,a_\alpha$ are said to ''create'' and ''annihilate'' some formal particles (not necessarily having a physical interpretation); from the commutation relation we see that they are two kinds ($\alpha=1,2$) of bosons.  Besides  $\,a^+_\alpha, \,a_\alpha$, there is also certain combination of them which deserves special mention here - the particle number operator
$$ N\,=\,a^+_\alpha a_\alpha \, , $$
$$ N |n_1,n_2\rangle\ = (n_1 +n_2) |n_1,n_2\rangle\ = n |n_1,n_2\rangle\ \, .$$
As we will see shortly, it has a strong connection also to  physically relevant issues.\\
\subsubsection{NC coordinates}
With this choice of Fock space,  $x_i$'s obeying (\ref{com}) are given as
\be\label{x_} \ x_j\ =\ \lambda\,a^+\,\sigma_j\,a\ \equiv\
\lambda\,\sigma^j_{\alpha\beta}\,a^\dagger_\alpha\,a_\beta,\
j\,=\,1,2,3\,.\ee
$\sigma^j $ are the Pauli matrices. In case the reader would wish to see the explicit form of each one $x_i$, here it goes:
$$x_1 \,= \,\lambda \left(a^\dagger_2a_1+a^\dagger_1a_2\right), $$
$$x_2 \, = \, i \lambda  \left(a^\dagger_2a_1-a^\dagger_1a_2\right),$$
$$x_3 \, = \, \lambda \left(a^\dagger_1a_1-a^\dagger_2a_2\right).$$

\noindent Now to the spherical symmetry. We hope to be able to employ it; clearly having the  NC analog of the Euclidean distance from the origin would help:
$$r\,=\,\lambda\,(a^\dagger_\alpha a_\alpha + 1) \,=\,\lambda\,(N + 1).$$
By straightforward calculation it can be proved that
\be\label{[x,r]}[x_i,r]\,=\,0\,,\ \ \ \ r^2 - x_j^2\,=\, \lambda^2\,.\ee
We will provide a strong argument supporting the exceptional role of
$r$ later. For now let us just point out that while the difference $ r^2 - x_j^2$ is of the order $\lambda^2$, the difference $(\lambda N)^2-  x_j^2$  involves also the first order contribution, so it would be a choice worse by one order.
The  definition of $r$  makes $N$ very important to us. It will appear frequently, after all, it is also its simple action on the vectors of ${\cal F}$ which makes the calculations manageable. \\
It is useful to introduce one more operator,
$$\rho\,=\,\lambda\,a^\dagger_\alpha a_\alpha\,=\,\lambda\,N .$$
 Obviously it is so simply related to both $r$ and $N$ that it may seem unnecessary to have a keep symbol for its sake, but it will prove to be quite convenient to have it.

\subsubsection{Hilbert space ${\cal H}_\lambda $ of NC wave functions}
\noindent One hopefully superfluous note: Although ${\cal F}$ mathematically is a Hilbert space, it is not the analog of the Hilbert space ${\cal H}$ of physical states. Similarly to the wave functions in QM being defined on ${\bf R}^3_0 $, our NC wave-functions are to be defined on ${\bf R}^3_\lambda $. The Hilbert space whose elements they are 
 will be denoted as ${\cal H}_\lambda $.\\
\\
As to the argument of NC wave functions,  we may write  $\Psi(x_i)$, since  $x_i$ are the NC coordinates in  ${\bf R}^3_\lambda $. However, it is often more convenient to express $\Psi$ as  $\Psi(a_\alpha, \, a_\alpha^\dagger)$. Of course not every array of c/a operators can represent a physical state. For example, the number of creation and annihilation operators should be equal in the expression for $\Psi$ - since it is a feature of all the $x_i$'s.  So let us specify what belongs to  ${\cal H}_\lambda $: It is the linear space of normal ordered analytic functions containing the same number of creation and annihilation operators
\be\label{Hilb} \Psi\ =\ \sum\, C_{m_1 m_2 n_1 n_2}\,(a^\dagger_1
)^{m_1}\,(a^\dagger_2)^{m_2}\,(a_1)^{n_1}\,(a_2)^{n_2}\,,\ee
where the summation is finite over nonnegative integers satisfying
$m_1+m_2 \,=\,n_1+n_2$.  ''Normally ordered'' means that all $a^+$'s are gathered to the left and all $a$'s to the right. \\
\\
QM is about probabilities, NCQM should be too, so we need some kind of norm in our Hilbert space - this one is the weighted Hilbert-Schmidt norm defined via finite weighted trace over the basis of Fock space:
\be\label{whs1} \| \Psi \|^2\ =\ 4\pi\,\lambda^3\,\mbox{Tr}
[(N+1)\,\Psi^\dagger\,\Psi]\ =\ 4\pi\,
\lambda^2\,\mbox{Tr}[\Psi^\dagger\,r\,\Psi] \,.\ee
The rotationally invariant weight $w(r)\,=\,4\pi\,
\lambda^2\,r$ is determined by the requirement that a ball
in ${\bf R}^3_\lambda $ with radius $r$ should
have a standard volume in the limit $r\,\to\,\infty$. The
projector $P_n$ on the subspace ${\cal F}_0
\oplus\,\dots\,\oplus {\cal F}_n$, corresponds to the
characteristic functions of a ball with the radius $r = \lambda (N+1)$.
Therefore, the volume of the ball  in question is
\be V_r\ =\ 4\pi\,\lambda^3\,\mbox{Tr}[(N+1)\,P_n]\ =\
4\pi\,\lambda^3\,\sum_{k=0}^n (k+1)^2\ =\
\frac{4\pi}{3}\,r^3\,+\,o \left(\frac{\lambda}{r} \right)\,.\ee
Thus, the chosen weight $w(r)\,=\,4\pi\,\lambda^2\,r$
has the desired property.

\subsection{Packing a mathematical toolkit}
The previous section contains all the basic information about the mathematics which will be behind our calculations in NC space. This section is in some way similar to preparing for a backpacking trip. There is a long journey ahead of us, a few chapters long. Some mathematical equipment will be required along the way, and this chapter is a fine place to prepare it, get acquainted with it and pack it in some concise form  before hitting the road.  \\
\\
We need roughly three things to consider: What are  the basic features of NC Schr\"{o}dinger equation, what do we expect from its solution (i.e. what ansatz would be sensible), and having those things, how do we make the comparison of QM and NCQM?\\
We probably will not be able to prepare three separate packets of mathematical tools, nicely one-by-one. The issue is, we are not only after solving some given equation. In this case we also have to find one before solving it. It may be wise to avoid drawing some artificial line between the equation and the ansatz features.
This will probably result into the fact that the tools we are to prepare and pack with us will be  hard to place into a neat row for an one-by-one  exposition. Focusing solely on one item at a time helps to keep a paper nicely structured, but those items may end up too self-centered to fit together. And we will need them to be compatible. So the reader may forgive certain mess in the process.

\subsubsection{Crash-course in NC calculus }
\noindent Not that we would go deep into the explicit form of either equation or its solutions here, this is not the proper chapter for such a discussion. This is just a concise list of what an undergraduate physics student finds out when the hydrogen atom problem is presented to them - so that this way we obtain basic guidelines about what we need to learn to handle in the NC case.\\
\\
$\bullet $ In QM the Schr\"{o}dinger equation is differential. We should be able to handle the basics  of the NC differential calculus. \\
\\
$\bullet $ The whole problem, equation + ansatz for hydrogen is separable in spherical coordinates. This was actually one of the key features that enabled exact solution in QM. If we are to reproduce such a success in NCQM, our $\Psi$'s should better have this property too, whatever that means in our case. \\
\\
$\bullet $ In the first courses of QM students are said that the angular part has been already done before considering  hydrogen or any other physical system, and the result can be (with proper gratitude) simply copied from those mathematically oriented papers.  Believe it or not, this is also the case in NC version. There has been a good deal of work done as to the angular part of the problem; the NC analogs of spherical harmonics, with some $r$ raised to a suitable power were found and examined in \cite{GP2}. So the radial part seems to get considerably more attention in what follows. \\
\\
$\bullet $ The radial part has a lot to do with power series in the radial coordinate, and those who do not know how to differentiate them may leave the QM class ... meaning NC power series in $r$  and their derivatives should be familiar by the time we try to formulate ourselves the problem. \\
\\
\\
Let us get down to work before the to-do list grows any longer.
\\
First, what do we already have in our NC case? Let us start from the Hilbert space. Right now we do  not know about the coefficients in (\ref{Hilb});  the expression is valid for the whole ${\cal H}_\lambda $.  Our task will be, beside others, to find the subspace of eigenstates related to hydrogen atom. \\
\\
As to the arrangement of the c/a operators in $\Psi$ in (\ref{Hilb}),  all  $a^+_\alpha$,'s are on the left and all  $ a _\alpha$'s  on the right, in other words, we have here the so-called normal ordering. Sometimes it is called Wick's, but it has became a tradition with us to call it ''normal''. There are some objects (reader will encounter some later) related to various orderings, and referring to one possibility as ''normal'' and to the other one  as ''ordinary'' (or ''usual'', ''standard'', or whatever synonym one would normally use, except for ''normal'')  has been a source of certain  amusement during our discussions about the matter (as well as a supplier of some confusion when we failed to keep on our toes).\\
\\
So much for the name, now how do we make calculations with variously ordered arrays of  $a^+_\alpha$,'s and $ a _\alpha$'s. There is no need to be overly general just now - the primary concern will be with the radial Schr\"{o}dinger equation, the dependence on the NC coordinates (and consequently on  $a^+_\alpha$,'s and $ a _\alpha$'s) is therefore a very specific here. There is a good reason to expect (besides other things, and making allowance for the unexpected too) some power series in the radial coordinate. \\
\\
One of the many reason why power series are so widely used as ansatz in differential equations is the way they behave under differentiation. We would certainly appreciate having something similar; so let us try to find such power-expansions and such derivatives that it all plays well.
We have already made acquaintance with the NC analog of radial distance, we know it is closely related to the particle-number operator $N$, we know how this works on the vectors of Fock space, raising it to the $n$-th power simply means to apply it $n$-times in a row, so there seem to be little need for dwelling on it much further. However, we can hardly hope to guess the complete solution of the yet to be guessed equation. We are looking for an ansatz to insert it in  when we manage to figure it out. In other words, we may write a power series in ordinary powers of $\rho$, but it will do us no good without the knowledge of how to differentiate it - the NC way, that is. \\
\\
This leads us to the question of derivative in NC space. For now let us search for one with respect to  $r$, or $\rho$.   Derivative of a function defined on the continuous space $R^3$ is usually defined as the difference between two values of this functions in two infinitesimally close points of this space. Clearly such a definition cannot be used in our space, because we do not have ''two infinitesimally close points'' here. However, since derivative is much more general notion, we are not at a loss here; we need something linear, obeying Leibniz rule and if possible having some obvious by analogous properties to the continuous case. One of the first candidates  we can think of as  linear and Leibniz-respecting is a commutator with one entry fixed; since for now everything relevant for our purposes is expressible via  $a^+_\alpha$,'s and $ a_\alpha$'s, it is not hard to guess that such things may prove to be also the right choice for occupants of that ''fixed'' entry. So NC derivatives will probably have something to do with objects like $[a^+_\alpha, \,.\,]$ and $[a_\alpha, \,.\,]$. Maybe it will prove necessary to multiply these by some additional things. The dot in the second entry indicates the place where we put whatever is going to be differentiated. Let us try how we are doing so far. How does our proposed operator act on, say, power series in $r$? This is one the most basic thing we can think of to try:
$$[a^+_\alpha, r^k]= \lambda^k [a^+_\alpha, \, (a^+_{\beta_1} a_{\beta_1}) \,...\,(a^+_{\beta_k} a_{\beta_k})]=... \,\,\,\, .$$
 We will not even bother the reader with the result; suffice it to say, it does not resemble that simple formula from ''continuous case'': $\partial_r r^k = kr^{k-1}$
Actually this is one of the  reasons we will deal with the above mentioned normal ordering.
Let us first compare what is meant by the ordinary $N^k$ and normal powers $:N^k:$ (note that the latter  are denoted by colon marks)
$$
N^k\,=\, (a^+_{\beta_1}\,a_{\beta_1}) \, ...\, (a^+_{\beta_k} \,a_{\beta_k}) \, ,    \,\,\,\,\,\,\,\,\,\, \,\,\,\,\,
:N^k:\,=\, (a^+_{\beta_1} ...\, a^+_{\beta_k}) \, (a_{\beta_1} \, ... \,a_{\beta_k}) \, .
$$
Obviously the ordinary power $N^k$ is simply meant as the particle number operator acting $k$-times in a row. Normal power $:N^k:$ means that we let act $k$ annihilation operators first, followed by $k$ creation ones thereafter. Important thing is, that normal powers show an appealing behavior in commutators with $a^+_\alpha$ or   $a_\alpha$.
$$[a_\alpha, :N^k:]=\,k\,:N^{k-1}:\, a_\alpha  \, , \,\,\,\,\,\,\,\,\,\,\,\,\,\,[a^+_\alpha, :N^k:]=\, -k \, a^+_\alpha :N^{k-1}: \mbox{ \ .} $$
We can hardly request  better  analogy with the continuous case. Normal ordering definitely helped here. However, there are also other issues to consider: At the end of the day, we plan to make comparison of QM and NCQM. So if we even had certain idea about how do NC wave functions look like, how would  we find the connections between them and those from the standard QM? Where should we actually make that correspondence limit $\lambda \rightarrow 0$ to find out whether the results of QM and NCQM are compatible? NC coordinates are not c-numbers, NC wave functions dependent on them are not c-number-valued by themselves either.  This is where the traces over Fock space basis come in to provide for some number-valued expressions comparable  to those from QM. And we know $|n_1,n_2\rangle\ $ are both elements of orthonormal basis in  ${\cal F}$ and eigenvectors for  ordinary powers $N^k$. These facts suggest there are c-numbers to be easily gained here. All this makes us a little reluctant to give up on ordinary powers completely. \\
\\
This dilemma between normal and ordinary is fortunately solvable without sacrificing either kind of powers - there is a way
 to relate them. All one need is certain affection for playing with combinatorics. However, before disclosing it, let us return to physics for a while. We are primarily interested in power series in radial coordinate $r=\lambda(N+1)$. So we need to consider that $\lambda$ multiplication and shift as well. As to the $\lambda$-shift, if it threatens to cause any annoyance, it is easy to dispose of the problem by simply considering powers of $\rho$ instead of $r$. That is what binomial theorem is here for. Indeed $\rho$ is more practical for the technical purposes. The multiplicative issue is more important. So here is the promised relation:
  $$
 :\rho^{n}:=\sum^{n}_{k=0} \lambda^{n-k} s(n,k)\rho^k .
 $$
 The coefficients $ s(n,k)$ are called  Stirling numbers of the first kind. The same numbers are used to express a falling factorial as a power series:
  $$
 (x)_{n} \equiv x(x-1)(x-2).....(x-n+1)=\sum^{n}_{k=0}s(n,k)x^k .
 $$
The above relation clearly implies (after realizing that $s(n,n)=1$) that normal and ordinary powers of $\rho$ coincide for $\lambda = 0$. It is expected, but welcomed nevertheless.\\
Putting it all together, a very useful formula is gained:
\be\label{nk} :(\lambda N)^k:\,|n_1,n_2\rangle\ =\ \lambda^k \frac{n!}{(n-k)!}\
|n_1,n_2 \rangle,\ \ \ n\,=\,n_1 + n_2. \ee
Note that for $k>n$ we obtain zero. Indeed, in such a case there are too many annihilation operators acting on  $|n \rangle $ for it to survive.
While we are at it, let us also add the negative power version (it may come handy sometime in the future):
\be\label{n^-k} :(\lambda N)^{-k}:\,|n_1,n_2\rangle\ =\ \lambda^{-k} \frac{n!}{(n+k)!}\
|n_1,n_2 \rangle,\ \ \ n\,=\,n_1 + n_2. \ee
These simple formulas will have  a lot to say in a good deal of our calculations. Here is some sample; at this phase it may be viewed as a mathematical toy; perhaps later it will be also of some more serious use.\\
\\
It has been mentioned that we try to keep track of analogies with QM throughout our NC dealing with the hydrogen atom problem. All the power series issue was motivated thereby. Let us bear for a while longer, this time saying a little more about the functions we encounter in QM. They are expressible via generalized hypergeometric series. This is not exactly a statement which would deserve the adjective ''specific'', since one can hardly think of any physically relevant function that could not be described by it. Exponentials, Laguerre, Legendre, etc. polynomials all belong here. So much the better, since there is some common pattern related to the kind of powers used. An example may clear up more:
\\
Suppose we had Taylor series corresponding to an exponential function :$e^{\beta \varrho}$:, where $\beta$ is an arbitrary constant and the colon marks tell us that for whatever reason all the powers are the normal ones. How would we translate it in  terms of the ordinary powers? We just have to rewrite  $:\varrho^n:$  in terms of  $\varrho^n$, which leads to
\be
\begin{array}{cl}
:e^{\beta \varrho}: & = \sum^\infty_{k=0} \frac{\beta^k}{k!}:\varrho^k:  \ = \sum^\infty_{k=0} \frac{(\beta\lambda)^k}{k!}
\frac{N!}{(N-k)!} = \\
& \\
& = \left(1+ \lambda \beta \right)^{N} = \left(  1+ \lambda \beta \right)^{\frac{\varrho}{\lambda}}. \\
\end{array}
\ee
Note that the sum is finite due to (\ref{nk}).
Considering the limit $\lambda\rightarrow 0$,  the above equation corresponds to the known Euler's formula. (If the colon marks on the left hand side cause some doubt, one just needs to keep in mind that the limit tells us to  dismiss them due to the fact that $:\varrho^n: \, \rightarrow \varrho^n$ if $\lambda\rightarrow 0$). 
\\
Many more formulas for limit relations of hypergeometric functions are known, and some of them are possible to be re-explored while playing around various NC calculations. Since there is one whole section coming later  dealing with this and related issues, the reader is invited to look them up there and the above example may be regarded as the conclusion of  our  preliminary tour of NC space.

\subsubsection{Our NC model as a deformation}
NC space is sometimes referred to as ''quantized space''. In this case, it is quite natural. Complex spaces ${\bf C}^n$ are even dimensional and possess Poisson structure $\{z_i, \bar{z_j}\} \propto \delta_{ij}$, which can be easily quantized by replacing Poisson brackets with commutators $\{. , .\}\, \to\, -i [ . , . ]$. However, we want a theory in ${\bf R}^3$ so we need to connect it with some ${\bf C}^n$ space. As it turns out, ${\bf C}^2$ is the right choice. It seems to have one more degree of freedom, which is however lost in the translation (as a Hoppf fibration). 
\noindent The points of the standard configuration space of the Coulomb problem can be parametrized in terms of two complex variables (Caley parameters)
\be \nonumber \vec{x} = (x_1, x_2, x_3) \in {\bf R}^3_0 = \{ \vec{x}\in{\bf R}^3 |\vec{x} \neq 0 \}\  \mapsto z = {z_1\choose z_2} \in {\bf C}^2_0 = \{ z\in{\bf R}^3 |z \neq 0 \} . \ee
The parametrization in question reads:
\be \label{XiZZ} x_i \ =\ \sigma^i_{\alpha \beta }\,z_\alpha^* z_\beta\ \equiv \ z^+ \,\sigma_i\,z\, \ \ i\,=\,1,2,3,\ \ \alpha ,\beta \,=\, 1,2 \,,    \ee
where $\sigma_i$, $i\,=\,1,2,3$, denote Pauli\ matrices.\\
Caley parameters can be written using Euler angles as
\begin{eqnarray}\nonumber
z_1 &=& \sqrt{r} \cos \left(\frac{\theta}{2}\right) e^{\frac{i}{2}\left(\varphi+\gamma \right)}, \\ \nonumber
\bar{z}_1 &=& \sqrt{r} \cos \left(\frac{\theta}{2}\right) e^{-\frac{i}{2}\left(\varphi+\gamma \right)}, \\ \nonumber
z_2 &=& \sqrt{r} \sin \left(\frac{\theta}{2}\right) e^{\frac{i}{2}\left(-\varphi+\gamma \right)}, \\ \label{CP}
\bar{z}_2 &=& \sqrt{r} \sin \left(\frac{\theta}{2}\right) e^{-\frac{i}{2}\left(-\varphi+\gamma \right)} . 
\end{eqnarray}
Plugging \eqref{CP} into \eqref{XiZZ} gives us spherical coordinates of $x_i$ in terms of $r,\theta, \varphi$, with the angle $\gamma$ left out.\\
Let us now consider the second order differential operator $ \partial_{z_\alpha} \partial_{z^*_\alpha }$ acting upon  functions of the form $f(\vec{x})$, $\vec{x}\,=\, z^+\,\vec{\sigma}\,z$. Simple calculation gives:
\begin{eqnarray}
\nonumber \partial_{z_\alpha} \partial_{z^*_\alpha }\,f(\vec{x})\ &=&   \ \partial_{z_\alpha} \left( \sigma^i_{\alpha \beta }\,z_\beta\,\partial_{x_i}f(\vec{x})\right) \ =\ \sigma^i_{\alpha \beta}\,\sigma^j_{\gamma \alpha}\,z_\beta\,z^*_\gamma \,\partial_{x_i}\partial_{x_j}f(\vec{x})\   \\
\label{partial_z} &=& \ z^+ z\,\partial_{x_i}\partial_{x_i}f(\vec{x}) \,,
\end{eqnarray}
where we have used the fact that the trace of Pauli matrices is vanishing $\sigma^i_{\alpha\alpha }\,=\,0$, and  that
\[ z^+ \sigma_j\,\sigma _i\,z \ =\ z^+ (\delta_{ji}{\bf 1}\,+\,i\varepsilon_{jik} \sigma_k )\,z \,.\]
The first term gives the contribution on the r.h.s. of ($\ref{partial_z}$), while the second one does not contribute since the derivatives $\partial_{x_i}$, $\partial_{x_j}$ commute. Thus we see that the standard Laplace operator can be expressed in terms of Caley parameters in the following way:
\be \Delta\,f(\vec{x})\ =\ \frac{1}{r}\,\partial_{z_\alpha} \partial_{z^*_\alpha }\,f(\vec{x})\,,\ \ \ r\,=\,  z^+ z\,.\ee
As was already mentioned ${\bf C}^2_0$ space allows introduction of elementary Poisson brackets
\be \{ z_\alpha ,z^*_\beta \}\ =\ - i\,\delta_{\alpha\beta}\,,\ \  \{ z_\alpha ,z_\beta \}\ =\ \{ z^*_\alpha ,z^*_\beta \}\ =\ 0. \, \ee
It is easy to check that they lead to the following  relations for coordinates $x_i$, $i\,=\,1,2,3$:
\be \{ x_i, x_j\}\ =\ \varepsilon_{ijk}\,x_k\,, \ \ i,j,k\,=\,1,2,3\,. \ee
We point out that the Laplace operator can be expressed also in the following way:
\be\label{z-formalism} \Delta\,f(\vec{x})\ =\ \frac{1}{r}\,\{z^*_\alpha, \{z_\alpha,f(\vec{x})\}\}\,,\ \ \ r\,=\,  z^+ z\,.\ee \\
Based on what has been given above,  the space ${\bf R}^3_0$ is obviously a {\it Poisson manifold} (since it is possible to introduce the relevant bracket relations there). In \cite{kontsevich} it was proved that any Poisson manifold can be {\it deformed} or {\it quantized}, i.e., one can introduce an associative noncommutative star-product for any two functions $f(\vec{x})$ and $g(\vec{x})$:
\be (f\star g)(\vec{x})\ =\ f(\vec{x})g(\vec{x})\ +\ i\lambda \,\omega (f,g)(\vec{x})\ +\ o(\lambda^2)\, , \ \ee
where the first term denotes the point-wise product of $f$ and $g$, $\lambda$ is a {\it deformation parameter} and the full expression is a formal power expansion in $\lambda$. The symbol $\omega (f,g)$ is a bilinear expression in $f$ and $g$ and their derivatives such that the star-commutator of $f$ and $g$ is given as
\be  [f,g]_\star \equiv f\star g\,-\, g\star f\ =\ 2i\,\lambda \,\{f,g\}\ +\,\ o(\lambda^2)\ . \ee
This condition and the requirement of associativity allows to (non-uniquely) reconstruct the associative star-product on any Poisson manifold which defines the deformed manifold ${\bf R}^3_*$.\\
In our particular case the deformed space ${\bf R}^3_*$ can be obtained simply by the replacement
\be z_\alpha \ \mapsto \sqrt{\lambda}\,a_\alpha \,,\ \ \ z^*_\alpha \ \mapsto \sqrt{\lambda}\,a^\dagger_\alpha \,, \ee
which is equivalent to the (particular) star-product manifold ${\bf R}^3_*$.\\
The star-product approach can be useful for the investigation of commutative limits $\lambda \to 0$. The deformed space ${\bf R}^3_*$ is formally equivalent to the fuzzy space ${\bf R}^3_\lambda $ (the products of polynomials in $x_i$ are isomorphic). However, as the $\lambda$-expansions are only formal, a situation may arise when some issues need to be attended to in the commutative limit. Therefore, in what follows we have  restricted ourselves to the operator realization of the fuzzy space  ${\bf R}^3_\lambda $.

\section{Hydrogen atom in standard QM}
This chapter is not supposed to provide the reader with something new; the ideas are certainly familiar to them. However, sometimes details are to be reminded, and the following  is written under the assumption that turning a few pages  back  to this chapter when the need arises disturbs the comfort of the reader less than searching for albeit notoriously known, but too-long-to-be-kept-in-memory formulas elsewhere, with the additional tedious task to adjust the notation. If there is a wish for more detail, then for example the standard books like \cite{Schiff} or \cite{LL} are to be consulted.\\
\\
We will briefly mention some aspects of both ''differential'' (Schr\"{o}dinger's) and ''algebraic'' (Pauli's) way of addressing the problem here.
Here is in a few lines  Schr\"{o}dinger's idea:\\
\\
The  ''rules of the game'' are described by the  equation
\be\label{Schr0} - \frac{\hbar^2}{2m_e}\,\Delta \psi({\bf
x})\,-\,\frac{q}{r} \psi({\bf x})\,=\,E \psi({\bf x}),\ \ r =
|{\bf x}| > 0 \,. \ee
This is a game with probabilities; they are connected with wave functions from the Hilbert space ${\cal H}_0$ specified by the norm
\be\label{norm0} \|\psi\|^2_0\,=\,\int\,d^3{\bf x}\ |\psi({\bf
x})|^2 \,. \ee
Important (for the future reference) note about considered potential:
In QM it is just that Coulomb potential which, in the usual (commutative) setting, is a solution of the Laplace equation vanishing at infinity:
\be\label{coul0} \Delta U(r)\,=\,0\ \ \Rightarrow\ \ U(r)\ =\
-\,\frac{q}{r}\,. \ee
In a Gaussian system of units, $q$ is a square of electric charge $q\,=\,\pm\,e^2$: $q>0$ or $q<0$ corresponding to Coulomb attraction or repulsion respectively. There is an obvious rotational invariance of this potential and hence the whole Hamiltonian. It seems that Nature likes to reward (often with analytic solutions) those who pay respects to symmetry when meeting one. So it is reasonable to write the sought-for wave function in a  separable form, notice that angular momentum operator accounts for a good deal of what the free part of the Hamiltonian does, remind ourselves that mathematicians had been so kind as to provide us with the related eigenfunctions long ago, and write
\be\label{psi0} \psi({\bf x})\,=\,R_j(r)\,H_{jm}({\bf x}),\ \
H_{jm}({\bf x}) \,\sim\,r^j\,Y_{jm}(\vartheta,\varphi)\,, \ee
where $j$ is the quantum number corresponding to the angular momentum, while $m$ does this job for its third component and $Y_{jm}(\vartheta, \varphi)$ is the standard spherical function.\\
The equation for the radial part is
\be\label{rSchr00}  r\,R^{\prime\prime}_j(r) + (2j+2) R^\prime_j(r)
+ 2 \alpha R_j(r)\,=\,-\,k^2 r\,R_j(r)\,, \ee
where $\alpha = m_e q/\hbar^2$ and $k^2 = 2m_e E /\hbar^2$ have been used to avoid too much constant clutter.
The parameter $\alpha$ is simply related to the H-atom Bohr radius $a_0 = \hbar^2/(m_e e^2) = |\alpha|^{-1} $. The solutions of (\ref{rSchr00}) are given in terms of solutions of the confluent hypergeometric equation (see, e.g. \cite{Schiff}).
\be\label{solQM} R_j(r)\,=\, e^{\pm\sqrt{-2E}r} \, _1F_1\left(  j+1 \pm \frac{\alpha}{\sqrt{-2E}} \,,\, 2j+2 \,,\,
                                                                     \mp 2\sqrt{-2E} r \right). \ee
(We often  put  $m_e/\hbar^2 = 1$ so that we  do not have to worry about confusing $q$ with $\alpha$.)
$ _1F_1 (a,b,x)$ is confluent hypergeometric function,  regular at the origin. The features of the solution  depend crucially on the sign of the energy, so there are two main areas to explore in QM: bounded states labeled by negative discrete energy eigenvalues, and  scattering states related to positive continuous spectrum.\\
\\
\textbf{Discrete spectrum of hydrogen}\\
\\
Hydrogen accounts for a better part of the ''known'' matter in the universe. We know so from examination of its unique signature (the red H-alpha  of Balmer series being probably the most noticeable character) in spectral lines.  These are easy to calculate once one has the solution of Schr\"{o}dinger equation and knows how to interpret the fact that energy is negative - the wave function has to be decently integrable, therefore the $_1 F_1$ function in (\ref{solQM}), which has been looked at as a power series so far, has to reduce to a polynomial, what happens if the first argument of $_1 F_1$ is a non-positive integer. This requirement gives the formula for discrete energy levels:
\be\label{ener0} E_n\,=\,
-\frac{m_e\,e^4}{2\hbar^2 n^2}\ ,\ \ \ n\,=\,j+1,\,j+2,\,\dots\,. \ee
There is something peculiar about this spectrum, namely its degeneracy. It is not only degenerate with respect to quantum number $m$, what happens whenever we deal with rotationally invariant potential; in this case the degeneracy is also with respect to $j$.  So there seem to be more symmetry waiting behind and worth exploring.\\
\\
\textbf{Scattering}\\
\\
The information about this process is concisely stored in the $S$-matrix. It can be gained by rewriting (\ref{solQM}) via the fundamental system of solutions of the equation (\ref{rSchr00}). These two correspond to in- and out- going spherical waves, and by comparing factors multiplying the two parts (for each entering  momentum separately), one finds the contribution from each of the  partial waves (here labeled by $j$). Let us see it more closely:
\\
Radial part of solution with the angular momentum $j$ and energy $E > 0$, regular in  $r\to 0$ is given as
\be\label{schiff1} R_j\ =\ e^{ikr}\,_1F_1\left(j+1-i\frac{\alpha}{k},\,2j+2,\,-2ikr \right)\,,\ \ \ k\, =\, \sqrt{2E}\,>\,0\,.\ee
The solution is real and for  $r\rightarrow  \infty$ it can be written as a sum of two complex conjugated parts. In the following formula a real factor common for both parts is left out, having no influence on the S-matrix.
\be\label{schiff2} R_j\ \sim \ \frac{i^{j+1}}{\Gamma (j+1 +i\frac{\alpha}{k})}\,e^{ikr +i\frac{\alpha}{k} \ln(2kr)}\ +\ \frac{i^{-j-1}}{\Gamma (j+1 -i\frac{\alpha}{k})}\,e^{-ikr -i\frac{\alpha}{k} \ln(2kr)} \,.\ee
The $S$-matrix for the $j$-th partial wave is defined as the ratio of the  $r$-independent factors multiplying the exponentials with the kinematical factor $(-1)^{j+1}$ left out.
\be\label{schiff3} S_j(E)\ =\ \frac{\Gamma (j+1 -i\frac{\alpha}{k})}{\Gamma (j+1 +i\frac{\alpha}{k})}\,,\ \ \ \ E = \frac{1}{2}\,k^2 > 0 \,.\ee
$S$-matrix has poles just in those points of the complex plane which correspond to  the bound states, so the way this works makes one really appreciate the charm of complex analysis.\\
In the case of an attractive potential ($\alpha > 0$)  the $S$-matrix (\ref{schiff3}) has poles in the upper complex  $k$-plane for
\be\label{schiff4} k_n\ =\ i\,\frac{\alpha}{n}\,, \ \ \ n\, =\, j+1,\ j+2,\ \dots \ee
The wave function (\ref{schiff1}) is integrable  for $k\,=\,k_n$:
\be  R_{nj}\ =\ e^{-\alpha\frac{r}{n}}\,_1F_1 \left(n,\,2j+2,\,2 \alpha\frac{r}{n} \right)\,.\ee
It is obvious that the energy levels correspond to the poles of the $S$-matrix:
\be\label{schiff5}  E_n\ =\ - \frac{\alpha^2}{2n^2}\,, \ \ \ n\, =\, j+1,\ j+2,\ \dots\ee
The fact that the bound states energies appear as poles of the S-matrix on the upper  imaginary half-axis in  the complex $k$-plane is a deep consequence of causality principle in QM, see \cite{Schiff}, \cite{LL}.
\\
\\
\textbf{More symmetry}\\
\\
The degeneracy mentioned above deserves  attention, since it indicates more symmetry and  the penalty for a ''so-what''-attitude about this would be missing some beautiful aspect of the problem. 
The following few remarks on the subject have been well known  since 1926, when  W. Pauli published his paper on the subject, \cite{Pa}.  He found QM analog of the classical Laplace-Runge-Lenz (LRL) vector, recognized the  ''hidden'' dynamical symmetry of the problem, and consequently arrived at the spectrum of hydrogen, without the knowledge of the explicit solution of  Schr\"odinger equation. Here is a sketch of the motivation and the method:\\
\\
It has been known that there are  conserved vectors for problems with Newton's gravitational force law, accounting for closed orbits. Coulomb and gravitational forces are so  much alike that it was natural to look for deeper analogies, the hydrogen atom being the first thing to begin with. Of course due to the need of QM treatment it is not appropriate to speak of orbits or perihelia, but still. The motivation is quite straightforward - there are two three-component vectors (angular momentum $\vec{L}$ and the LRL vector $\vec{A}$) conserved for a planet orbiting the Sun, we can try to find the quantum mechanical analogs for electron ''orbiting''  proton. \\
\\
It turned out that the LRL vector can be found among the hermitian operators acting in the considered Hilbert space in an almost complete analogy with the classical case. One subtlety occurs (due to the ordering issues one often encounters when dealing with noncommuting objects) - the cross product needs to be properly symmetrized, resulting into
\be
A_k = \frac{1}{2}\varepsilon_{ijk} (L_i v_j+ v_j L_i ) + q\frac{x_k}{r},
\ee
where $v_j = - \frac{i\hbar}{m_e}\, \partial_j $ stands for velocity operator. Vital information  is that the operators $L_i$ and $A_i$ commute with  the Hamiltonian, i.e. are conserved with respect to the  time evolution, and as to their mutual commutation relations, there would be pretty good views of them forming a closed algebra, if it were not for the commutator $[A_i,A_j] = -2 i \varepsilon_{ijk} HL_k$. This is not a Lie algebra relation, due to the presence of $H$ on the r.h.s. . However, restricting ourselves to ${\cal H}_E$ spanned by eigenvectors of the Hamiltonian corresponding to certain energy, we can replace $H$ by its eigenvalue $E$, which is a c-number.
For $E<0$ we recover the $so(4)$ algebra relation $[A_i,A_j] = -2 i \varepsilon_{ijk} EL_k$, for $E>0$ the relevant algebra is $so(3,1)$.
Besides enabling the algebra to close (possibly with $A_i$ being rescaled suitably), the relation between  energy and the symmetry generators has been somewhat clarified. This, together with the theory related to the relevant Casimir operators generates the energy spectrum.\\
To sum up, the components of $\vec{L}$ and $\vec{A}$  form a representation of generators of a dynamical group on the subspace  $\mathcal{H}_E$. There are many subspaces $\mathcal{H}_E$ for any admissible eigenvalue $H$ for which $\vec{L}$ and $\vec{A}$ are given as hermitian operators, i.e., $\mathcal{H}_E$ is a carrier space of unitary representation of the dynamical group. As to the classification of these representations, the Casimir operators are of vital importance. Here we will leave out the theory behind, we just provide an example: for the Coulomb attractive potential and negative energies the symmetry group is $SO(4)$, its Casimir operators have integer eigenvalues that are related to energy through the generators;  a discrete energy spectrum is to be expected. Pauli  worked it out and obtained the correct formulas for the hydrogen spectrum even before Schr\"{o}dinger.\\
\\
This amount of recollection related to the standard QM version of handling the hydrogen atom is perhaps enough, so we may move on to search for analogies in the NC case.\\

\section{NC hydrogen atom: Schr\"{o}dinger-like differential  approach}
Having the introductory parts behind us, we can finally move on to our task to solve hydrogen atom problem in the NC setting and compare the results with those from standard QM. ''NC setting'' means there are modified versions of configuration space (${\bf R}^3_0  \rightarrow {\bf R}^3_\lambda $) and Hilbert space of states (${\cal H} \rightarrow {\cal H}_\lambda $) to be considered.\\
\\
Just to avoid not seeing the forest because of the trees, let us remind that we do this to explore the possibility that there might be some aspects of space that the model ${\bf R}^3_0 $ fails to catch - at least to show  that  the by experiments so well approved  quantum theory may be consistent with such an assumption.  This ''bigger picture'' will be recalled again and  the results discussed over in the conclusions, but those are  long way from here yet.\\
\\
We are after analogies; so let us try to find those corresponding to all the objects appearing in the previous chapter (about ordinary QM). Maybe not quite in the same order, the main ideas remain nevertheless. \\
This chapter focuses mainly on the ''differential'' approach \`{a} la  Schr\"{o}dinger, the following one will be more in Pauli's algebraic style. Needless to say, the results should better agree ...
\subsection{Search for NC Schr\"{o}dinger equation}
First let us talk about physical quantities - after all, whatever the model, it always boils down to them. In analogy with the ordinary QM case, they will be represented by hermitian operators acting on Hilbert space of states  ${\cal H}_\lambda $. So the next few paragraphs should serve as an introduction of some of them, maybe together with some features of the related eigenfunctions.\\
We have got a bit tricky situation here, since the NC coordinates (in ${\bf R}^3_\lambda $, on which NC wave functions $\Psi$ are defined) and the NC wave functions $\Psi$ (elements of ${\cal H}_\lambda $), are composed of operators acting in the auxiliary Fock space. Now, we want to define operators acting on those $\Psi$ from ${\cal H}_\lambda $.  It has been suggested that the following notation should prevent  confusion: to leave the NC coordinates and the NC wave functions $\Psi$  as they are,  and to denote the operators acting on $\Psi$ with a hat.
\\
Now to the matter at hand - which operators will be needed for treating the hydrogen atom problem, and what do we expect from their eigenfunctions? Obviously a prominent role will be played by the Hamiltonian, since it appears directly in the Schr\"{o}dinger equation; to keep the analogy as close as possible we expect it to consist of the free part (up to some multiplicative constant some  NC version of Laplacian) and the potential, which should be a solution to the Laplace equation - in this instance, the NC analog thereof. Since we want to employ the spherical symmetry, NC representatives of rotation generators will be necessary, and  the corresponding eigenfunctions are needed as well. There will be more operators coming later; for now let us see what can be done with these mentioned.
\subsubsection*{Orbital momentum in ${\cal H}_\lambda $}
\noindent In ${\cal H}_\lambda $ we define
orbital momentum operators, the generators of rotations $\hat{L}_j$, $j\,=\,1,2,3$, as follows
\be\label{L0} \hat{L}_j\,\Psi\ =\ \frac{1}{2}\,[a^+\,\sigma^j\,a,\Psi] \, = \frac{1}{2\lambda}\,(x_j \Psi -\Psi x_j),\,\, \ j\,=\,1,2,3\,.\ee
They are hermitian (self-adjoint) operators in ${\cal H}_\lambda $ and obey the commutation relations
\be [\hat{L}_i,\hat{L}_j] \Psi\,\equiv\,(\hat{L}_i \hat{L}_j\,-\,\hat{L}_j \hat{L}_i)
\Psi\,=\, i\,\varepsilon_{ijk}\, \hat{L}_k \Psi\,.\ee
The  eigenfunctions $\Psi_{jm}$, $j =
0,1,2,\,\dots\,, \,m = -j,\,\dots,\,+j$, satisfying
\be\label{Hjm1} \hat{L}^2_i\,\Psi_{jm}\ =\ j(j+1)\,\Psi_{jm},\ \ \
\hat{L}_3\,\Psi_{jm}\ =\ m\,\Psi_{jm}\ ,\ee
are given by the formula
\be\label{harm2} \Psi_{jm}\ =\ \sum_{(jm)}\
\frac{(a^\dagger_1)^{m_1}\,(a^\dagger_2)^{m_2}}{m_1!\,m_2!}\ R_j(\varrho
)\ \frac{a^{n_1}_1\,(-a_2)^{n_2}}{n_1!\ n_2!}\,, \ee
where $\varrho = \lambda a^\dagger_\alpha a_\alpha = \lambda N$.
The summation goes over all nonnegative integers satisfying
$m_1+m_2 \,=\,n_1+n_2\,=\,j$, $m_1-m_2-n_1+n_2\,=\,2 m$. Thus
$\Psi_{jm} = 0$ when restricted to the subspaces ${\cal
F}_n$ with $n < j$ (what corresponds to the fact that in the standard QM the
first $j-1$ derivatives of $\Psi_{jm}$ vanish at the
origin). For any fixed $R_j(\varrho)$ equation
(\ref{harm2}) defines a representation space for a unitary irreducible
representation with spin $j$. \\
It has been mentioned that there is an amusing analogy present in this instance: in the basic undergraduate courses of quantum mechanics the angular part of the solution is often dealt with by simply quoting some mathematical ''guide book'' and copying the results therefrom. Here something similar is being done,  the angular part of the solution is taken from \cite{GP2}.
\subsubsection*{Radial part and normalization in ${\cal H}_\lambda $}
\noindent The two wave functions $\Psi_{jm}$ and $\tilde{\Psi}_{j'm'}$, with
$(j,m) \neq (j',m')$ and arbitrary factors $R_j(\varrho)$ and
$\tilde{R}_{j'}(\varrho)$, are orthogonal in ${\cal H}_\lambda $. Thus, when evaluating the norm of $\Psi_{jm}$, it is sufficient
to calculate  $\|\Psi_{jm}\|^2\,=\,\|\Psi_{jj}\|^2$ (this equality
follows from the rotational invariance of the norm in question):
\be\label{psi2} \|\Psi_{jm} \|^2\ =\ 4\pi \lambda^3\,\sum_{n=j}^\infty
\,\sum_{k=0}^n (n+1)\ \langle k,n-k|\Psi_{jj}^\dagger
\,\Psi_{jj}\,|k,n-k\rangle\,.\ee
We benefit from the fact that $\Psi_{jj}$ has a simple form
\be\label{psijj} \Psi_{jj}\ =\ \frac{\lambda^j}{(j!)^2}\
(a^\dagger_1)^j \,R_j(\varrho)\,(-a_2)^j \,.\ee
The matrix element we need to calculate is
\[ \langle k,n-k|\ (a^\dagger_2)^j \,R_j(\varrho)\,a^j_1\
(a^\dagger_1)^j\,R_j(\varrho)\,a^j_2\ |k,n-k\rangle \]
\be\label{mel}  =\ \frac{(k+j)!(n-k)!}{k!\,(n-j-k)!}\
|{\cal R}_j(n-j)|^2\,, \ee
where
\be R_j(n)\,=\,\langle k,n-k|R_j(\varrho)
|k,n-k\rangle \ee
(the expression on the r.h.s. is $k$ - independent). Inserting (\ref{psijj}),
(\ref{mel}) into (\ref{psi2}) and using the identity (see \cite{Pr})
\[ \sum_{k=0}^{n-j}\ {{k+j}\choose j}\ {{n-k}\choose j}\ =\
{{n+j+1}\choose {2j+1}}\,,\]
we obtain
\be\label{Psi2} \|\Psi_{jm} \|^2\ =\ \frac{4\pi
\lambda^{3+2j}}{(j!)^2}\ \sum_{n=0}^\infty\ (n+j+1)\
{{n+2j+1}\choose {2j+1}}\ | R_j(n)|^2 \,. \ee
This expression represents, up to an eventual normalization, the
square of a norm of the radial part of the wave function.
\subsubsection*{NC analog of Laplace operator in ${\cal H}_\lambda $}
\noindent The NC analog of the usual \textbf{Laplace operator }is
\be\label{Lapl} \hat{\Delta}_\lambda\,\Psi\ =\ -\,\frac{1}{\lambda r}\
[\hat{a}^+_\alpha,\,[\hat{a}_\alpha,\,\Psi]]\ =\ -\,\frac{1}{\lambda^2
(N+1)}\ [\hat{a}^+_\alpha ,\,[\hat{a}_\alpha ,\,\Psi]]\ .\ee
Choosing this NC version of Laplacian is crucial for obtaining a decent Schr\"{o}dinger equation. This choice is motivated by the following facts: (i) A double commutator
is an analog of a second order differential operator, (ii) the factor
$r^{-1}$ guarantees that the operator $\hat{\Delta}_\lambda$ is self-adjoint in ${\cal H}_\lambda $, and finally, (iii) the factors
$\lambda^{-1}$, or $\lambda^{-2}$ respectively, guarantee the correct
physical dimension of $\hat{\Delta}_\lambda$ and its correct commutative
limit.\\
The mathematical ansatz can be also inferred from (\ref{z-formalism}). Anyway, calculating the action of (\ref{Lapl}) on  $\Psi_{jm}$ given in
(\ref{harm2}) we can check whether the postulate (\ref{Lapl}) is a
reasonable choice.\\
\\
The operator $R_j(\varrho)$ in (\ref{harm2}) can be represented as a normal ordered expansion of an analytic function ${\cal R}_j(\varrho)$ :
\be\label{harm3} R_j(\varrho)\ = \,:{\cal R}_j(\varrho): \, = \sum_k c^j_k\,:
\varrho^k:\ =\ \sum_k c^j_k \lambda^k\,\frac{N!}{(N-k)!}. \ee
The last equality follows from what has been explained in the chapter devoted to NC space.\\
Since $:N^k:\,
|n_1,n_2\rangle\,=\,0$ for $k\,>\,n_1 + n_2$, the summation in
(\ref{harm3}) is effectively restricted to $k \le n$ on any
subspace ${\cal F}_n$.\\
\\
The following formula follows from commutation relations (\ref{A2 }); the proof is given in the Appendix:
\begin{eqnarray} \nonumber 
[\hat{a}^\dagger_\alpha ,\,[\hat{a}_\alpha,\,\Psi]]\ &=&\ \lambda^j\ \sum_{(jm)}\
\frac{(a^\dagger_1)^{m_1}\,(a^\dagger_2)^{m_2}}{m_1!\, m_2!} \\
& \times & :[ - \lambda \varrho\,{\cal R}''(\varrho )\,-\,2(j+1)\lambda\,
{\cal R}'(\varrho )]:\
\frac{a^{n_1}_1\,(-a_2)^{n_2} }{n_1!\ n_2!}\,.
\end{eqnarray}

Here ${\cal R}'(\varrho )$ denotes the usual derivative
 defined as:
\be\label{derivative0} {\cal R}(\rho )\,=\,\sum_{k=0}^\infty\,c_k\,\rho^k\ \ \ \Rightarrow
\ \ \  {\cal R}'(\varrho)\,=\,\sum_{k=1}^\infty\, k\,c_k\,\varrho^{k-1}\,,\ee
and ${\cal R}''(\varrho )$ is defined as the derivative of ${\cal R}'(\varrho )$.
Thus, the prime corresponds exactly to the usual derivative
$\partial_\varrho$. In the commutative limit
$\lambda\,\rightarrow\,0$ operator $\varrho$ formally reduces to the usual
radial $r$ variable in ${\bf R}^3$, and we see that $\hat{\Delta}_\lambda$ just
reduces  to the standard Laplace operator in ${\bf {R}}^3$. \vskip0.2cm
\subsubsection*{The potential term in  ${\cal H}_\lambda $}
\noindent The operator $\hat{U}$ corresponding to a  central potential is defined simply as the
multiplication of the NC wave function by $U(r)$:
\be (\hat{U} \Psi)(r)\ =\ U(r)\,\Psi \ =\ \Psi\,U(r)\,.\ee
Since any term of $\Psi\,\in\,{\cal H}_\lambda $ contains the same number of
creation and annihilation operators (any commutator of such a term with $r$ is zero), the left and right multiplications by $U(r)$ are equal.
\\
\\
In the commutative case the Coulomb potential is a radial solution
of the Laplace equation (\ref{coul0}) vanishing at infinity. Due to our
choice of the NC Laplace operator $\Delta_\lambda$ the
 NC analog of this equation is
\[  \hat{\Delta}_\lambda\,U(r)\ =\ 0\ \ \ \Leftrightarrow \ \ \
[\,\hat{a}^\dagger_\alpha,\, [\,\hat{a}_\alpha\,,U(N)\,]\,]\ =\ 0\,. \]
Last equation can be rewritten as a simple recurrent relation
\be\label{rec} (N+2)\,U(N+1)\,-\,(N+1)\,U(N)\ =\
(N+1)\,U(N)\,-\,N\,U(N-1)\,.\ee
Putting $(M+1)\,U(M)\,-\,M\,U(M-1)\,=\,q_0$, $U(0)\,=\,q_0\,-\,\frac{q}{\lambda}$
and summing up the first equation over $M\,=\,1,\,\dots\,N$, we
obtain the general solution:
\be\label{nccoul1} U(N)\ =\ -\,\frac{q}{\lambda\,
(N+1)}\ +\ q_0\ =\ -\,\frac{q}{r}\ +\ q_0\,,\ee
where $q$ and $q_0$ are arbitrary constants ($\lambda$ is
introduced for the future convenience). Thus the NC analog of the Coulomb potential
vanishing at infinity is given by
\be\label{nccoul}(\hat{U}\Psi)(r)\ =\ -\frac{q}{r}\,\Psi\,. \ee
We see that the $\frac{1}{r} = \frac{1}{\lambda (N+1)}$ dependence  of the NC Coulomb potential is inevitable, and that our NC analog of radial coordinate has been chosen suitably.
\subsubsection*{NC Hamiltonian}
\noindent Now we can put together the left hand side of the NC version of the stationary Schr\"{o}dinger equation:
\be\label{NCHamiltonian} \hat{H} \Psi\ =\ \frac{\hbar^2}{2m_e\lambda r}\ [\hat{a}^+_\alpha ,\,
[\hat{a}_\alpha ,\,\Psi]]\ -\ \frac{q}{r}\, \Psi \ .\ee
To make the notation more concise, we often choose the units so that $m_e=1$, $\hbar =1$.
\subsection{Solving NC Schr\"{o}dinger equation}
Based on (\ref{Lapl}) and (\ref{nccoul}) we postulate the NC analog of the
Schr\"odinger equation (\ref{Schr0}) with the Coulomb potential in ${\bf R}^3_\lambda$ as
\be\label{ncsch1} \frac{\hbar^2}{2m_e\lambda r}\,[\hat{a}^\dagger_\alpha ,
[\hat{a}_\alpha ,\Psi]] - \frac{q}{r}\, \Psi = E\,\Psi\ \ \
\Leftrightarrow\ \ \ \frac{1}{\lambda}\,[\hat{a}^\dagger_\alpha,
[\hat{a}_\alpha ,\Psi]] - 2\alpha\,\Psi = k^2\,r \Psi\,,\ee
\newline
These two equations are proved in the Appendix (the first one has been mentioned already):
\[ [\hat{a}^\dagger_\alpha ,\,[\hat{a}_\alpha ,\,\Psi_{jm}]]\ =\ \sum_{(jm)}\ \dots\
:[ -\varrho \lambda \,{\cal R}_j^{\prime\prime}\,-\,2(j+1) \lambda \,{\cal R}_j^\prime]:\ \dots \ ,\]
\be\label{appen} r\,\Psi_{jm}\ =\ \sum_{(jm)}\ \dots\
:[(\varrho+\lambda j+\lambda )\,{\cal R}_j\,+\,\lambda\,\varrho\,{\cal R}_j^\prime]:
\ \dots \ ,\ee
where ${\cal R}_j \equiv {\cal R}_j(\varrho )$ and similarly for derivatives. The dots on
the left and right in (\ref{appen}) denote the products in $\Psi$ containing
respectively creation and annihilation operators together with the factor
$\lambda^j$, that represent the angular dependence of $\Psi$
and remain untouched as the operators in question are
invariant under rotations. Inserting (\ref{appen}) into (\ref{ncsch1}) we obtain the NC
analog of the radial Schr\"odinger equation:
\be\label{radNC} : \varrho\,{\cal R}_j^{\prime\prime}
\,+\,[k^2 \lambda \varrho + 2j+2]\,{\cal R}_j^\prime\,+\,[k^2 \varrho\,+
\,k^2 \lambda (j+1)\,+\,2\alpha]\,{\cal R}_j:\ =\ 0\,.\ee
(The constants $\alpha = m_e q/\hbar^2$ and $k^2 = 2m_e E /\hbar^2$ are defined as they have been in QM case, and as it has been said there, we often put $m_e /\hbar^2 = 1$, so that confusing $q$ and $\alpha$ does not matter.) \\
\\
We claim (\ref{radNC})  to be an NCQM analog of the usual radial Schr\"{o}dinger equation (\ref{rSchr00}) known from QM. There definitely is a  resemblance; in the limit $\lambda \rightarrow 0$ the terms in (\ref{radNC}) proportional to $\lambda $  representing the NC corrections disappear. Considering the same limit we see that presence of the colon marks denoting the normal ordering should not worry us either; recall that for zero $\lambda$ it makes no difference whatsoever whether we care for the ordering or not.  This is a good news to start with, leaving us, however, with the task to solve (\ref{radNC}) for nonzero $\lambda$, which means that both the extra terms proportional to $\lambda$ and the normal ordering are to be taken at a face value. If it was not for the ordering issues, the solution would be quite straightforward - the extra terms would mean just adding some more work needed to complete the calculation, but it is known how to solve the problems of this kind. In fact this is precisely what we are going to do:
We associate the following ordinary differential equation to the
mentioned operator radial Schr\"odinger equation (\ref{radNC}):
\be\label{rad} \varrho\,{\cal R}_j^{\prime\prime}
\,+\,[k^2 \lambda \varrho + 2j+2]\,{\cal R}_j^\prime\,+\,[k^2 \varrho\,+
\,k^2 \lambda (j+1)\,+\,2\alpha]\,{\cal R}_j\ =\ 0\,,\ee
with $\varrho$ being real variable, and we will solve this one. But how come we expect this step to be of any use to us, when we actually \textit{do} have to care about the ordering? One should notice the following: whatever appears in the equations, it can be expressed in terms of powers in $\varrho$ - it is just that they are normal powers in one case and the usual powers in the other. Next, we have some operators in both equations; normal derivatives in one case and the usual ones in the other. The key information is, that the derivative defined in (\ref{derivative0}) acts on the normal powers just like a carbon copy of usual  derivative, see (\ref{A2 }).
\\
Now bearing this in mind, we expect $R \, =\, :{\cal R}:$, the solution of (\ref{radNC}), to be of the same form as $\cal R $, the solution of (\ref{rad}), except for the nature of the powers involved. So a brief summary goes like this: The solution of (\ref{rad}) with all the usual powers replaced by the normal ones is the solution of (\ref{radNC}).  However, the form of the solution is not the best one yet. We have already mentioned the relation between the equation given by QM (\ref{rSchr00}) and (\ref{radNC}), the one supplied by NCQM. Of course we would like to compare the corresponding solution as well, but this is rather a difficult task as long as we have normal powers in the first  and the usual ones in the latter one. Fortunately we have the formula (\ref{nk}) relating $:\varrho^n:$ and $\varrho^n$, it has been mentioned in the chapter devoted to NC space.
%
%
All we need is to rewrite $:{\cal R} :$ using those relations. Then the above mentioned comparison of QM and NCQM will be obtained.
\\
Perhaps is has been made clear enough what is to be done, so let us get started with the solution of equations (\ref{radNC}) and (\ref{rad}). Some mathematic theory is to be studied here, useful enough to deserve a subsection on its own.
\subsubsection{Higher transcendental functions}
\noindent This section is an attempt to provide some pieces from the theory of ''higher transcendental function'', focusing on (generalized) hypergeometric series. Obviously we cannot contain it all here, we just cherry-pick some parts needed for that follows.The theory is rich, and exploring it is extremely instructive when dealing with differential equations in physics. It is also a fine place to get lost among  appealing, useful, but overwhelmingly numerous identities and symmetries (not to mention subtleties that should be attended to). Fortunately there are guide books written by mathematicians like \cite{Bat} and \cite{Abr}, which help a physicist to survive the trip to this pretty mathematical wilderness, and to leave with useful tools to handle whatever problem it had been that motivated the journey therein. (Besides those tools, one may also leave with increased respect for the authors of the guide books, and with own ego slightly bruised.) Part of the information provided here is known (at least among mathematicians) for a long time, the person who should be given the due credit is sometimes indicated by the name of the considered object. The by us most frequently used source has been \cite{Bat} or \cite{Abr}.\\
\\
First of all, let us settle on some notation, which is going to be used frequently. Caution is recommended, because some authors prefer different conventions. \\
$(a)_m$ is the  Pochhammer symbol: $(a)_0\,=\,1 $ and
$$ (a)_m = a (a+1)\ \dots\ (a+m-1)\,=\,\frac{\Gamma(a+m)}{\Gamma(a)}\,, \ \ \ \ m\ =\ 0, 1,2,\dots \ . $$
Generalized hypergeometric series is an object on the right hand side of the below equation. If it converges, we can call it generalized hypergeometric function, and use the symbol on the left-hand side
\be\label{hpg_gen}  _p F_q (a_1,\,...,\,a_p ; b_1,\,...,\,b_q ; x)\ =\ \sum^\infty_{m=0}\ \frac{(a_1)_m\,...\,(a_p)_m }{(b_1)_m\,...\,(b_q)_m }\, \frac{{x}^m}{m!}\,, \ee
Considering the definition of Pochhammer symbol, the series may reduce to a polynomial if some of the ''upstairs arguments'' is a non-positive integer $a_i= -n$,  (supposing that  any of the ''downstairs arguments'' $b_j$ does not spoil it by being equal to a non-positive integer $-k$, $k<n$; in such a case the series  would diverge.)\\
\\
The cases $_1 F_1$ and  $ _2 F_1$, called confluent and (Gauss) hypergeometric functions respectively, are the  most relevant for our purposes, so let us pay them more attention. We will also mention Bessel functions below.\\
\\
\textbf{Confluent hypergeometric equation}\\
\\
The confluent hypergeometric equation reads
\be\label{confl} x\,y''(x)\,+\,(c-x )y'(x) \,-\,a\,y(x)\ =\ 0\,. \ee
Solution of (\ref{confl}) regular at the origin is just the $_1 F_1$  mentioned above:
\be\label{y1}  _1F_1 (a;c;x )\ =\ \sum^\infty_{m=0}\ \frac{(a)_m }{(c)_m}\, \frac{{x}^m}{m!}\,. \ee
Sometimes  it is called ''Kummer's confluent hypergeometric function'', and the notation varies according to the taste of the author and the part of the alphabet yet not appearing in their calculations. Symbols $\phi$ and $M$ are frequently used.
The fundamental system of solutions of (\ref{confl}) consists of the following functions:
\be\label{y5-7} U(a,c;x )\,,\ \ \ \mbox{and\,} \ \ \ e^{x }U(c-a,c;-x )\,,\ee
 $U(a,c;x )$ is often called  ''Tricomi's confluent hypergeometric function'', and symbol $\psi$ is sometimes used to denote it. The following expression of $U$ is used when asymptotic expansion for $ x \,\to\,\infty$ \, is needed:
\be\label{psi}U(a,c;x )\ =\ \sum^\infty_{m=0} (-1)^m\frac{(a)_m (a-c+1)_m} {m!}\, {x}^{-a-m}\,. \ee
Every solution (e.g. also (\ref{y1}), the one regular at the origin) can be expressed as a suitable linear
combination of (\ref{y5-7}). This possibility is useful when treating the scattering processes.\\
\\
There is a vast number of identities related to the above introduced functions. The following one, called Kummer identity, is of particular interest to us:
\be\label{Kid} e^{-x/2}\,_1 F_1(a,c;x)\, =\, e^{x/2}\,_1 F_1 (c-a,c; -x)\,.\ee

\textbf{(Gauss) hypergeometric equation }\\
\\
Let us move to $_2 F_1$.  This one is the  solution (regular at the origin) of the hypergeometric equation
\be\label{hypgeoeq} x (1-x)y''(x) + [c - (a+b+1)x]\, y'(x) - a b \,\, y(x)=0 . \ee
$_2 F_1$ is called (Gauss) hypergeometric function. Sometimes the subscripts are left out and the symbol is thus reduced to just $F$, and if a function is said to be hypergeometric without any additional adjectives, this one is probably meant by it:
\be\label{hypgeo}_2 F_1 (a,b;c;x)=\sum^\infty_{m=0} \frac{(a)_m(b)_m}{(c)_m} \frac{x^m}{m!}. \ee
Equation (\ref{hypgeoeq}) has the fundamental system consisting of
\beqa\label{FF}
\nn &(-x)^{-a}\,_2F_1(a,\, a+1-c; \, a+1-b; \,x^{-1} ) & \\
\nn &\mbox{\,\,and \,\,}& \\
 \nn &(x)^{a-c}(1-x)^{c-a-b}\,_2F_1(c-a,\, 1-a; \, c+1-a-b; \,x^{-1}(x-1) ). & \\
  & &
\eeqa
There are again many identities, among them for example this one:
\be\label{Eid} _2F_1 (a,b;c;x)\, =(1-x)^{-b} \,_2F_1(c-a,\, b;\, c;\,x(x-1)^{-1} )\,.\ee
%
%
%
Confluent and (Gauss) hypergeometric function are related by the following limit formula:
\be\label{confluence} _1 F_1 (a;c;x)= \lim_{b\rightarrow \infty} \, _2F_1(a,\, b; \, c; \,b^{-1}x ). \ee
\\
\textbf{Bessel equation}\\
\\
Besides those mentioned above, there is another point of interest deserving special attention - Bessel equation (as usual, in the equation below $y=y(x)$)
\be\label{bessel} x^2 y_\nu  '' + x \, y_\nu '   +(x^2 -\nu^2)y_\nu =0. \ee
If we require regularity in the origin, our solution is the Bessel function:
\be\label{form bessel} J_{\nu}(x)=\sum^\infty_{m=0}(-1)^m \left( \frac{x}{2}\right)^{2m +\nu}\frac{1}{m!\Gamma(m+\nu+1)}. \ee
\textbf{Solving more general equations}\\
\\
It is all well and good, but  we do not encounter equations which would be just exactly in the form (\ref{confl}) or (\ref{bessel}) on a daily basis. Equations like
\be\label{general}
\left( a_0 x+ b_0\right)y''(x) +\left( a_1x+ b_1\right)y'(x) +\left( a_2x+ b_2\right)y(x) =0
\ee
are  much more common - obviously, since this is much more general case. The good news is, that in many cases (\ref{general}) has solutions expressible via those of (\ref{confl}) and (\ref{bessel}).  \\
\\
Depending on whether the quantity defined as $ D^2 \equiv a_1^2 -4 a_0 a_2 $ is zero or not, the regular solutions of (\ref{general}) are given in terms of Bessel or confluent hypergeometric functions respectively. We will restrict ourselves to the case $a_0 = 1, \,\, b_0 =0 $ at the price of some generality loss -but generality is not what we are after in the first place.\\
\\
\textbf{(i)} $a_0 = 1$, \, $b_0=0$, \,  $ D^2 =a_1^2 -4 a_0 a_2 \neq 0$ \\
In this case, the solution of (\ref{general}) is of the form
\be\label{form} y(x) \ = \ e^{\frac{D-a_1}{2}x} \zeta(a, c, -Dx) ,\ee
where $ a=\frac{1}{D}\left( \frac{D-a_1}{2}b_1 +b_2    \right)\mbox{,\,\,}c=b_1 ,$ and  $\zeta$  is  some solution of the confluent hypergeometric equation. Note that $D$ is fixed up to the sign, since the coefficients in the equation determine the value of $D^2$ only. In fact it does not matter which possibility is preferred in (\ref{form}). Replacing $D$ with $-D$  makes no difference  because of the Kummer identity (\ref{Kid}).\\
\textbf{(ii) } $a_0 = 1$, \, $b_0=0$, \,  $ D^2 =  a_1^2 -4 a_0 a_2 =0$ \\
This time the solution of (\ref{general}) has the following  form:
\be\label{formbessel} y(x)\ = \ e^{-\frac{a_1}{2}x}        x^{\frac{1-b_1}{2}}  C_{1-b_1} \left( \sqrt{(-2a_1b_1 +4b_2)x}\right), \ee
$C_{1-b_1}$ is any solution of the Bessel equation.
\\
\\
\textbf{Old acquaintances in the hypergeometric frame}\\
\\
Equation (\ref{general}) is of a pretty general form; so considering all that has been said above the fact that so many decent functions from QM can be written in terms of hypergeometric series should not take us by surprise.  For example $e^x$ can be written as $_1F_1(a;a;x)$, generalized Laguerre polynomials (appearing as a part of radial solution for hydrogen atom bound states) can be written in  terms of  $_1F_1(-n;b;x)$, Legendre polynomials (so well known ingredient of spherical harmonics that are independent of the azimuthal angle) can be expressed via $_2F_1(-n, n+1;1;(1-x)/2)$ etc. ($n$ stands for non-negative integer.)

\subsubsection{Solving the radial equation}
\noindent As to solving our equation (\ref{rad}),  that  mathematical interlude above provides the necessary tools. Regarding the translation from normal powers  to ordinary ones, the equalities appearing below are to be either believed or verified; consulting the Appendix  may prove to be useful in the latter case.\\
\\
For the sake of brevity it is suitable to introduce a new parameter $\eta$ defined as
\be \eta=\frac{k \lambda}{2}=\frac{\sqrt{2m_eE}\lambda}{2\hbar} \ee
One note regarding the  notation: In chapter 6 there will be \emph{another }$\eta$ used to shorten formulas. Hopefully it will not lead to confusion, since they are both used only as auxiliary shorthands, each in the separate chapter and final formulas at the end of these chapters are not expressed in terms of either $\eta$. \\
\\
The following connection holds between (\ref{rad}) and (\ref{general}):
\be\label{coef}
\begin{array}{lll}
a_0 =1, & \,\,\,\, a_1 =\lambda k^2 =\frac{4\eta^2}{\lambda},& \, a_2 =k^2=\frac{4\eta^2}{\lambda^2}, \\
b_0 =0, & \,\,\,\, b_1 =2(j+1), & \, b_2 =\lambda k^2  (j+1)+\alpha =\frac{4}{\lambda}(j+1)+2\alpha . \\
& & \\

\end{array}
\ee
$$ \Rightarrow \ \ \ \ D =\pm(\lambda^2 k^4 -4 k^2 )^{1/2} = \pm \frac{4}{\lambda}\sqrt{\eta^2\left(  \eta^2 -1 \right)} .
$$
For $D\neq 0$, or equivalently $\eta \neq 0, \, \eta\neq 1$ the solution of (\ref{radNC}) is
\be\label{reg3}
\begin{array}{lll}
R_{j\pm} & =& \, :{\cal R}_{j \pm}:\\
          & = & \, :\exp\left[\left(\pm\frac{2\eta\sqrt{\eta^2-1}}{\lambda}-\frac{2\eta^2}{\lambda}\right)\varrho\right] \times \\
   & & \times \, _1F_1 \left(j+1\pm\frac{\alpha\lambda}{2\eta\sqrt{\eta^2-1}}; \, 2j+2; \, \mp 4\eta\sqrt{\eta^2-1}\ \frac{\varrho}{\lambda}\right):\\
   &=& ...\mbox{see Appendix}...\\
   & =& \left[1\pm 2\eta\sqrt{\eta^2-1}-2\eta^2\right]^N \times \\
   & & \times  \, _2F_1 \left(j+1\pm\frac{\alpha\lambda}{2\eta\sqrt{\eta^2-1}},\,-N ;\, 2j+2; \, \pm\frac{4\eta\sqrt{\eta^2-1}}{1\pm 2\eta\sqrt{\eta^2-1}-2\eta^2}\right).\\
\end{array}
\ee
The calculations needed to get rid of the normal ordering  in the above equation are briefly sketched in Appendix.
The $\pm $ signs that emerged as a lower index in $R_{\pm}$  spring from the two possible choices of the sign of $D$. We have mentioned that the choice of sign is completely arbitrary due to the Kummer identity which holds for the confluent hypergeometric function. This fact survives the process of rewriting the normal powers in terms of the usual ones and is reflected in the  analogous identity for the (Gauss) hypergeometric functions.
\\
\\
If $\eta=0$, the solution of (\ref{radNC}) is
\be\label{gsol2}
\begin{array}{lll}
R_{j} & =& \, :{\cal R}_{j}:\\
 & =& :\varrho^{-j-1/2}\,J_{-2j-1}(\sqrt{8\alpha\varrho}): \\
   &= & ...\mbox{see Appendix} ... \\
   &=& -\frac{(2\alpha)^{j+1/2}}{(2j+1)!}\, _1F_1(-N; \, 2j+2; \, 2\alpha\lambda).
\end{array}
\ee
\\
And finally for $\eta=1$
\be\label{gsol3}
\begin{array}{lll}
R_{j} & =& \, :{\cal R}_{j }:\\
& =& :e^{2\varrho/\lambda}\varrho^{-j-1/2}\,J_{-2j-1}(\sqrt{8\alpha\varrho}): \\
   &= & ...\mbox{see Appendix}... \\
   &=& -(-1)^N\frac{(2\alpha)^{j+1/2}}{(2j+1)!}\, _1F_1(-N;\, 2j+2; \, -2\alpha\lambda).
\end{array}
\ee
\\
To sum up, the solution of (\ref{radNC}) typically consists of an exponential factor (or its NCQM analog,  see Appendix for further explanation) multiplied by certain power series. Depending on the energy, the exponent may be real or imaginary.
We are supposed to investigate  the following cases:\\
\\
(i) $\eta \in i \textbf{R} \, \, \,\mbox{or} \,\,E<0$;\\
(ii)  $\eta \in (1, + \infty) \,  \, \, \mbox{or}\,\, E>\frac{2\hbar^2}{m\lambda^2}$;\\
(iii)  $\eta \in (0, 1 )  \, \, \, \mbox{or} \,\, 0<E< \frac{2\hbar^2}{m\lambda^2}$;\\
(iv)  $\eta=0 \,\, \,\mbox{or} \,\, E=0$;\\
(v)  $\eta =1 \,\, \, \mbox{or}\,\,  E= \frac{2\hbar^2}{m\lambda^2} $ \\
\\
In (i) and (ii)  the arguments in the exponentials are real - bound states may occur for certain energy values. The (iii) leads to the scattering, the exponential factor involves an imaginary part. The last two, (iv) and (v), are the  border points of the scattering interval (iii).
\subsubsection{Bound states}
\noindent At the very beginning it is suitable to remind us of the QM version, which predicts bound states to occur under the condition that the potential is attractive $(\alpha > 0)$ and the energy has some specific negative values mentioned in (\ref{ener0}).
Now to NCQM. In this section, the $R_{j+}$ form of the solution of (\ref{radNC}) will be suitable for our purposes. The reason is, that in $R_{j+}$ the absolute value of the factor multiplying the hypergeometric function is less than 1 for every $N$. Consequently, when looking for bound states, is is sufficient to check whether the power series terminates. In case of hypergeometric function this happens if the first argument is a negative integer. (Supposing the third one does not spoil it, as has been mentioned before. This is not the case here.) This leads to discrete energy values.
\\
\\
\textbf{Bound states for $E<0$ , \, $\eta = i \, |\eta|$}\\
\\
The equation (\ref{reg3}) in this case reads
\begin{eqnarray}\label{solbound1}
 R_j  &=& \left[1 - 2|\eta|   \sqrt{|\eta|^2+1}  +2|\eta|^2 \right]^N\times     \nonumber \\
   & &  \times _2F_1\left( j+1- \frac{\alpha\lambda}{2|\eta|\sqrt{|\eta|^2+1}} , -N; 2j+2; \frac{- 4|\eta|\sqrt{|\eta|^2+1}}{1 - 2|\eta| \sqrt{|\eta|^2+1}  +2|\eta|^2 }\right).\nonumber \\
   & &
\end{eqnarray}
Square integrable bound state solutions can be easily seen, since the hypergeometric function has to reduce to a  polynomial. This happens only if $\alpha > 0$, i.e. in the {\it Coulomb attractive} case, provided that
\be\label{BS}  \alpha > 0\ \ \ \ \ \mbox{and}\ \ \ \ \ j+1 - \frac{\alpha\lambda}{2|\eta|\sqrt{|\eta|^2+1}}\, =\,-n_r\,.\ee
(We have used $n_r$ to denote NC ''radial quantum number''.) This gives the bound state energies (with Planck constant $\hbar$ and mass $m_e$ explicitly introduced):
\be\label{energy}
E^{I}_{\lambda \,n} \ =\ \frac{m_ee^4}{2\hbar^2 n^2}\, \frac{2}{1+ \sqrt{1+(\alpha \lambda/n)^2}}\ =\
\frac{\hbar^2}{m_e\lambda^2}\left(1- \sqrt{1+(\alpha\lambda/n)^2}\right)\,, \ee
where $n=n_r +j+1$. Taking the limit $\lambda \to 0$ we recover the QM result. 
\\
\\
\textbf{Bound states for $E>2/\lambda^2$ , \, $\eta = |\eta|>1$}\\
\\
The equation (\ref{reg3}) in this case is
\begin{eqnarray}\label{solbound2}
 R_j  &=& \left[1 + 2|\eta|   \sqrt{|\eta|^2-1}  -2|\eta|^2 \right]^N\times     \nonumber \\
   & &  \times _2F_1\left( j+1+\frac{\alpha\lambda}{2|\eta|\sqrt{|\eta|^2-1}} , -N; 2j+2; \frac{4|\eta|\sqrt{|\eta|^2-1}}{1 + 2|\eta| \sqrt{|\eta|^2-1}  -2|\eta|^2 }\right).\nonumber \\
   & &
\end{eqnarray}
Since the absolute value of the prefactor preceding the hypergeometric function is less than 1, the whole solution can have a finite norm. There is a possibility for  the hypergeometric function to terminate, since the first argument becomes a negative integer for certain energy values, under the condition that $\alpha<0$ (so  the potential has to be repulsive this time).
\be  \alpha <0\ \ \ \ \ \mbox{and}\ \ \ \ \ j+1+\frac{\alpha}{k\sqrt{|\eta|^2-1}} =\,-n_r\,.\ee
In this case the bound state energies read:
\be\label{energy1}
E^{II}_{\lambda \, n} \ =\ \frac{\hbar^2}{m_e\lambda^2}\left(1 + \sqrt{1+(\alpha\lambda/n)^2}\right)\
=\ \frac{2\hbar^2}{m_e \lambda^2} - E^{I}_{\lambda \, n} . \ee
These are very unexpected bound solutions for Coulomb {\it repulsive} force, $E^{II}_{\lambda \,n}$ being a mirror of $E^I_{\lambda \, n}$ with respect to the ''critical energy'' $E_{crit} = 2\hbar^2/(m_e\lambda^2)$. However, the states corresponding to the energies   $E\geq E^{II}_{\lambda \, n}$ disappear from the Hilbert space since $E^{II}_{\lambda \,n} \to \infty$ in the commutative limit $\lambda \to 0$. For energies $E=E^{II}_{\lambda \,n}$ the solution (\ref{solbound2}) has the same finite norm as (\ref{solbound1}) for $E=E^{I}_{\lambda \,n}$.
\subsubsection{Scattering }
\noindent In this case the following intervals are relevant: $E\in(0, \, 2/\lambda^2)$, \, $\eta \in (0,1)$.
Let us have a look at the Coulomb scattering in NCQM, considering the  $j$-th partial wave and the energy $E\,\in\,(0,\,2/\lambda ^2)$. For the future convenience it is suitable to distinguish  the following quantity:
\be\label{NCries2}  p\, =\, \sqrt{2E \left(1 -\frac{\lambda ^2 E }{2}\right) }\ .\ee
The formula above represents a conformal map from the upper  complex $E$-plane, denoted as $\bf{C}_+$, on a right complex  $p$-plane with a branch cut $p\,\in\,(0,\,1/\lambda )$. For $E \in \bf{C}_+ $ we may introduce another useful complex variable
\be\label{OmegaCont}  E \in  {\bf C}_+  \, \longrightarrow \, \Omega\, =\, \frac{p-i \lambda E}{p + i \lambda E} \in D ,\ee
mapping $\bf{C}_+$ onto the unit disc $D$ in the complex $\Omega$-plane. The scattering energies $E \in (0,\, 2/\lambda^2)$ are  mapped onto the  unit circle $|\Omega| = 1$ in the $\Omega$-plane, whereas the bound state energies have their images on the real axis in the $\Omega$-plane, so that $E^{I, II}_{\lambda \,n} \mapsto \Omega^{I, II}_n$  given below, see (\ref{NCries9}), (\ref{NCries9mirror}).

The solution regular in the origin is given in terms of the hypergeometric function (\ref{hypgeo}):
\be\label{NCries1} R_{E j}\ =\ \Omega^{-N}\, _2F_1 \left(j+1-i\frac{\alpha }{p},\,-N;\,2j+2;\, 2i\lambda p\, \Omega \right).\ee
We choose the positive  square root (\ref{NCries2}) for $E\,\in\,(0,\,2/\lambda ^2)$.
The radial  dependence of  $R_j$ is present in the hermitian operator  $N$: $r = \varrho +\lambda $, $\varrho  = \lambda N$. In analogy with   the method used in standard QM (\ref{schiff2}) we will rewrite also the NC solution as a sum of two terms corresponding to the in- and out- going spherical wave. Again, leaving out the common hermitian factor which is irrelevant regarding the $S$-matrix, we can write (see Appendix):
\beqa\nn R_{Ej} &\sim &  \frac{(-1)^{j+1}e^{\alpha\pi/p}}{\Gamma (j+1 +i\frac{\alpha}{p})}\, \Omega^{-N-j-1+i\frac{\alpha}{p}}\,\frac{\Gamma(2j+2)\Gamma(N+1)}{\Gamma(N+2+j-i\alpha /p)}\,(2\lambda p)^{-1-j+i\frac{\alpha}{p}}\\
&\times & \, _2F_1 \left(j+1-i\frac{\alpha }{p},\,-j-i\frac{\alpha }{p}; \,N+2+j-i\frac{\alpha }{p};\,\frac{-i}{2\lambda p}\, \Omega^{-1} \right) \nn\\
&+ &  \frac{(-1)^{j}e^{\alpha\pi/p}}{\Gamma (j+1 -i\frac{\alpha}{p})}\, \Omega^{N+j+1+i\frac{\alpha}{p}}\,\frac{\Gamma(2j+2)\Gamma(N+1)}{\Gamma(N+2+j+i\alpha /p)}\,(2\lambda p)^{-1-j-i\frac{\alpha}{p}} \nn \\
&\times & \, _2F_1 \left(j+1+i\frac{\alpha }{p},\,-j+i\frac{\alpha }{p}; \,N+2+j+i\frac{\alpha }{p};\,\frac{i}{2\lambda p}\,  \Omega \right)\,. \label{NCries3} \nn \\
& & \eeqa
To enable better comparison with (\ref{schiff2}) let us  rewrite also (\ref{NCries3}) as a sum of two complex conjugated parts. Some sort of sketch of the calculation leading to it is to be found in the Appendix.
\beqa\label{rozklad} R_{Ej} &\sim &  (-1)^{j+1}i^{-j-1}e^{-\alpha\pi/2p}\ \frac{\Gamma(2j+2)}{\Gamma (j+1 -i\alpha/ p)} \frac{e^{-i\frac{\alpha}{p}\ln(2pr)}}{(2pr)^{j+1}}\nn \\
&\times & \exp \left[-\left(r / \lambda+j+i \alpha/ p \right)\, \ln (\Omega^{-1})\,\right] \nn \\
&\times & \exp \left[- \sum_{n=1}^\infty (\lambda / r)^n\, \frac{B_{n+1}(j+1+i\alpha /p)\,-\,B_{n+1}(0)}{n(n+1)} \right] \nn \\
&\times & \, _2F_1 \left(j+1+i\frac{\alpha}{p},\,\,\,-j+i\frac{\alpha}{p} ;\,\,\,\frac{r}{\lambda} +j+1+i\frac{\alpha}{p};\,\,\,-\frac{1}{2i\lambda p}\, \Omega \right) \nn \\
& & \nn \\
& + &  (-1)^{j+1}i^{j+1}e^{-\alpha\pi/2p}\ \frac{\Gamma(2j+2)}{\Gamma (j+1+i\alpha/ p)} \frac{e^{i\frac{\alpha}{p}\ln(2pr)}}{(2pr)^{j+1}}\nn \\
&\times & \exp \left[-\left(r/ \lambda +j - i \alpha/ p \right)\, \ln (\Omega)\,\right] \nn \\
&\times & \exp \left[- \sum_{n=1}^\infty (\lambda / r)^n\, \frac{B_{n+1}(j+1-i\alpha /p)\,-\,B_{n+1}(0)}{n(n+1)} \right] \nn \\
&\times & \, _2F_1 \left(j+1-i\frac{\alpha}{p},\,\,\,-j-i\frac{\alpha}{p} ;\,\,\,\frac{r}{\lambda}+j+1-i\frac{\alpha}{p};\,\,\,\frac{1}{2i\lambda p}\, \Omega^{-1}\right). \nn \\
& &  \label{NCries4}\eeqa
%
The $S$-matrix  is the ratio of the  $r$-independent factors :
\be\label{NCries5} S^\lambda_j(E)\ =\ \frac{\Gamma (j+1 -i\frac{\alpha}{p})}{\Gamma (j+1 +i\frac{\alpha}{p})}\ ,\ee
where
\be\label{NCries6} E\ =\ \frac{1}{\lambda^2}\, \left(1\,+\,i\,\sqrt{\lambda^2 p^2 - 1} \right) \ee
is the  conformal map inverse to (\ref{NCries2}), which maps the  cut $p$ right-half-plane into the  $E$ upper-half-plane. We take the positive square root in  (\ref{NCries6}) for $p\,\in\,(1/\lambda ,\,+\infty)$.
\\
The physical values of the $S$-matrix are obtained as $S^\lambda_j(E+i\varepsilon )$ in the limit $\varepsilon\,\rightarrow \,0_+ $. The interval corresponding to the scattering $E\,\in\,(0,2/\lambda ^2)$ is mapped onto the branch cut in the $p$-plane as follows:
\beqa\nn E \in (0,1/\lambda ^2)\ &\mapsto &  \mbox{upper edge of the branch cut} \ p\in\,(0,\,1/\lambda ),\\
\nn E \in (1/\lambda ^2,2/\lambda ^2)\ &\mapsto &  \mbox{lower edge of the  branch cut} \ p\in\,(0,\,1/\lambda ).\\
& & \label{NCries7}
\eeqa
\subsubsection{Bound states revisited - poles of the S-matrix }
\noindent In NCQM there is an analogy with QM,  the poles of the S-matrix  occur in the case of attractive potential $(\alpha>0)$ for some special values of energy below $0$. However, poles can be also found in the case of repulsive potential $(\alpha<0)$ for particular values of energy above $2/\lambda^2$ ).
\\
\\
 \textbf{ Poles of the $S$-matrix for attractive potential}\\
 \\
\[ p^\lambda_n\,=\,i\,\frac{\alpha }{n}\,,\ \ \ \alpha > 0\ \ \ \Leftrightarrow\ \ \ E^{I}_{\lambda \, n}\,=\,\frac{1}{\lambda^2}\, \left(1\,-\,\sqrt{1+(\lambda\alpha/n )^2 } \right)\, < 0,\]
\be\label{NCries8}  n\, =\, j+1,\ j+2,\ \dots \ee
In the limit $\lambda \rightarrow 0$ this coincides with the standard self-energies of the hydrogen atom (\ref{schiff5}).
Let us denote
\be\label{NCries9} \kappa_n = \frac{\lambda \alpha}{n},\ \ \ \Omega^I_n\ =\  \frac{\kappa_n-\sqrt{1+\kappa_n^2 }+1}{\kappa_n +\sqrt{1+\kappa_n^2  }-1}\,.\ee
Then the solution   (\ref{NCries1}) is
\be\label{NCries10} R^I_{nj}\ =\ (\Omega^I_n)^N\, _2F_1(-n,\,-N,\,2j+2;\,-2 \kappa_n (\Omega^I_n)^{-1})\,.\ee
It is integrable since $\Omega_n\,\in\,(0, 1)$ for positive $\kappa_n$ and under given conditions the hypergeometric function is a polynomial. The norm (\ref{Psi2}) of $R^I_{nj}$ is finite and given in terms of a generalized hypergeometric function. We do not present the corresponding cumbersome formula as it is not needed for our purpose.
\\
\\
\textbf{ Poles of the $S$-matrix for repulsive potential}\\
\\
(These disappear from the Hilbert space of the physical states in the limit $\lambda \rightarrow 0$ )
\[ p^\lambda_n\,=\,i\,\frac{\alpha }{n}\,,\ \ \ \alpha < 0\ \ \ \Leftrightarrow\ \ \ E^{II}_{\lambda \, n}\,=\,\frac{1}{\lambda^2}\, \left(1\,+\,\sqrt{1+(\lambda\alpha/n )^2 } \right)\,>2/\lambda^{2}, \]
\be\label{NCries11}  n\, =\, j+1,\ j+2,\ \dots \ee
Now (\ref{NCries1}) has the form
\be\label{NCries12} R^{II}_{nj}\ =\ (-\Omega^{II}_n)^N\, _2F_1(-n,\,-N,\,2j+2;\,2 \kappa_n (\Omega^{II}_n)^{-1} \,), \ee
where
\be\label{NCries9mirror}
\Omega^{II}_n\ =\  -\frac{\kappa_n+\sqrt{1+\kappa_n^2 }+1}{\kappa_n -\sqrt{1+\kappa_n^2  }-1}\,.\ee
The definition of $\kappa_n$ is the same as in (\ref{NCries9}) (note that it is negative
this time). Since $\Omega^{II}_n = \Omega^I_n \in (0,1)$  the solution (\ref{NCries1})
is integrable because the hypergeometric function terminates like in the
previous case.\\
\\
\\
Here we are done with the NC version of ''Schr\"{o}dingerian'' treatment of hydrogen atom. Later we are going to examine  solution of the same problem from a slightly different point of view. So the summary may be postponed until we see things also from that another angle, providing the possibility to discuss both.

\section{Velocity operator and uncertainty relations}
This chapter is some kind of an intermezzo, when we take a break from focusing mainly on the Coulomb problem.  Using what we have learned in the previous chapter, important broader aspects of NCQM can be shown and further examined, including analogs of Heisenberg's uncertainty relations; the reader certainly appreciates their role in QM and suspects the NC version to be equipped with something similar.
As long as we were  occupied only with Schr\"{o}dinger's  approach (finding the equation and solving it), it was not inevitable to deal with the velocity operator explicitly (implicitly it is deeply involved via hamiltonian). However,  the velocity operator will have a lot to say later - it provides insight into the theory, and is a vital part of the algebraic approach based on the ''hidden'' dynamical symmetry, so it deserves a chapter of its own.
\subsection{Definition}
Let us define the velocity operator using the Heisenberg equation
\begin{equation}
\hat{V}^j \Psi= -i [ \hat{X}^j , \hat{H}]\Psi .
\end{equation}
Here $\hat{X}_j$ is the coordinate operator that acts on $\Psi$ symmetrically as
\begin{equation}
 \hat{X}_j\Psi= \frac{1}{2} (x_j\Psi + \Psi x_j) .
\end{equation}
The relation for $\hat{V}^j \Psi$ can be simplified,  if we consider potentials with radial dependence only: $U=U(r)$ (in fact building our fuzzy space in a rotationally invariant fashion has been based on the intention to deal with such potentials). If that is the case, we can write  (since $[x^j, r]=0$)
\begin{eqnarray}
 \label{V}  \hat{V}^j \Psi &=& -i [ \hat{X}^j , \hat{H}_0]\Psi \\
\label{V1} &=& -\frac{i}{2r}\sigma ^j _{\alpha \beta} (a^+_\alpha [a_\beta , \Psi] - a_\beta [ a^+_\alpha ,\Psi]) \\
\label{V2.0} & = & \frac{i}{2r} \sigma^i_{\alpha \beta} (a^+_\alpha \Psi a_\beta - a_\beta \Psi a^+_\alpha) \\
\label{walphabeta} & = & \frac{i}{2r} \sigma^i_{\alpha \beta} \hat{w}_{\alpha \beta} \Psi .
\end{eqnarray}
The quantity $\hat{w}_{\alpha \beta}$ has been introduced here just as a shorthand; it will appear again, so we suggest the reader does not forget its relation to the velocity operator.
\subsection{Some basic properties of $\hat{V}^j$}
From its construction it is obvious that $\hat{V}^j$ inherits the hermicity of operators $\hat{X}^i$ and $\hat{H}_0$. Now we shall investigate its action upon some basic objects to prove that it really does the job of a  gradient operator.
\begin{equation} \label{VX}
\hat{V}^i x^j = -\frac{i}{2r}\sigma ^i _{\alpha \beta} ( a^+_\alpha [a_\beta , x^j] - a_\beta [ a^+_\alpha , x^j])=
\end{equation}
\begin{equation*}
= -\frac{i}{2r}\sigma^i_{\alpha \beta} (a^+_\alpha [ a_\beta , \lambda \sigma^j_{\gamma \delta} a^+_\gamma a_\delta] - a_\beta [ a^+_\alpha , \lambda \sigma^j_{\gamma \beta} a^+_\gamma a_\delta]) =
\end{equation*}
\begin{equation*}
= - \frac{i \lambda}{2r} \sigma^i_{\alpha \beta} \sigma^j_{\gamma \beta} (a^+_\alpha a_\delta \delta_{\beta \gamma} + a_\beta a^+_\gamma \delta_{\alpha \delta})=
\end{equation*}
now using $a_\beta a^+_\gamma = a^+_\gamma a_\beta + \delta_{\beta \gamma}$ we get
\begin{equation*}
=-\frac{i\lambda}{2r} (a^+_\alpha a_\beta ((\sigma^i \sigma^j)_{\alpha \beta}+(\sigma^j \sigma^i)_{\alpha \beta} + Tr(\sigma^j \sigma^j)) =
\end{equation*}
and finally using the anticommutation relation of Pauli matrices $\{\sigma^i, \sigma^j\}_{\alpha \beta} = 2 \delta^{ij}\delta_{\alpha \beta}$, $Tr(\sigma^i \sigma^j) = 2\delta^{ij}$
\begin{equation*}
= -\frac{i}{2r}(\lambda(a^+_\alpha a_\alpha+1)\delta^{ij}) = -i \delta^{ij} .
\end{equation*}
Similarly, one can verify, that
\begin{equation} \label{Vr}
\hat{V}^j f(r)=-i \frac{x^j}{r} f_\lambda'(r) .
\end{equation}
where
\begin{equation} \label{f'}
 f_\lambda'(r)=\frac{f(r+\lambda)-f(r-\lambda)}{2\lambda} .
\end{equation}
So $i \hat{V}^i$ makes  as fine a gradient operator as  the NC deformation allows. There is one crucial aspect brought by the NC modification - the Lebnitz rule  gets the following correction:
\begin{equation} \label{Leibnitz}
\hat{V}^j (AB)=(\hat{V}^j A)B+A(\hat{V}^jB)+{\cal K}^j(A,B).
\end{equation}
where the ${\cal K}^j (.,.)$ is the correction term
\begin{equation} \label{kor}
{\cal K}^i(A,B)=- \frac{i}{2r}\sigma ^i _{\alpha \beta}([a^+ _\alpha ,A][a_\beta ,B]-[a_\beta ,A][a^+ _\alpha ,B]).
\end{equation}
This correction vanishes in the limit of $\lambda \rightarrow 0$.\\
One can ask if the operator defined by \eqref{V} transforms as a vector (under some rotation generated by $\hat{L}$). This may be verified using the commutation relations for $a_\alpha, a_\alpha^+$, definition of $\hat{L}$ and \eqref{V}, and turns out to be true
\begin{equation}
[\hat{L}^i, \hat{V}^j] = i \varepsilon^{ijk} \hat{V}^k .
\end{equation}
\subsection{Uncertainty relations}
After examining the basic properties of the velocity/gradient operator $\hat{V}^j$ we can move on to something more interesting. In the ordinary QM, quite a prominent role is played by the Heisenberg's relations $[\hat{p}^i,\hat{X}^j]=-i \delta^{ij} \hbar$. Recalling that $\hbar=m_e=1$, this corresponds to $[\hat{V}^i , \hat{X}^j]=-i \delta_{ij}$.\\
In  QM this is easily calculated using  the fact that $\partial_i x^j = \delta^{ij}$ and the Leibnitz rule. The former claim is true in NC QM as well (see \eqref{VX}), but the latter brings a tiny (but important) modification (see \eqref{Leibnitz}). Combining those two equations, we get
\begin{equation} \label{UR}
[\hat{V}^i , \hat{X}^j] \Psi= -i \delta^{ij} \Psi + \frac{1}{2}({\cal K}^i(x^j, \Psi)+{\cal K}^i(\Psi,x^j)) .
\end{equation}
By evaluating the correction terms
\begin{eqnarray}
{\cal K}^i (x^j,\Psi) &=& -\frac{i}{2r}\sigma ^i _{\alpha \beta} ([a^+_\alpha , x^j][a_\beta , \Psi]- [a_\beta , x^j][a^+_\alpha , \Psi]),\\ \nonumber
{\cal K}^i (\Psi,x^j) &=& -\frac{i}{2r}\sigma ^i _{\alpha \beta} ([a^+_\alpha ,\Psi][a_\beta , x^j]- [a_\beta , \Psi][a^+_\alpha ,  x^j]),
\end{eqnarray}
one finds quite a surprising result, that together they are equal to $2 i \delta^{ij} \lambda^2 \hat{H}_0 \Psi $. Inserting this into the uncertainty relation in the form \eqref{UR} we obtain
\begin{equation} \label{uncertainty}
[\hat{V}^i , \hat{X}^j] = -i \delta^{ij} (1- \lambda^2 \hat{H}_0) .
\end{equation}
Let us analyze this result a little. First, it is exact (we have not neglected any terms of higher orders in $\lambda$). Obviously, the correction is small for energies $E \ll \lambda^{-2}$ and vanishes for $\lambda=0$, \footnote{If $\lambda$ is tiny, then $\frac{1}{\lambda^2}$, which corresponds to the energy scale on which the correction term starts being dominant, is immense. This is probably true for all NC corrections to QM. They become important (or measurable) for energies, for which QM is not a suitable theory (but instead a relativistic theory is needed). The corrections still might be true for smaller energies, being relics of the higher theories.}. It also defines two important energy scales: $E_0 = \frac{1}{\lambda^2}$, for which the uncertainty vanishes and $E_1 = \frac{2}{\lambda^2}$, for which the principle gets negative sign, with respect to the ordinary version ($-i \delta^{ij}$). Both these scales appeared in the study of scattering/bound states of NC QM hydrogen atom.
\subsection{Commutator $[\hat{V}^i,\hat{V}^j]$}
Here we are going to use the notation introduced in (\ref{walphabeta}), so that
 the commutator has the form
\begin{equation}
[\hat{V}^i , \hat{V}^j]\, \Psi = \left(\frac{i}{2} \right)^2 \sigma ^i _{\alpha \beta} \sigma ^j_{\gamma \delta} \left[ \frac{1}{\hat{r}} \hat{w}_{\alpha \beta} ,\frac{1}{\hat{r}} \hat{w}_{\gamma \delta}\right] \, \Psi .
\end{equation}
Using the rule for $[AB,CD]$ and the obvious fact $[\hat{r},\hat{r}]=0$, this can be split into 3 separate terms: one of the form $\frac{1}{r^2}[w,w]\, \Psi$ and two of the form $\frac{1}{r}[\frac{1}{r},w]\, \Psi$. We have evaluated them separately, the first term showed up to be equal to $\frac{2i}{\lambda \hat{r}^2} \varepsilon^{ijk} (x^k \Psi - \Psi x^k)$, the other two, together, are equal to $\frac{2i}{\lambda \hat{r}^2}\varepsilon^{ijk} (-x^k \Psi + \Psi x^k)$. Combining those two results, quite a non-trivial zero is obtained:
\begin{equation} \label{VV}
[\hat{V}^i,\hat{V}^j] \,\Psi\,=\, 0.
\end{equation}
One may ask, if this zero isn't somehow obvious. To answer this question, we investigated this commutator for generalized states  $\Psi$ containing a different number of c/a operators, yielding highly non-trivial (and interesting) result. However, presenting it here would be at the cost of wandering too far away from the main problems this paper is dealing with.
\subsection{Relation between  $\hat{H}_0$ and  $\hat{V}^2=\hat{V}^i \hat{V}^i$}
Another important relation is the one between the velocity operator (its square) and the free Hamiltonian. This calculation is a bit tricky, one has to evaluate both
\begin{equation}
\hat{V}^2 = \hat{V}^j \hat{V}^j = -\frac{1}{4} \sigma ^j_{\alpha \beta} \sigma^j_{\gamma \delta}\frac{1}{\hat{r}} \hat{w}_{\alpha \beta} \frac{1}{\hat{r}}w_{\gamma \delta}
\end{equation}
and
\begin{equation}
\left(\frac{1}{\lambda}^2 - \hat{H}_0 \right)^2.
\end{equation}
These two turn out to be related in the following fashion:
\begin{equation} \label{VVH}
\left(\frac{1}{\lambda^2}-\hat{H}_0 \right)^2 = \frac{1}{\lambda ^2} \left(\frac{1}{\lambda^2}-\hat{V}^2 \right) ,
\end{equation}
which can be rewritten as
\begin{equation} \label{V^2=H}
\frac{1}{2}\hat{V}^2=\hat{H}_0 \left(1-\frac{1}{2}\lambda ^2 \hat{H}_0 \right),
\end{equation}
\begin{equation}\label{H=V^2}
\hat{H}_0 = \frac{1}{\lambda ^2}\left(1-\sqrt{1-\lambda^2 \hat{V}^2}\right) .
\end{equation}
Equation \eqref{V^2=H} implies, that the kinetic energy of a particle may not be infinitely large, instead, it has a natural cut-off at $E_1 = \frac{2}{\lambda^2}$. Alternatively, from \eqref{V^2=H} follows that for $E_0 =\frac{1}{\lambda^2}$ we achieve the maximal velocity square $\hat{V}^2 = \frac{1}{\lambda^2}$.
\subsection{Kinematic symmetry in NC QM}
Symmetry seems to be a recurring theme throughout this whole paper - quite a heartwarming fact for a physicist. As to the particular symmetry examined in this section, we suspect there is more depth to it than we are going to present here. For the time being let us at least mention it as something to think about. \\
As has already become a tradition, let us start with reminding ourselves of the standard QM, where the configuration space is ${\bf R}^3_0$, and the following kinematic commutation relations among basic operators corresponding to rotation generators $L_k$, coordinates $x_k$ and momenta $p_k$, $k = 1,2,3$, hold (regardless of the form of potential):
\[ [x_i, x_j]\ =\ 0,\ \ \  [p_i, p_j]\ =\ 0,\ \ \ [p_i, x_j]\ =\ - i \delta_{ij}\,, \]
\be \label{kinsym} [L_i, L_j]\ =\ - i \varepsilon_{ijk}\,L_k, \ \  [L_i, x_j]\ =\ - i \varepsilon_{ijk}\,x_k, \ \ [L_i, p_j]\ =\ - i \varepsilon_{ijk}\,p_k . \ee
Due to the nature of the problems dealt with in this paper we will consider the case of a {\it central} potential; this assumption is followed by the well-known conservation laws. As to the above equations, the first line represents the canonical commutation relation in QM; for the sake of brevity we have set  $\hbar=1$. The second line tells us that the triplets $x_k$ and $p_k$ are $SO(3)$-vectors. Operators $L_k$, $x_k$, $p_k$  form a Lie algebra of the 3D Euclidean group $E(3) = SO(3)\,\triangleright H(3)$ - the semi-direct product of rotation group $SO(3)$ and 3D Heisenberg-Weyl group.    \\
\\
Now to NC QM. An alternative  notation can be of use here:
\begin{equation} \label{L-SO4}
\hat{L}_{ij}\ =\ \varepsilon_{ijk}\,\hat{L}_k,\ \ \ \hat{L}_{k4}\ =\ -\,\hat{L}_{4k}\ =\ \frac{1}{\lambda}\, \hat{X}_k\,.
\end{equation}
The commutation relations among $L_k$ and $x_k$, $k = 1,2,3$, take the explicit $SO(4)$ invariant form
\begin{equation}
[\hat{L}_{ab},\hat{L}_{cd}]\ =\ i\,(\delta_{ac}\,\hat{L}_{bd}\,-\, \delta_{bd}\,\hat{L}_{ac} )\,.
\end{equation}
This relations express the requirements that $\hat{L}_k$, $k = 1,2,3$, are $SO(3)$ and $x_k$, $k = 1,2,3$ are $SO(3)$ vectors.\\
\\
The triplet of NC velocity operators $\hat{V}_k$, $k = 1,2,3$, may be joined by one additional component, whose form can be inferred from the r.h.s. of the uncertainty relations, so that defining
\begin{equation} \label{V4}
\hat{V}_4\,\Psi\ =\ \left( \frac{1}{\lambda}\,-\,\lambda\, \hat{H}_0\right)\,\Psi\ =\ \frac{1}{2r} \left( a^+_\alpha\, \Psi\, a_\alpha\, +\, a_\alpha\, \Psi\, a^+_\alpha \right)\,,
\end{equation}
we obtain a remarkable result that the four commuting operators $\hat{V}$, $c = 1,2,3,4$, form an $SO(4)$-vector:
\begin{equation} \label{VSO4}
[\hat{V}_a,\,\hat{V}_b] \ =\ 0\,,\ \ \ [\hat{L}_{ab},\,\hat{V}_c] \ =\ i\,(\delta_{ac}\,\hat{V}_b\,-\, \delta_{bc}\,\hat{V}_a )\,,\ \ \ a,b,c\,=\,1,\,...\,,4\,.
\end{equation}
The second commutator combines the fact that $\hat{V}$, $k = 1,2,3$ is an $SO(3)$-vector and that the NC uncertainty relations hold.  Equations (\ref{VSO4}) represent Lie algebra commutation relations of the 4D Euclidean group $E(4)\,=\,SO(4)\,\triangleright T(4)$ - the semi-direct product of 4D orthogonal group $SO(4)$ with 4D translations generated by four velocity operators. \\
\\
Adding now to \eqref{V^2=H} the square of the fourth component of the velocity operator \eqref{V4} we obtain the quadratic Casimir operator of the  $E(4)$ group:
\begin{equation} \label{casim2}
\hat{C}_2\ =\ \hat{V}^2_a \ =\ \hat{V}^2_j \,+\, \hat{V}_4^2 \ =\
2\,\hat{H}_0\,-\,\lambda ^2\, \hat{H}^2_0 \,+\, \left( \frac{1}{\lambda}\,-\,\lambda\, \hat{H}_0\right)^2 \ =\ \frac{1}{\lambda^2} \ .
\end{equation}
The second quartic Casimir operator is given as a square of the $E(4)$ Pauli-Lubanski vector
\begin{equation} \label{PL}
\hat{\Lambda}_d\ =\ \frac{1}{2}\,\varepsilon_{abcd}\,\hat{L}_{ab}\,\hat{V}_d \ \ \ \Longleftrightarrow \ \ \ \hat{\Lambda}_i\ =\ \hat{V_4}\,\hat{L}_i \,+\, \varepsilon_{ijk}\,\hat{V}_j\,\hat{L}_{k4}, \ \hat{\Lambda}_4\ =\ \hat{L}_j\,\hat{V}_j \,.
\end{equation}
Particularly, the action of the fourth component of the Pauli-Lubanski vector is evaluated in the Appendix, the result is
\begin{equation} \label{4th}
\hat{\Lambda}_4\,\Psi\ =\ \hat{L}_j\,\hat{V}_j\,\Psi\ =\ 0\,.
\end{equation}
It follows from the $SO(4)$ invariance  that all four components of Pauli-Lubanski vector vanish, and consequently the quartic Casimir operator vanishes too:
\begin{equation} \label{casim4}
\hat{\Lambda}_a\,=\,0,\ \ a\,=\,1,\,..., \,4,\ \ \ \Longrightarrow \ \ \ \hat{C}_4\ =\ \hat{\Lambda}^2_a \ =\ 0\,.
\end{equation}

Thus, the NC QM in question is specified by a scalar $E(4)$ representation specified by the values of Casimir operators: $\hat{C}_2\,=\, 1/\lambda^2$ and $\hat{C}_4\,=\,0$. In such representation the common eigenvalues of velocity operators form a 3-sphere $S^3_\nu$ with radius $\nu\,=\,\frac{1}{\lambda}$ for any central potential $\hat{U}(\hat{r})$. \\

Finally, inverting equations (\ref{L-SO4}) as
\be \hat{L}_k\ =\ \frac{1}{2}\,\varepsilon_{ijk}\,\hat{L}_{ij} ,\ \ \  \hat{X}_k\ =\ \lambda\, \hat{L}_{k4}\,, \ee
and taking the commutative limit $\lambda \to 0$ we recover the kinematic symmetry relations (\ref{kinsym}). Thus the kinematic symmetry in NC QM is a $\lambda $-deformation  of the kinematic symmetry in the standard QM, or equivalently, the kinematic symmetry in QM is a $\lambda $-contraction of the kinematic symmetry in NC QM (within the framework of Lie algebra deformations and contractions).

{\it Note}: In the commutative limit $\lambda \to 0$ the fourth component of the velocity $\hat{V}_4$ diverges, and consequently, the quadratic Casimir operator $\hat{C}_2$ diverges too. In this limit we have now no restrictions on the remaining three components of velocity $\hat{V}_k$, $k = 1,2,3$: the 3-sphere $S^3$ of NC velocities goes to 3-plane of standard velocities (momenta). In the commutative limit the role of the Casimir operator overtakes the factor $\hbar$ at $-i\,\delta _{ij}$ on the r.h.s. of the uncertainty relations (we have chosen $\hbar = 1$). In addition, in the commutative limit just the fourth component of Pauli-Lubanski vector given in (\ref{4th}) persists: the condition $\hat{L}_j\,\hat{V}_j\,=\,0$ is the well-known condition valid for scalar particles in standard QM.
\subsection{Ehrenfest theorem}
In Newtonian mechanics the change of particle's momentum is governed by the equation $ \dot{\vec{p}} = - \vec{\nabla} U(\vec{x},t)$. In  QM, this is replaced with the Ehrenfest theorem $\frac{d}{dt}\langle \vec{p}\rangle = \langle -\vec{\nabla}U(\vec{x},t)\rangle$. One may ask how this changes in NC QM. The result for $U=U(r)$ is easily calculated using the previous results (\eqref{H=V^2} combined with \eqref{VV} tells us that $[\hat{V}^i, \hat{H}_0]=0$), so
\begin{equation} \label{ehren}
\dot{\hat{V}}^i  = -i [\hat{V}^i , \hat{H}] = -i [ \hat{V}^i , \hat{H}_0 + \hat{U}(\hat{r})] = -i [ \hat{V}^i ,\hat{U}(\hat{r})] ,
\end{equation}
which, due to the modified Leibnitz rule \eqref{Leibnitz}, equals to
\begin{equation} \label{Acc}
\dot{\hat{V}}^i  = -i (\hat{V}_i\hat{U}(\hat{r})) + \hat{U}_\lambda'(\hat{r}) \left(\frac{\lambda}{\hat{r}}\hat{L}^i + \lambda^2 \hat{W}^i\right)
+ \frac{\lambda^2}{2} \hat{U}_\lambda''(\hat{r})\hat{V}^i ,
\end{equation}
where $\hat{U}_\lambda'(\hat{r})$ is defined in \eqref{f'} and
\begin{eqnarray}
\hat{U}''_\lambda(\hat{r})&=&\frac{1}{\lambda^2}(\hat{U}(\hat{r}+\lambda) - 2\hat{U}(\hat{r}) + \hat{U}(\hat{r}-\lambda)) ,\\
\hat{W}^i \Psi&= &\frac{1}{2r}\sigma ^i _{\alpha \beta} [a_\beta, [a^+_\alpha , \Psi]] .
\end{eqnarray}
The NC Ehrenfest theorem contains new terms, which vanish in the commutative limit $\lambda \rightarrow 0$. The vector $\hat{W}^i$ has not been introduced yet, but it turns out to be closely related to the Laplace - Runge - Lenz vector and we shall meet it again later, \eqref{A_k}.

\section{NC hydrogen atom: Pauli-like approach via dynamical symmetry }
Now we are going to investigate the existence of dynamical symmetry connected with the conservation of  Laplace-Runge-Lenz vector (LRL)  of the Coulomb-Kepler problem in NCQM and possibly to try to find the NC version it.\\
This chapter  is going to deal with the same problem as that Schr\"{o}dinger-like one; it is just a different approach. Symmetry was very important in the previous  discussion, now it is about to be twice as important. ''Twice '' is not just a figure of speech; the symmetry group that is going to be exposed has twice as much generators as $SO(3)$ employed before. Of course, since  we  plan to spend this chapter with the same hydrogen that has been our companion so far, the $SO(3)$ group is not leaving; and the same holds the other way round, the ''new'' symmetry has not arrived  just now;  only  it has been at work in an inconspicuous manner. (Its footprints were visible - the spectrum has showed more degeneracy that could be accounted for by only $SO(3)$.)\\
One  note at the beginning: To live up to our plan (''to examine a hidden symmetry in the hydrogen atom problem solved by means of NCQM.''), this chapter could be a little shorter than it is. However, a few notes have been inserted  which aim to show the problem in a bigger picture. A comparison of this few chapters to a hiking trip has been mentioned. One usually does not mind doing some sightseeing for its own sake and enjoying views also of the more distant peaks which may be a little out of the set-out track. \\
In this case there is conserved LRL vector  to be examined, the related law  being   important comparably to the conservation of angular momentum.  This quantity does not enter physics exclusively in the realm of quantum mechanics; on the contrary, the huge systems accessible to  bare-eye-observations have a strong connection  with it. It is  appealing to see patterns in Nature play in such a harmony, notwithstanding our present disability to find a description that would have the same quality.
\subsection{Laplace-Runge-Lenz vector enters}
Coulomb-Kepler problem corresponding to the motion of a particle in a field of a central force proportional to $r^{-2}$, was one of the main issues which stood in the centre  of attention at the very beginning of the modern physics.  Newton equation of motion for a particle of mass $m$  is
\be \label{CKP} m\,\dot{\vec{v}}\ =\ -\ q\,\frac{\vec{r}}{r^3}\,. \ee
Here $q$ is just a constant which has to do with the magnitude of the force applied. The system with a central force definitely \textit{is} symmetric, at least the rotational invariance is striking; and the  orbital momentum
\be\label{LRLclas} \vec{L}\ =\ m\,\vec{r}\,\times\,\vec{v} \,\ee
is conserved in any central field.
However, we are on lookout for\textit{ all} possible symmetries; and there are indeed more of them, because the force is not only central, but also falls off with distance as $r^{-2}$.
It turns out that such system is in a certain way equivalent to a harmonic oscillator moving in the four dimensional space $\mathbb{R}^4 \simeq {\bf C}^2$, which possesses eight integrals of motion. By fixing two of them one obtains system equivalent to Coulomb-Kepler problem  (\ref{CKP}). Thus, we should be able to find six integrals of motion. Three of them have been introduced already, namely the components of $\vec{L}$, the remaining three form the constant vector
\be \vec{A}\ =\ \vec{L}\,\times\,\vec{v}\ + q \,\frac{\vec{r}}{r}, \ee
which  is defined only up to an inessential multiplicative constant. Below we will use its normalization as indicated above. That $\dot{\vec{A}}=0$ follows directly from \eqref{CKP} and \eqref{LRLclas}.\\
\\
Since gravity was the first interaction people dealt with in modern physics, and because according to Newton the force keeping the solar system together goes as $r^{-2}$, it is natural that the first interpretation of the conserved quantity was provided by astronomy. Just briefly: The vector pointing towards the perihelion of the orbit is conserved; the fixed direction tells us that the perihelion does not precess, while the constant magnitude means that the eccentricity does not change. We know it is not quite like that in our solar system; but that is due to the fact that the force does not exactly meet the requirement of the $r^{-2}$ dependence, mainly due to 
the influence of other planets. This corrections have been estimated, and a tiny difference with the observed precession of Mercury's perihelion was explained by general relativity.\\
\\
A little historical note should be put here. We call the mentioned quantity Laplace-Runge-Lenz vector, as it is commonly done nowadays. This is, however,  more for the sake of convention than a historical correctness. The vector had been forgotten and brought to light again several times and neither of the three gentlemen was the first one to think of it. We have to go back into the earlier days of physics.\\
No such quantity was  found among Newton's publications, so he might really not have been aware of the related conservation law.  As far as we know, the first ones to make a mention of it were  Jakob Hermann and  Johann Bernoulli in the letters they exchanged in 1710, see \cite{He}, \cite{Be}. So the name ''Hermann-Bernoulli vector'' would probably be far more just. It was much later in  1799 that the vector was rediscovered by Laplace  in his Celestial mechanics \cite{La}. Then it appeared as an example  in a popular German textbook on vectors by C. Runge \cite{Ru}, which was referenced by W. Lenz in his paper on the (old) quantum mechanical treatment of the Kepler problem or hydrogen atom \cite{Le}.  After Pauli's publication \cite{Pa}, it became known mainly as the Runge-Lenz vector, nowadays this vector usually carries the name which we use - with certain objections - also here.\\
\\
As to the QM version of the story, it has been summed up already in the related chapter how Pauli found a QM analog of LRL vector and  how this lead him along a beautiful way to derive the hydrogen spectrum. Now we would like to examine whether, and if so, then how, is this affected by the noncommutativity of the space.
\subsection{Dynamical symmetry in NCQM}
Let us move on to the NCQM version of the Coulomb-Kepler problem. The question is, whether we can find sensible analogs of the three components $A_i$ of the LRL vector in such a way, that  all the requirements regarding commutation relations are satisfied (the commutator with the Hamiltonian has to be zero because of the conservation law and relations among all components of $\vec{A}$ and $\vec{L}$ are supposed to correspond to the relevant symmetry).
\\
We are going to answer this question by actually finding the NC version of $A_i$. There is no prescribed way to launch the search for it - in this case it was done by a combination of an educated guess and a piece of good luck. The ''educated'' part involves trying to keep the procedure  as close to the classical case as possible. To build up the ''original'' classical  LRL vector one just has to suitably combine the components of velocity, angular momentum and position vectors (it is no big deal  to replace momentum by velocity multiplied by mass in the classical mechanics, but an important step to introduce the idea that we actually have the NC analogs of all that is needed and the task reduces to finding the proper way of combining it together).\\
\\
Recall the lesson taken from QM: When constructing the QM version of LRL vector, the cross product of momentum and angular momentum needed to be symmetrized due to their non-vanishing commutator. The NC operators we are going to use when constructing the analog of the cross product part, i.e. $\hat{V}_i$, $\hat{L}_i$ do not commute either, so some symmetrization of the mentioned sort is supposed to take place as well.   Yet this is not all,  an additional subtlety  is to be taken into account: there is another, ''potential'' part of the LRL vector, in both classical and  quantum mechanics proportional to $ \vec{r}/ r$.  In QM $x_i/r$ was supposed just to multiply the wave function, but since the corresponding NC analogs of $x_i$ and $\Psi$ do not commute, as well the ordering in the product makes a difference and there is no reason to prefer either of the two possibilities.  We resolve this the same way we did in the cross product case - we choose both and take the average:
\be\label{A}
\hat{A}_k\ =\ \frac{1}{2}\,\varepsilon_{ijk}\, (\hat{L}_i \hat{V}_j\, +\, \hat{V}_j \hat{L}_i)\, +\, q\, \frac{\hat{X}_k}{r}\ ,
\ee
where $\hat{X}_k$ acts as $\hat{X}_k \Psi = (1/2)\, (x_k \Psi + \Psi x_k)$.\\
\\
Now comes the  ''lucky'' part of the story - besides coping with the ordering  dilemma, nothing  more needs to be done, except for actually working out the calculations to justify our definition of $\vec{A}$. That requires more work than  the  previous sentence may suggest, and the better part of this chapter deals with it.
\\
So now we are going to take the NC analogs of the Hamiltonian, velocity, angular momentum and position operators, mix them together according to the recipe similar to that known from the  QM case, and symmetrize what should be symmetrized.\\
\\
Then the main work will follow - evaluating the commutator $[\hat{A}_i,\hat{H}]$, examining the commutation relations between $\vec{A}$ and $\vec{L}$, rescaling $\vec{A}$ by a suitable numerical factor if needed... all in all,  searching for the signs of a higher dynamical symmetry. Once the symmetry group is known to be present, we can construct the corresponding Casimir operators. We know how Casimirs for $so(4)$ look like, and their prescribed eigenvalues  are ''responsible'' for the discrete energy spectrum.\\
All those that play  important roles: the Hamiltonian, velocity, angular momentum and position operators, have been defined already in terms of creation and annihilation operators $a^+_\alpha$, $ a_\alpha$;  we know commutation relations for these, so we should be able to calculate all that is needed. However, after writing it all down and trying to make heads and tails of it, one quickly comes to the conclusion that the problem is not assigned in the most friendly way. It is better to think twice before introducing any new symbols, but this is definitely the case when it helps. There are certain combinations of $a^+_\alpha$, $ a_\alpha$ that our expressions have in common, so separating them the right way makes the calculations easier.
\subsubsection{Auxiliary operators}
\noindent We are interested in the way in which the considered operators act on the wave functions $\Psi$.
They are expressed in terms of $a^{+}_\alpha$, $ a_\alpha$. Generally it matters whether the creation and annihilation operators act from the right or the left and the following notation will turn out to be useful.
\begin{eqnarray}
\hat{a}_\alpha \,\Psi = a_\alpha\, \Psi\,, &\,&\hat{b}_{\alpha }\,\Psi = \Psi\, a_{\alpha}\,, \\
\hat{a}^+_\alpha\, \Psi = a^+_\alpha\, \Psi\,, &\,&\hat{b}^{+}_{\alpha }\,\Psi = \Psi\, a^{+}_{\alpha }\,.
\end{eqnarray}
One obvious virtue of this notation is the fact that from now on we do not have to drag $\Psi$ into the formulas just to make clear which side do the operators act from. The relevant commutation relations are
\begin{equation} \label{kom}
[\hat{a}_\alpha,\, \hat{a}^+_\beta]\ =\ \delta_{\alpha \beta}\,,\ \ \
[\hat{b}_\alpha,\, \hat{b}^+_\beta]\ =\ -\,\delta_{\alpha \beta}\,.
\end{equation}
The other commutators are zero. Having this last sentence in mind spares a lot of paper while doing the calculations.\\
As already mentioned, we will use the position operator in the form
\begin{eqnarray} \label{xbar}
 \nonumber \hat{X}_i\,\Psi &=& \frac{1}{2}\,(x_i\, \Psi\, +\, \Psi\, x_i)\,=\, \frac{\lambda}{2}\,\sigma^i_{\alpha \beta}\,(\hat{a}^+_\alpha\, \hat{a}_\beta \, +\, \hat{b}_\beta\, \hat{b}^+_\alpha )\, \Psi\,,\\
  \hat{r}\,\Psi &=&\frac{1}{2}\,(r\, \Psi\, +\, \Psi\, r)\,=\,\frac{\lambda}{2}\,((\hat{a}^+_\alpha\, \hat{a}_\alpha +1)\,  +\, (\hat{b}_\alpha\, \hat{b}^+_\alpha\, +\,1) )\, \Psi \,.
\end{eqnarray}
Since our wavefunction $\Psi$ commutes with $r$ we have $\hat{r} \Psi = r \Psi = \Psi r$. Similarly $\hat{f}(r) \Psi = f(r) \Psi = \Psi f(r)$ for any (reasonable) function, e.g. $f(r) = \frac{1}{r}$.
\\
The following sequences of operators will occur so often and their
role is going to be so important that they deserve to have notation on their own:
$$
\begin{array}{lllclll}
\hat{w}_{\alpha \beta} &=& \hat{a}^+_\alpha \hat{b}_\beta - \hat{a}_\beta \hat{b}^+_\alpha\,,  &\,\,\,\, & \hat{\zeta}_{\alpha \beta} &=& \hat{a}^+_\alpha \hat{b}_\beta + \hat{a}_\beta \hat{b}^+_\alpha\,, \\
\hat{w} &=& \hat{w}_{\alpha \alpha} \,,                                    &\,\,\,\, & \hat{\zeta} &= &\hat{\zeta} _{\alpha \alpha} \,,   \\
\hat{w}_k  &=&  \sigma^k_{\alpha \beta}  \hat{w}_{\alpha \beta}\,,           & \,\,\,\,&  \hat{\zeta} _k &=& \sigma^k_{\alpha \beta} \hat{\zeta}_{\alpha \beta} \,.   \\
\end{array}
$$
\begin{eqnarray}
\label{W_k}\hat{W}_k &=&\frac{2\hat{X}_k }{\lambda}-\hat{\zeta} _k  \,\,= \,\, \sigma^k_{\alpha \beta} (\hat{a}^+_\alpha \hat{a}_\beta - \hat{a}^+_\alpha \hat{b}_\beta -\hat{a}_\beta \hat{b}^+_\alpha + \hat{b}^+_\alpha \hat{b}_\beta)\,, \label{W} \\
\hat{W} &=& \frac{2\hat{r}}{\lambda}-\hat{\zeta}   \,\, = \,\, \hat{a}^+_\alpha \hat{a}_\alpha - \hat{a}^+_\alpha \hat{b}_\alpha - \hat{b}^+_\alpha \hat{a}_\alpha +\hat{b}^+_\alpha \hat{b}_\alpha \,,\nonumber\\
&& \nonumber \\
\hat{W}'_k &=& \hat{W}_k + \omega \hat{X}_k =\eta \hat{X}_k - \hat{\zeta}_k \,,\nonumber \\ \nonumber
\hat{W}'& =& \hat{W} + \omega \hat{r} = \eta \hat{r} - \hat{\zeta}\,.
\end{eqnarray}

\noindent $ \sigma^k_{\alpha \beta}$ denotes the Pauli matrices and the  new letters which appeared in the  last two lines are just shorthands:
 $$
\begin{array}{lcl}
\omega\, =\, -2 \lambda E \,,& \mbox{\,\,\,\,}& \eta\, =\, ( \frac{2}{\lambda}\, +\, \omega)\,. \\
  \end{array}
$$
The reader perhaps remembers the note about another auxiliary $\eta$ which appeared in the Schr\"{o}din\-ger-like chapter. Hopefully just the first half of the rule ''forgive and forget'' is being followed in this case.
$E$ is energy and $\lambda$ is the NC parameter mentioned in the introduction. Note that the only difference between   $\hat{W}'_k$  and $\hat{W}_k$ is the constant multiplying one of their terms, $\hat{W}'$  and $\hat{W}$ are related in the same way.

\subsubsection{NC operators revisited}
\noindent We have introduced  new auxiliary operators which should  make the calculations more manageable, now we will rewrite everything relevant in their terms - the Hamiltonian, the velocity operator and the NC LRL vector.
\begin{eqnarray}
\hat{H} &= &\frac{1}{2\lambda \hat{r}}(\hat{a}^+_\alpha \hat{a}_\alpha + \hat{b}^+_\alpha \hat{b}_\alpha - \hat{a}^+_\alpha \hat{b}_\alpha -  \hat{a}_\alpha \hat{b}^+_\alpha ) -\frac{q}{\hat{r}}\ \nonumber \\
\label{H} &= &\frac{1}{2\lambda \hat{r}}(\frac{2\hat{r}}{\lambda}-\hat{\zeta}) -\frac{q}{\hat{r}} \,\,\,=\,\,\, \frac{1}{2\lambda \hat{r}} \hat{W}   -\frac{q}{\hat{r}} \,,
\end{eqnarray}

\begin{eqnarray}
 \label{V_i} \hat{V}_i &=& -i \left[\hat{X}_i, \hat{H} \right] = \frac{i}{2\hat{r}} \hat{w}_i \,,
\end{eqnarray}

 \begin{eqnarray}
\label{A_k} \hat{A}_k &= &\frac{1}{2}\varepsilon_{ijk} (\hat{L}_i \hat{V}_j + \hat{V}_j\hat{L}_i) + q \frac{\hat{X}_k}{\hat{r}} \,\, = \,\, -\frac{1}{2\lambda \hat{r}}(\hat{r}\hat{\zeta} _k - \hat{X}_k \hat{\zeta}) + q \frac{\hat{X}_k}{\hat{r}} \nonumber \\
&=& \frac{1}{2\hat{r}\lambda}(\hat{r}\hat{W}_k - \hat{X}_k \hat{W}) + q \frac{\hat{X}_k}{\hat{r}} \,\,=\,\, \frac{1}{2\hat{r}\lambda}(\hat{r}\hat{W}'_k - \hat{X}_k \hat{W}') + q \frac{\hat{X}_k}{\hat{r}} \\ \nonumber
&=& \frac{1}{2\hat{r}\lambda}(\hat{r}\hat{W}'_k - \hat{X}_k (\hat{W}' - 2 \lambda q)) \,.
\end{eqnarray}
Deriving equations (\ref{V_i}) and (\ref{A_k}) involves  somewhat  laborious calculations, many steps have been skipped here. However, all of them can be reconstructed from the  definitions given in the preceding paragraphs. Some more details are given in the Appendix, see (\ref{Acalc}). In the second line of (\ref{A_k}) we have used the equality of $(\hat{r}\hat{W}_k - \hat{X}_k \hat{W})=(\hat{r}\hat{W}'_k - \hat{X}_k \hat{W}')$.\\
\\
Now let us rewrite the NC Schr\"{o}dinger equation in the following way:
\be \label{SchEworkingform}
\left(\frac{1}{2\lambda \hat{r}} \hat{W}   -\frac{q}{\hat{r}} - E \right)\Psi_E =\frac{1}{2\lambda \hat{r}}(\hat{W}' - 2 \lambda q) \Psi_E= 0\,.
\ee
\\
$\Psi_E$ belongs to $\mathcal{H}_{\lambda}^E$, i.e. to the subspace spanned by the eigenvectors of the Hamiltonian.\\
\\
It makes things easier if we  recognize a zero when  stumbling across one. As for (\ref{SchEworkingform}),  it provides a funny form of zero - the calculations, however,  may get less amusing if this point is overlooked.
\\
Here is how does it pay off: Let us examine $\hat{A}_k |_{\mathcal{H}_{\lambda}^E}$, that is, the LRL vector as it works on the solutions of the Schr\"{o}dinger equation. (And once again: we do not have to know the $\Psi_E$ explicitly, they are just supposed to exist and form a basis for the subspace  $\mathcal{H}_{\lambda}^E$.)
\be\label{Aeigen}
\hat{A}_k |_{\mathcal{H}_{\lambda}^E} = \frac{1}{2\hat{r}\lambda}(\hat{r}\hat{W}'_k - \hat{X}_k \underbrace{(\hat{W}' - 2 \lambda q)}_{\mbox{see Eq. (\ref{SchEworkingform}) }})=\frac{1}{2\lambda}\hat{W}'_k \,.
\ee
It is nice to know the ''full form'' of $\hat{A}_k$, but  if we are interested in the actual physical bound  states, we do not need to take care of the whole abstract Hilbert space. We did not expect the $SO(4)$ symmetry to show elsewhere than $\mathcal{H}_{\lambda}^E$. So when dealing with calculations related to conservation of $A_k$, we just need to find out whether the following commutator with the Hamiltonian vanishes.
\begin{eqnarray}
\dot{\hat{W}}'_k &=&  i\left[\hat{H}_0 - \frac{q}{\hat{r}} , \hat{W}'_k\right] = i\left[\frac{1}{2\hat{r}\lambda}\hat{W} - \frac{q}{\hat{r}},\hat{W}'_k\right] \nonumber \\
&= & i\left[\frac{1}{2\hat{r}\lambda}\hat{W}' - \frac{q}{\hat{r}},\hat{W}'_k\right] = i\left[\frac{1}{2\hat{r}\lambda}\hat{W}', \hat{W}'_k\right] - iq\left[\frac{1}{\hat{r}},\hat{W}'_k\right]  \\ \nonumber
\label{dotA} &= & \frac{i}{2\hat{r}\lambda}\left[\hat{W}', \hat{W}'_k\right] + i \left[\frac{1}{\hat{r}}, \hat{W}'_k\right]\left(\frac{\hat{W}'}{2\lambda}-q\right)=0 \,.
\end{eqnarray}
In the second line, $\hat{W}'$ appears instead of $\hat{W}$. It is a legal step to do, since it does not change the commutator, and it is a sensible step too, since we obtained the (\ref{SchEworkingform})-style zero in the second term in the last line (it vanishes  when acting on vectors from $\mathcal{H}_{\lambda}^E$ and we are not interested in the rest of $\mathcal{H}_{\lambda}$). To prove that the first term  proportional to $[\hat{W}', \hat{W}'_k]$ also does not contribute requires calculations lengthy enough to be placed in the Appendix. \\
The equation above apparently conveys an encouragement to search for the underlying $SO(4)$ symmetry, since the LRL vector conservation makes its components suitable candidates for a half of its generators, the remaining three consisting of the components of the angular momentum.
So let us check it (for detailed derivation see (\ref{tildWW}) in the Appendix)
 \begin{equation}\label{AiAj}
[\hat{A}_i,\hat{A}_j]  =\frac{1}{4\lambda^2} [\hat{W}'_i, \hat{W}'_j]  = i\frac{\omega}{\lambda}\left(1+\frac{\omega\lambda}{4}\right)\varepsilon_{ijk} \hat{L}_k ,
\end{equation}
or to see explicitly the energy dependence ($\omega = - 2E \lambda$):
\begin{equation}
[\hat{A}_i,\hat{A}_j] = i \varepsilon_{ijk} \left(- 2E + \lambda^2 E^2\right) \hat{L}_k.
\end{equation}
There is nothing but a constant in the way, as long as we let the operator $[\hat{A}_i,\hat{A}_j]$ act upon the states from $\mathcal{H}_{\lambda}^E$ with the energy fixed. Eq. (\ref{AiAj}) and
\begin{equation}\label{LA}
[\hat{L}_i, \hat{L}_j] = i \varepsilon_{ijk} \hat{L}_k, \ \ \ [\hat{L}_i,\hat{A}_j] = i \varepsilon_{ijk} \hat{A}_k ,
\end{equation}
define  Lie algebra relations corresponding to a particular symmetry group which actual form  depends on the sign of the $E$ dependent factor in (\ref{AiAj}). The relevant relations for $\hat{L}_i$ have been already mentioned, so we just need to check the mixed commutator $[\hat{L}_i, \hat{A}_j]$. This is a long process again, so to not distract our attention from what is going on (besides reshuffling operators here and there), we relocated the calculation to the Appendix, see eq. (\ref{liwj}) and the related ones. \\
\\
There are three independent cases\\
$\bullet $ $SO(4)$ symmetry: $ -2E + \lambda^2 E^2 > 0\ \Longleftrightarrow \ E < 0 $ or $E > 2/\lambda^2 $,\\
$\bullet $ $SO(3,1)$ symmetry: $ - 2E + \lambda^2 E^2 < 0\ \Longleftrightarrow \ 0 <  E < 2/\lambda^2 $,\\
$\bullet $ $E(3)$ Euclidean group:  $ - 2E + \lambda^2 E^2 = 0\ \Longleftrightarrow \ E = 0 $ or $E = 2/\lambda^2 $.\\
\\
The admissible values of $E$ should correspond to the unitary representations of the symmetry in question. This requirement guarantees that the generators $\hat{L}_j$ an  $\hat{A}_j$ are realized as hermitian operators, and consequently correspond to physical observables. The Casimir operators in all mentioned cases are
\begin{eqnarray}
\nn \hat{C}^\prime_1 &=& \hat{L}_j \hat{A}_j\,,\\
\hat{C}^\prime_2 &=& \hat{A}_i\hat{A}_i+(- 2E + \lambda^2 E^2)(\hat{L}_i\hat{L}_i+1) \\
\nn \label{CAS} &=& \frac{1}{4\lambda^2} \left(\hat{W}'_i\hat{W}'_i +(\eta^2\lambda^2-4)(\hat{L}_i\hat{L}_i+1) \right) \,.
\end{eqnarray}
The prime indicates that we are not using the standard normalization of Casimir operators.\\
\\
Now, we need to calculate their values in $\mathcal{H}_{\lambda}^E$. The first Casimir  is vanishing in all cases due to the fact that $\hat{C}^\prime_1 \Psi_E \sim r \Psi_E - \Psi_E r = 0$. (See (\ref{zeroCas1}), (\ref{zeroCas2}) in the Appendix.) The second Casimir operator is somewhat more demanding, and either believe it or convince yourself (consulting the part of the Appendix beginning with (\ref{id}) may help in the latter case), the terms in the bracket add up  exactly to  $(\hat{W}')^2$. According to the Schr\"{o}dinger equation, $(\hat{W}')^2\Psi_E = 4\lambda^2 q^2 \Psi_E $, and we are left with
\begin{equation} \label{CAS2}
\hat{C}^\prime_2\, \Psi_E \ =\ \left( \hat{A}_i\hat{A}_i\,+\,(- 2E + \lambda^2 E^2) \, (\hat{L}_i\hat{L}_i+1) \right) \Psi_E \ =\ q^2\, \Psi_E \,.
\end{equation}
Since both Casimir operators take  constant values $\hat{C}^\prime_1 = 0$ and $\hat{C}^\prime_2 = q^2$ in $\mathcal{H}_{\lambda}^E$, we are dealing with irreducible representations of the dynamical symmetry group $G$ that are unitary for particular values of energy. In all considered cases, $G = SO(4), SO(3,1), E(3)$, the unitary irreducible representations are well known. The corresponding systems of eigenfunctions that span the representation space have been found in chapter 4. Here we shall  restrict ourselves to brief comments pointing out some interesting aspects.
\subsubsection{SO(4) symmetry \& bound states}
\noindent {\bf 1. Bound states -} $\boldsymbol{SO(4)}$ {\bf symmetry:} $\boldsymbol{- 2E + \lambda^2 E^2 > 0}$. In this case we rescale the LRL vector as
\be
\hat{K}_j \ = \frac{\hat{A}_j}{\sqrt{- 2E + \lambda^2 E^2}}\ =\  \frac{\hat{W}'_j}{\sqrt{\eta^2\lambda^2-4}}\,.
\ee
After this step Eqs. (\ref{AiAj}), (\ref{LA}) turn into the following relations:
\begin{equation}\label{lilj}
[\hat{L}_i, \hat{L}_j] = i \varepsilon_{ijk} \hat{L}_k \,,\ \ \
[\hat{L}_i, \hat{K}_j] = i \varepsilon_{ijk} \hat{K}_k \,, \ \ \
[\hat{K}_i, \hat{K}_j] = i \varepsilon_{ijk} \hat{L}_k \,.
\end{equation}
Thus we have got the representation of the $so(4)$ algebra. The relevant normalized Casimir operators read
\begin{equation}
\hat{C}_1\ =\ \hat{L}_i\, \hat{K}_i\,,\ \ \  \hat{C}_2\ =\ \hat{K}_i\,\hat{K}_i\,+\,\hat{L}_i\hat{L}_i\,+\,1\,.
\end{equation}
As we have stated already, the $\hat{C}_1$ acting on an eigenfunction of the Hamiltonian returns zero. As to $\hat{C}_2$, we know that for $so(4)$,  under the condition that the first Casimir is zero, the second one  has to equal to $(2j+1)^2$ for some integer or half-integer $j$. At the same time, according to (\ref{CAS2}) it is related to the energy:
\begin{eqnarray}\label{cas2}
 \hat{K}_i\,\hat{K}_i\,+\,\hat{L}_i\,\hat{L}_i+1 &=&  \frac{4\lambda^2 q^2}{\eta^2\lambda^2-4}\,=\, \frac{ q^2}{\lambda^2 E^2 - 2E}\,, \nonumber \\
 & &    \nonumber \\
(2j+1)^2 &= & \frac{ q^2}{\lambda^2 E^2 - 2E}\ .
\end{eqnarray}
We will write $n^2$, $n= 1,\, 2,\, ...$ instead of $(2j+1)^2 $.  Now solving the quadratic equation for energy  we obtain two discrete sets of solutions depending on $n$:
\begin{equation}\label{spectrum+-}
E\ =\  \frac{1}{\lambda^2}\, \mp\, \frac{1}{\lambda^2} \sqrt{1+\kappa_n^2}\,,\ \ \ \kappa_n\ =\ \frac{q\lambda}{n} \,.
\end{equation}
\\
Now let us show the connection between the present chapter and that  one dealing with the problem in Schr\"{o}dinger's style. The first set of eigenfunctions of the Hamiltonian  for energies $E < 0$ (i.e. negative sign in front of the square root in (\ref{spectrum+-})) has been found for Coulomb {\it attractive} potential, i.e. $q>0$. The eigenvalues possess smooth standard limit for $\lambda\rightarrow 0$.
\begin{eqnarray}
E^I_{\lambda n} &=&  \frac{1}{\lambda^2}\, -\, \frac{1}{\lambda^2} \sqrt{1+\kappa_n^2}\, =\,  \frac{1}{\lambda^2}\, -\, \frac{1}{\lambda^2}\, \left(1\,+\, \frac{1}{2}\, \kappa_n^2\, -\, \frac{1}{24}\, \kappa_n^4\, +\, ...\right)\nonumber \\
 & &    \nonumber \\
& =&   -\frac{q^2}{2n^2}\, +\, \lambda^2\, \frac{q^4 }{24 n^4}\, +\, ... \nonumber \\
 & &    \nonumber
 \end{eqnarray}
Recall that we usually have $m_e=1, \, \hbar=1 $ so as to not care whether we write $q$ or $\alpha = m_e q\hbar^{-2}$. If we give up those units, we have
\begin{eqnarray}\label{spectrum}
 E^I_{\lambda n}&= &   -\frac{q^2m_e}{2n^2\hbar^2}\, +\, \lambda^2 \frac{q^4 m_e^3}{24 n^4\hbar^6} + ...
\end{eqnarray}
The ''...'' stands for the terms proportional to higher (even) powers of $\lambda$. The first term corresponds to the energy spectrum of H-atom in the ordinary QM, remaining are $\lambda$-dependent corrections.
\\
The full set of eigenfunctions of Hamiltonian for energies $E < 0$ has been constructed in Chapter 4. by explicitly solving the NC Schr\"odinger equation.  The radial NC wave functions  are
\begin{eqnarray}
\nn R^I_{\lambda n} &=& (\Omega_n)^N\, F(-n,\, -N,\, 2j+2,\, -2 \kappa_n\, \Omega_n^{-1})\,,\\
\label{radI} \Omega_n &=& \frac{\kappa_n - \sqrt{1+\kappa_n^2}+1}{\kappa_n + \sqrt{1+\kappa_n^2} -1}\,,
\end{eqnarray}
\\
The second set of very unexpected solutions corresponds to energies (\ref{spectrum+-}) with positive sign
\begin{equation}\label{spectrum'}
E^{II}_{\lambda n}\ =\  \frac{1}{\lambda^2}\, + \,\frac{1}{\lambda^2}\, \sqrt{1+\kappa_n^2}\, >\, \frac{2}{\lambda^2} \,.
\end{equation}
The corresponding radial NC wave functions has been found in Chapter 4. solving NC Schr\"odinger equation for a Coulomb {\it repulsive}  potential, $q<0$. These radial NC wave functions are closely related to those given above
\begin{equation}\label{radII}
R^{II}_{\lambda n}\ =\ (-\Omega_n)^N\, F(-n,\, -N,\, 2j+2,\, 2 \kappa_n\, \Omega_n^{-1})\,.
\end{equation}
Let us denote the first set of eigenfunctions for the Coulomb attractive potential with radial part $R^I_{\lambda n}$  as $\Psi^I_{nlm}$, and similarly, the second set of solutions with radial part $R^{II}_{\lambda n}$ for Coulomb repulsive potential by $\Psi^{II}_{nlm}$. The mapping
$$ \Psi^I_{nlm}\ \longmapsto\  \Psi^{II}_{nlm} $$
is obviously a unitary transformation in ${\cal H}_\lambda$. Thus both $SO(4)$ representations, the one for Coulomb attractive potential with $E^I_{\lambda n} < 0$ and that for ultra-high energies $E^{II}_{\lambda n} > 2/\lambda^2$ for Coulomb repulsive potential, are unitary equivalent. That is natural, since in both representations the Casimir operators take the same values, $ 0$ and $ n^2$. However, in the commutative limit $\lambda\rightarrow 0$ the extraordinary bound states at ultra-high energies disappear from the Hilbert space.
\subsubsection{SO(3,1) symmetry \& scattering}
\noindent {\bf 2. Coulomb scattering:} $\boldsymbol{2E - \lambda^2 E^2 > 0$ $\Longleftrightarrow$ $0 < E < 2/\lambda^2}$. In this case we rescale the LRL vector as
\be
\hat{K}_j =\frac{\hat{A}_j}{\sqrt{2E - \lambda^2 E^2}} = \frac{\hat{W}'_j}{\sqrt{4-\eta^2\lambda^2}},
\ee
After this step we obtain equations
\begin{equation}\label{lilj'}
[\hat{L}_i,\, \hat{L}_j]\, = \,i\, \varepsilon_{ijk}\, \hat{L}_k \,, \ \ \
[\hat{L}_i,\, \hat{K}_j]\, =\, i\, \varepsilon_{ijk}\, \hat{K}_k \,, \ \ \
[\hat{K}_i,\, \hat{K}_j]\, =\, - i\, \varepsilon_{ijk}\, \hat{L}_k \,.
\end{equation}
So this time we have got the representation of the $so(3,1)$ algebra. The relevant normalized Casimir operators read
\begin{equation}
\hat{C}_1\ =\ \hat{L}_i\, \hat{K}_i\,,\ \ \  \hat{C}_2\ =\ \hat{K}_i\,\hat{K}_i \,-\, \hat{L}_i\,\hat{L}_i\,.
\end{equation}
In our case $\hat{C}_1 = 0$, so we are dealing with $SO(3,1)$ unitary representations that are labeled by the value of second Casimir operator $\hat{C}_2 = \tau $, see e.g \cite{BarutRanczka}:

$\bullet $ Spherical principal series for $\tau > 1$;

$\bullet $ Complementary series for $ 0 < \tau < 1 $.\\
\\
Let us point out that for integer $\hat{C}_1 \neq 0$ and arbitrary real $\hat{C}_2$ a (non-spherical) remainder of principal series appears that completes the set of unitary representations of $SO(3,1)$ group.\\
\\
Rewriting (\ref{CAS2}) in terms of $\hat{K}_j$ we obtain relation between energy $E$ and the parameter $\tau$:
\begin{equation}
\hat{K}_i\,\hat{K}_i\,-\,\hat{L}_i\,\hat{L}_i\ =\ 1\, +\, \frac{q^2}{2E - \lambda^2 E^2}\ =\ \tau\ >\ 1\,.
\end{equation}
Thus we are dealing with the principal series $SO(3,1)$ unitary representations. The scattering NC wave functions have been constructed in Chapter 4. For any admissible energy $E \in (0,\,2/\lambda^2)$, and from their asymptotic behavior the partial wave $S$-matrix has been derived
\begin{equation}
S^\lambda_j (E)\ =\ \frac{\Gamma (j + 1 - i \frac{q}{p})}{\Gamma (j + 1 + i \frac{q}{p})}\,,\ \ \ p\ =\ \sqrt{2E - \lambda^2 E^2 }\ .
\end{equation}
(Again, in our favourite units  $m_e=1, \, \hbar=1 $  we have $\alpha = m_eq\hbar^{-2}=q$).
It can be easily seen that such $S$-matrix possesses poles at energies $E = E^I_{\lambda n}$ for Coulomb attractive potential and  poles at energies $E = E^{II}_{\lambda n}$ for Coulomb repulsive potential, where both $E^I_{\lambda n}$ and $E^{II}_{\lambda n}$ coincide with (\ref{spectrum}) and (\ref{spectrum'}) given above. As for energies
\begin{equation}
E_\mp\ =\ \frac{1}{\lambda^2} \left( 1\, \mp\, \sqrt{1 - \frac{\lambda^2 q^2}{\tau -1}}\right),
\end{equation}
the values of Casimir operators coincide, the corresponding representations are unitary equivalent. This relates the scattering for energies $0 < E < 1/\lambda^2$ to that at high energies $1/\lambda^2 < E < 2/\lambda^2$.
\\
\\
We skip the limiting cases of the scattering at the edges $E = 0$ and $E = 2/\lambda^2$ of the admissible interval of energies, where the $SO(3,1)$ group contracts to the group $E(3) = SO(3)\triangleright T(3)$ of isometries of 3D space with generators $\hat{L}_j$ and $\hat{A}_j$ satisfying commutation relations (see \cite{BarutRanczka}):
\begin{equation}
[\hat{L}_i,\, \hat{L}_j]\ =\ i\, \varepsilon_{ijk}\, \hat{L}_k\,, \ \ \ [\hat{L}_i,\,\hat{A}_j]\, =\, i\, \varepsilon_{ijk}\, \hat{A}_k\,, \ \ \ [\hat{A}_i,\,\hat{A}_j]\, =\,0\,.
\end{equation}
The corresponding NC hamiltonian eigenstates are given in Chapter 4.\\
\\
We are done also with the NC version of Pauli's approach to hydrogen atom. It is time to discuss the results.

\section{Conclusions}
If we stick with our parable of a backpacking trip,  this chapter probably corresponds to taking a rest and maybe going through the most memorable entries in the travel diary or photographs taken along the way. So let us review them; sum up what has been done in the matter of hydrogen atom problem in NC setting. \\
The problem has been dealt with in a twofold way; however, the beginning was common to both: The NC rotationally invariant analog of the
QM configuration space and the Hilbert space of operator wave
functions have been introduced. The central point of the construction was the definition of $\hat{\Delta}_\lambda$,  the NC analog
of Laplacian, supplemented by a consequent definition of the weighted Hilbert-Schmidt norm and the definition of the Coulomb
potential satisfying NC Laplace equation.\\
Then the road was split up, offering two main directions - differential Schr\"{o}dinger approach and the higher-symmetry-based Pauli's method. There was an interesting intermezzo, however, related to the examining of the velocity operator - while  the above mentioned objects were included in the process, and the whole issue was to be deeply involved in what was about to follow,  the velocity operator provided a way to explore also some more general aspects of NCQM, not focusing solely on the hydrogen atom problem.
\subsection{Comments on NC version of Schr\"{o}dinger's approach}
The NC analog of the Schr\"{o}dinger equation was built, taking advantage of the spherical symmetry of the problem when separating the radial part. The knowledge of how are normally ordered powers of the ''NC radial variable'' correlated with the ''usual'' powers enabled us to solve the NC problem using the associated ordinary differential equation. Let us see how NCQM matches the ordinary QM in this instance.\\
\\
The quantity labeling the solutions of Schr\"{o}dinger equation is energy. Some of the labels are excluded in the sense
that they cannot be attributed to a physical state. In standard QM, the solutions with negative energy were dismissed as
lacking the physical interpretation, except for those with some special energy values for which the wave function was normalizable.
\\
The state of affairs seems to be a bit different in NCQM. Although the energy spectrum for an electron trapped in the atom
is predicted in agreement with QM with small correction of order $\lambda^2$ , there  are \textit{two} special values
of energy, $E=0$ and $E_{crit}=2\hbar^2/(m_e\lambda^2)$ with the following feature: certain energy values  below $E=0$  for an attractive Coulomb potential  and certain values above $E_{crit}$  for repulsive potential provide  normalizable states.
\\
There is a remarkable symmetry between normalizable states corresponding
to $E^I_n < 0$ and those corresponding to $E^{II}_n > E_{crit}$:
\\
The bound state energies are symmetric with respect to the energy
$\frac{1}{2} E_{crit}\,=\,\hbar^2/(m_e \lambda^2)$,
$$ E^{I,\,II}_{\lambda\,n}\ =\ \frac{1}{2} E_{crit}\left( 1\,\mp\,
\sqrt{1 + (\lambda\alpha/n)^2}\right)\,. $$
Moreover, the corresponding radial wave functions are equal up to the
change $\alpha\,\to\,-\alpha$ (which is equivalent to swapping attraction and repulsion) and the sign changing factor at each step
$\lambda$ in the radial direction,
$$ R^{II}_{nj}(-\alpha)\ =\ (-1)^N\ R^I_{nj} (\alpha)\,. $$
The same symmetry can be seen for scattering states for energies:
$$ E^I\,=\,\frac{1}{2} E_{crit}\,-\,\varepsilon\,,\ \ \ \
p^I\,=\,+\,\sqrt{\frac{1}{2} E_{crit} - \lambda^2}\,, $$
$$ E^{II}\,=\,\frac{1}{2} E_{crit}\,+\,\varepsilon\,,\ \ \ \
p^{II}\,=\,-\,\sqrt{\frac{1}{2} E_{crit} - \lambda^2}\,, $$
with $\varepsilon\,\in\,(0, \frac{1}{2} E_{crit})$. Namely,
$$ R^{II}_{\varepsilon j}(-\alpha)\ =\
(-1)^N\ R^I_{\varepsilon j} (\alpha)\,. $$
%
%
%
It would be highly desirable to see the background of this almost perfect
reflection symmetry with respect to $\frac{1}{2} E_{crit}$.
\\
\\
The magnitude of the NC corrections to the $H$-atom energy levels  given in this paper
depends on the magnitude of $\lambda$, and this can be small enough to produce effects which are out of reach of experiments at the moment. However, in the earliest stages of the Universe - this symmetry could have mattered (but for such hight energies NC QM has to be replaced with (general) relativistic theory).\\
\\
 Our investigation indicates that the non - commutativity of the configuration space is fully consistent with the general QM
axioms, since in the correspondence limit $\lambda \rightarrow 0$ NCQM and QM agree - at least as far as the hydrogen atom problem solved in Schr\"{o}dinger-like approach is considered.
\subsection{Comments on NC version of Pauli's approach}
That chapter has been  focused on the Coulomb-Kepler problem in NC space. We have found the NC analog of the LRL vector; its components,   together with those of  the NC angular momentum operator,  supply the algebra of generators of the symmetry group. It is remarkable that the  formula for the NC generalization of LRL vector looks like a mirror image of the standard formula when written in terms of the proper NC observables: NC angular momentum, NC velocity, symmetrized NC coordinate and NC radial distance.\\
\\
The group $SO(4)$ has appeared twice: Firstly,   we have been dealing with bound states for negative energies in the case of the attractive Coulomb potential, which have an analog in the standard quantum mechanics; and secondly,   we have found an unexpected set of bound states for positive energies above certain ultra-high value in the case the potential is repulsive.\\
\\
$SO(3,1)$ is the symmetry group to be considered when examining the scattering (relevant for the interval of energies between zero and the mentioned critical ultra-high  value).\\
\\
The conservation  of our NC LRL vector  has been shown, as well as the commutation relations  related to $so(4)$ and $so(3,1)$ algebra. In case of the bound states (of both kinds), the calculations of the relevant Casimir operators  have revealed the specific values that the energy can take. The result for negative energy bound states  coincides with the well known QM prediction in the \,''commutative'' limit  $\lambda \rightarrow  0 $.  Besides that, the NC corrections proportional to $\lambda^2$ and higher (even) powers occur. As to the unexpected ultra-high energy bound states, they disappear from the Hilbert space providing that $\lambda \rightarrow  0 $.  \\
\\
\\
So now we have NCQM predictions for  the hydrogen atom energy spectrum obtained in two ways:\\
$\bullet$ the results  we got using the dynamical symmetry in analogy with Pauli's method  (without explicitly solving the NC Schr\"{o}dinger equation for eigenstates of the Hamiltonian)\\
$\bullet$  result   acquired by explicitly solving  the NC analog of the Schr\"{o}dinger equation  (without using those dynamical symmetry arguments). \\

 \,\,\, \textbf{We are glad to find out that these outcomes \textit{do }agree.}\\
 \\
 \\
Now it is time to ''step back'' and conclude our trip with a  look at the ''bigger picture'':
\subsection{What was it all about?}
If the answer to this question should be inferred from the expressions that occurred  most frequently, one would hesitate whether it is hydrogen, noncommutative geometry, quantum mechanics... So let us put things into perspective. What kind of a trip has all this been?\\
\\
Does our  reader still recall the metaphor about the world theatre from the introduction?  \emph{There are certain peculiarities about the drama that Nature performs... this is one strange theatre we are dealing with}, that was roughly the (rather indistinct) introductory statement.  Could it be the stage - space-time - that we understand so poorly? Could it be ''different'' in ways that avoided detection carried out by theories so successful as quantum mechanics or general relativity? Could this theories  have blind spots in just those respects in which this otherness of space-time can manifest itself? Those were the (still somewhat vague) questions. And finally, there was a resolution to try it - using a toy model that is clearly missing some features of the ''real stage'', but chosen so that its saving graces may be enough to make the play meaningful. It is quite an irony that we have moved from a vague statements to a concrete  model, which, however, is by its own definition a bit blurred. It would seem like we are destined for certain haziness.\\
\\
Have we proved the stage structure is what makes the theatre so peculiar? No. It has not been our ambition anyway. However, we have been playing on a toy stage for a few years, and this seems (to us at least, for what it is worth) to resemble the real one in certain aspects. The rather surprising results was, that after we spoilt the stage (quite dramatically), QM survived with an elegance. There may be some sense in  games of this kind after all.\\
\\
This trip has been a tour of the stage, with hydrogen providing guidance during the excursion.  Being itself an actor, our guide showed the expected flair for answering ''what if'' questions: \emph{''Supposing you would have to play your part once on the stage  being this way, and then on the stage being different; how like or unlike would your performances be? How would your lines change? Could that alteration be so subtle that we might hardly even notice?'' }\\
\\
In fact our excursion covered only of a fraction of the stage; only that area in which we were able to follow our guide.
There are parts which we may hope to try exploring in the future - for example, the stage is space-time, and  we have been dealing only with the stationary aspect here. Maybe Dirac equation could serve similarly as Schr\"{o}dinger one. That is just one of many paths we can choose. It may prove to be impassable or not. In any case, we have been told fractions of how the stage might be - our guide has  probably been telling much more than we have heard, and presumably also  things we never thought of asking about. Assuming from  the appealing nature of those fractions that we were able to catch, told by spectral lines and symmetries, the rest is in all likelihood worth hearing as well. Maybe it will require learning some new vocabulary, set of notions or a completely new language to be able to even ask about those things and be able to understand the answer.  \\
\\
 It is time to say farewell.  Does the reader know that old saying, often referred to as ''Chinese curse''? Under certain interpretation it can be considered to be more of a blessing. \textit{''May you live in interesting times.''} Hopefully they will be interesting because of the yet unexplored ways to wander and wonder about the Nature's  theatre. Why use the future tense, anyway. Supposedly there really is no time like the present. Photons even claim to have no other time at all and that attitude seems to pay off, since their lot surely is light and bright. So may the reader live in interesting and light bulb moments ... all the time there is.

\section{Appendix}
\noindent Some more detailed calculations follow that have been skipped in order to avoid overloading the previous chapters with profusion of technical details.
\subsection*{Appendix A}
Here we prove two formulas (\ref{appen}) we need for the
calculation of the NC Coulomb Hamiltonian. (Indices $j$ and $m$ are skipped here.) Let us begin with the first formula:
\beqa\nn [a^\dagger_\alpha ,[a_\alpha
,\Psi_{jm}]] &=& \lambda^j [a^\dagger_\alpha
,\,[a_\alpha ,\sum_{(jm)}\frac{(a^\dagger_1
)^{m_1}\,(a^\dagger_2)^{m_2}}{m_1!\,m_2!}\ :{\cal R}:
\frac{a^{n_1}_1\,(-a_2)^{n_2}}{n_1!\ n_2!}]]\\
\nn &=&\ \lambda^j \sum_{(jm)}[a_\alpha
,\frac{(a^\dagger_1 )^{m_1}
(a^\dagger_2)^{m_2}}{m_1!\,m_2!}]\,
[a^\dagger_\alpha,:{\cal R}:]\,
\frac{a^{n_1}_1(-a_2)^{n_2}}{n_1!\ n_2!} \\
\label{A1}  &+&\ \lambda^j \sum_{(jm)}[a_\alpha
,\,\frac{(a^\dagger_1
)^{m_1}\,(a^\dagger_2)^{m_2}}{m_1!\,m_2!}]\,:{\cal R}:\,
[a^\dagger_\alpha,\,\frac{a^{n_1}_1
(-a_2)^{n_2}}{n_1!\ n_2!}] \\
\nn  &+& \lambda^j \sum_{(jm)}\frac{(a^\dagger_1
)^{m_1}\,(a^\dagger_2)^{m_2}}{m_1!\,m_2!}\,
[a^\dagger_\alpha,\,[a_\alpha ,:{\cal R}:]]\,
\frac{a^{n_1}_1\,(-a_2)^{n_2}}{n_1!\
n_2!} \\
\nn &+& \lambda^j \sum_{(jm)}\frac{(a^\dagger_1
)^{m_1}\,(a^\dagger_2)^{m_2}}{m_1!\ m_2!}\ [a_\alpha
,:{\cal R}:]\, [a^\dagger_\alpha,\frac{a^{n_1}_1
\,(-a_2)^{n_2}}{n_1!\ n_2!}] ,  \eeqa
where ${\cal R}\,=\,\sum_{k=0}^\infty\,c_k \varrho^k =\,\sum_{k=0}^\infty\,c_k \lambda^k N^k$. The second sum after the last $=$ vanishes.
Now we shall use the
following commutation relations
\beqa\nn [a^\dagger_\alpha,:N^k:]\,=- \,k\,
a^\dagger_\alpha\, :N^{k-1}:\ \ &\Rightarrow& \ \
[a^\dagger_\alpha,:{\cal R}:]\,=\, -\lambda a^\dagger_\alpha\,
:\partial_{\varrho}{\cal R}:\,,\\
\nn [a_\alpha,:N^k:]\,=\, k\,
:N^{k-1}:\,a_\alpha\ \ &\Rightarrow& \ \
[a_\alpha,:{\cal R}:]\,=\,\lambda :\partial_{\varrho}{\cal R}:
\,a_\alpha \,, \\
\label{A2 } & & \eeqa
where $\partial_{\varrho}$ denotes the standard derivatives of ${\cal R}$
$\varrho$: $\partial_{\varrho}{\cal R}\,=\, \sum_{k=1}^\infty
k\,c_k\, \varrho^{k-1}$.\\
The the first and third line give the same contribution
\be\label{A3 }  \sum_{(jm)}\,\frac{(a^\dagger_1
)^{m_1}\,(a^\dagger_2)^{m_2}}{m_1!\,m_2!}\
(-\lambda \, j\,:\partial_{\varrho}{\cal R}:)\
\frac{a^{n_1}_1\,(-a_2)^{n_2}}{n_1!\ n_2!}\,.\ee
From (\ref{A2 }) the double commutator
$[a^\dagger_\alpha,[a_\alpha,:{\cal R}:]]$ follows
directly, and this gives the value of the third line in (\ref{A1})
\be\label{A4 } \sum_{(jm)}\,\frac{(a^\dagger_1 )^{m_1}\,
(a^\dagger_2)^{m_2}}{m_1!\,m_2!}\ (-\lambda :\varrho \,
\partial^2_{\varrho} {\cal R}:\,+\,2 \lambda \,:\partial_{\varrho}{\cal R}:)\
\frac{a^{n_1}_1\,(-a_2)^{n_2}}{n_1!\ n_2!}\,.\ee
The last two equations yields the first
formula in (\ref{appen}).
Let us move to another relation. From equation (\ref{nk}) it follows easily
\be\label{A5} N\,:N^k:\ =\ :N^{k+1}:\,+\,k\,:
N^k:\ \ \ \Rightarrow\ \ \ \varrho\,:{\cal R}:\ =\
:\varrho \,{\cal R}:\,+\,\lambda :\varrho\,\partial_{\varrho}{\cal R}:\,.\ee
This relation directly gives the second formula in (\ref{appen}) :
\beqa \nn &(N +1)&\sum_{(jm)}\,\frac{(a^\dagger_1
)^{m_1}\,(a^\dagger_2)^{m_2}}{m_1!\,m_2!}\ :{\cal R}:\
\frac{a^{n_1}_1\,(-a_2)^{n_2}}{n_1!\ n_2!} \\
 \label{A6}&=& \ \sum_{(jm)}\,\dots\ [(N +j +1)\,:{\cal R}:]\
\dots \\
\nn &=& \ \sum_{(jm)}\,\dots\ :[(N +j
+1)\,{\cal R}\,+\,\varrho\,\partial_{\varrho}{\cal R}]:\
\dots\,,\eeqa
where we have replaced  both untouched factors containing
annihilation and creation operators by dots.
\subsection*{ Appendix B}
\noindent This part of the Appendix deals with how are normal and usual powers of $\varrho$ related, and how can this information be used to rewrite various kinds of power series. Recall that
\be
\begin{array}{lr}
:\varrho^k: \ = \lambda^k \frac{N!}{(N-k)!}=(-\lambda)^k(-N)_k, \ \ \  & \ \ \  :\varrho^{-k}: \ = \lambda^{-k} \frac{N!}{(N+k)!}= \lambda^{-k}
\frac{1}{(N)_k}.\\
\end{array}
\ee
$(N)_k$ stands for Pochhammer symbol. Since $\varrho = \lambda N$, there really are  powers of  $\varrho$ on the right-hand sides. \\
Now to the above mentioned power series. Let us start with finding out how does the function $:e^{\beta \varrho}:$ modify when we rewrite  $:\varrho^n:$  in terms of  $\varrho^n$  in the corresponding Taylor series  ($\beta$
denotes some arbitrary constant here):
\be
\begin{array}{cl}
:e^{\beta \varrho}: & = \sum^\infty_{k=0} \frac{\beta^k}{k!}:\varrho^k:  \ = \sum^\infty_{k=0} \frac{(\beta\lambda)^k}{k!}
\frac{N!}{(N-k)!} = \\
& \\
& = \left(1+ \lambda \beta \right)^{N} = \left(  1+ \lambda \beta \right)^{\frac{\varrho}{\lambda}} . \\
\end{array}
\ee
Considering the limit $\lambda\rightarrow 0$  the above equation corresponds to the knowm Euler's formula. A potential doubt arising from the colon
marks on the left hand side ought to be dismissed due to the fact that $:\varrho^n: \, \rightarrow \varrho^n$ if $\lambda\rightarrow 0$.
There is no way to distinguish between the normal and usual ordering in $\lambda=0$ world. \\
\\
We can move to a more complex tasks now. In the course of many calculations we need to handle expressions of the kind $:\varrho^n e^{\beta \varrho}:$
\be\nn
\begin{array}{cl}
 :\varrho^n e^{\beta \varrho}: &= \sum^\infty_{k=0}\frac {\beta^k :\varrho ^{n+k}:}{k!}=
\sum^\infty_{k=0}\frac {\beta^k \lambda^{n+k}}{k!} \frac{N!}{(N-(n+k))!} = \\
  & \\
&=  \lambda^n  \frac{N!}{(N-n)!}    \sum^\infty_{k=0}\frac {\beta^k \lambda^{k}}{k!}
\frac{(N-n)!}{(N-n-k)!}       =    \\
& \\
  &      = \ \lambda^n  \frac{N!}{(N-n)!}   \left(   1+ \beta \lambda \right)^{N-n} .
     \\
\end{array}
\ee
And in the case of negative powers:
\be\nn \begin{array}{cl}
 :\varrho^{-n} e^{\beta \varrho}: &= \sum^\infty_{k=0}\frac {\beta^k :\varrho ^{k-n}:}{k!}=
\sum^\infty_{k=0}\frac {\beta^k \lambda^{k-n}}{k!} \frac{N!}{(N-(k-n))!}  \\
  & \\
&=  \lambda^{-n}  \frac{N!}{(N+n)!}    \sum^\infty_{k=0}\frac {\beta^k \lambda^{k}}{k!}
\frac{(N+n)!}{(N+n-k)!}          \\
& \\
  &      = \  \lambda^{-n}  \frac{N!}{(N+n)!}    \left(   1+ \beta \lambda \right)^{N+n} .    \\
\end{array}
\ee
That is almost all that is needed to get rid of the normal powers in the solutions of the NCQM equations. Since hypergeometric functions are
often written in the form of power series with coefficients expressed in terms of the Pochhammer symbols, it is useful to take notice of  the following identities when it comes to handling factorials:
\be\nn \begin{array}{ll}
(a)_n &= a (a+1)...(a+n-1)=\frac{(a+n-1)!}{(a-1)!},\\
  &\\
  \frac{N!}{(N+n)!} &= \frac{N!}{(N+n)(N+n-1)...(N+1)N!}= \frac{1}{(N+1)_n },\\
  & \\
\frac{N!}{(N-n)!} &= \frac{N(N-1)...(N-n+1)(N-n)!}{(N-n)!}= (-1)^n (-N)(-N+1)...(-N+n-1)\\
  & =(-1)^n (-N)_n.\\
\end{array}
\ee
Now we are going to supply the reader with some of the  calculations which were skipped in the previous sections. To keep reasonable length of the formulas, we will replace the long expressions with $a,\,c, \, \beta, \, Q$... substituting the ''right arguments'' instead of our abbreviations and completing the calculation is straightforward...yet not so tempting.\\
\\
The following  derivation was  left out in (\ref{reg3}):
\be
\begin{array}{lcl}
 &:e^{\beta \varrho} \, _1F_1 \left( a; c; Q\varrho\right):&= \, :e^{\beta \varrho}
 \sum^\infty_{m=0}\frac{(a)_m}{(c)_m}\frac{(Q\varrho)^m}{m!}:   \\
           & = & \,  \sum^\infty_{m=0}\left[ 1+\lambda\beta\right]^{N-m}\frac{(a)_m}{(c)_m}(Q\lambda)^m\frac{N!}{(N-m)!m!}   \\
          &= & \left[ 1+\lambda\beta\right]^{N}  \sum^\infty_{m=0}\left[ \frac{Q\lambda}{
          1+\lambda\beta}\right]^{m}\frac{(a)_m}{(c)_m}\frac{1}{m!}(-1)^{m}(-N)_m  \\
                 & = & \left[ 1+\lambda\beta\right]^{N}  \sum^\infty_{m=0}\left[ \frac{-Q\lambda}{ 1+\lambda\beta}\right]^{m} \frac{(a)_m(-N)_m}{(c)_m
         }\frac{1}{m!}  \\
                   &= & \left[ 1+\lambda\beta\right]^{N}  \,_2F_1 \left(a,-N; c;  \frac{-Q\lambda}{ 1+\lambda\beta} . \right)
\end{array}
\ee
This is to be done and suitably used if one wishes to rewrite (\ref{reg3}) in terms of the fundamental system (\ref{y5-7}) and to get rid of the normal ordering only thereafter:
\be\label{psiodb}
\begin{array}{lcl}
&:e^{\beta \varrho} \, _1F_1 \left( a; c; Q\varrho \right):&= \, :e^{\beta\varrho} \times \\
                           & & \times \sum^\infty_{m=0}\frac{(-1)^{m}(a)_m(a-c+1)_m (Q\varrho)^{-a-m}}{m!}:   \\
           & = & \left[ 1+\lambda\beta\right]^{N}\left[\frac{1+\lambda\beta}{\lambda Q}\right]^a \frac{N!}{(N+a)!} \times \\
              && \times \sum^\infty_{m=0} \frac{(a)_m(a-c+1)_m }{m!}  \left[\frac{1+\lambda\beta}{-\lambda Q}\right]^m \frac{(N+a)!}{(N+m+a)!} \\
                    & =& \left[ 1+\lambda\beta\right]^{N}\left[\frac{1+\lambda\beta}{\lambda Q}\right]^a \frac{N!}{(N+a)!} \times \\
                    && \times \sum^\infty_{m=0} \frac{(a)_m(a-c+1)_m }{(N+a+1)_m m!}  \left[\frac{1+\lambda\beta}{-\lambda Q}\right]^m  \\
           & = & \left[ 1+\lambda\beta\right]^{N}\left[\frac{1+\lambda\beta}{\lambda Q}\right]^a \frac{N!}{(N+a)!} \\
            && \times \, _2F_1 \left( a,\, a-c+1; \, N+a+1 ; \,
          \frac{1+\lambda\beta}{-\lambda Q}\right).  \\
\end{array}
\ee
To complete (\ref{gsol2}), this was needed:
\be
\begin{array}{lcl}
 &:\varrho^{-j-1/2} J_{-2j-1}(\sqrt{8\alpha \varrho}):&= \, \sum^\infty_{m=0}\frac{(-1)^{m+1}
 (2\alpha)^{m+j+1/2}}{(2j+1+m)!m!}:\varrho^{m}:   \\
 &=&- \frac{(2\alpha)^{j+1/2}}{(2j+1)!}\sum^\infty_{m=0}\frac{(-1)^{m+1}  (2\alpha\lambda)^{m}(2j+1)!N!}{(2j+1+m)!m!(N-m)!}    \\
   &=&- \frac{(2\alpha)^{j+1/2}}{(2j+1)!}\sum^\infty_{m=0}\frac{ (-N)_m (2\alpha\lambda)^{m}}{(2j+2)_m m!}    \\
      &=&- \frac{(2\alpha)^{j+1/2}}{(2j+1)!} \, _1F_1 (-N; \, 2j+2; \,  2\alpha\lambda).    \\
\end{array}
\ee
The equation (\ref{gsol3}) required:
\be
\begin{array}{lcl}
 &:e^{-2\varrho/\lambda}\varrho^{-j-1/2} J_{-2j-1}(\sqrt{8\alpha\varrho}):&= \, \sum^\infty_{m=0}\frac{(-1)^{m+1}
 (2\alpha)^{m+j+1/2}}{(2j+1+m)!m!}:e^{-2\varrho/\lambda}\varrho^{m}:   \\
 &=&- \frac{(2\alpha)^{j+1/2}}{(2j+1)!}\sum^\infty_{m=0}\frac{(-1)^{m}(1-2)^N  (2\alpha\lambda)^{m}(2j+1)!N!}{(1-2)^m (2j+1+m)!m!(N-m)!}
          \\
           &=&\frac{(2\alpha)^{j+1/2}(-1)^{N+1}}{(2j+1)!}\sum^\infty_{m=0}\frac{ (-N)_m (-2\alpha\lambda)^{m}}{(2j+2)_m m!} \\
            &=&\frac{(-1)^{N+1}(2\alpha)^{j+1/2}}{(2j+1)!} \, _1F_1 (-N; \, 2j+2; \,  -2\alpha\lambda).    \\
\end{array}
\ee
\subsection*{Appendix C}
\noindent In this section we are going to look on the scattering case more closely, namely the part dealing with decomposition of the solution (\ref{NCries1}) into two parts corresponding to the in- and out- going spherical waves. Firstly let us remind the form in which we got  (\ref{NCries1}):
\be\nn R_{E j}\ =\ \Omega^{-N}\, _2F_1 \left(j+1-i\frac{\alpha }{p},\,-N ; \,2j+2;\, 2i\lambda p\,\Omega  \right). \ee
If we, for the sake of brevity, denote
\be\label{subst}
\begin{array}{llll}
a=j+1-i\frac{\alpha}{p} &, \,\,\, b=-N &, \,\,\, c=2j+2 &, \, \,\, z=2i\lambda p\, \Omega  , \\
\end{array}
\ee
then $R_{Ej}$ can be written as
\be R_{E j} = \left( 1-z\right)^{b/2}\, _2F_1 \left(a,\,b ;\,c;\, z\right). \ee
According to Kummer identities (see \cite{Bat}), $ \, _2F_1 (a,b;c; z)$ can be written as a linear combination of two other solutions of the hypergeometric equation (\ref{hypgeoeq}), namely
\beqa\label{FF}
\nn &(-z)^{-a} \, _2F_1 (a,\, a+1-c ; \, a+1-b; \,z^{-1} ) & \\
\nn &\mbox{\,\,and \,\,}& \\
\nn &(z)^{a-c}(1-z)^{c-a-b} \, _2F_1 (c-a,\, 1-a ; \, c+1-a-b; \,z^{-1}(z-1) ). \eeqa
Take notice of the fact that if we  decompose our radial solution (\ref{reg3}) before handling the normal ordering, i.e. if we write the confluent hypergeometric function (\ref{y1}) in terms of the fundamental system (\ref{y5-7}) of confluent hypergeometric equation (\ref{confl}), and deal with the normal ordering  only thereafter (see Appendix B, eq. (\ref{psiodb})), we will  end up with (\ref{FF}) again.
So after using one of the numerous Kummer relations listed in \cite{Bat}, we have
\be
\begin{array}{lll}
R_{Ej} & =& \left( 1-z\right)^{b/2}\, _2F_1 \left(a,\,b;\,c;\, z\right) = \\
        & & \\
   & =&  e^{i\pi(c-a)}\frac{\Gamma(c)\Gamma(1-b)}{\Gamma(a)\Gamma(c+1-a-b)} \, (z)^{a-c}(1-z)^{c-a-b/2}\\
   & & \times   \, _2F_1 (c-a,\, 1-a; \, c+1-a-b; \,z^{-1}(z-1) ) \\
   & & \\
   & +&   e^{i\pi(1-a)}     \frac{\Gamma(c)\Gamma(1-b)}{\Gamma(c-a)\Gamma(a+1-b)} \, (-z)^{-a}\left( 1-z\right)^{b/2}  \\
   & & \times  \, _2F_1 (a,\, a+1-c; \, a+1-b; \,z^{-1} ). \\
\end{array}
\ee
Substituting back (\ref{subst}) leads to
\be\label{rozklad1}
\begin{array}{lll}
R_{Ej} & =& \Omega^{-N}\, _2F_1 \left(j+1-i\frac{\alpha }{p},\,-N; \,2j+2;\, 2i\lambda p\,\Omega \right)\\
        & & \\
   & =&  e^{i\pi(j+1+i\frac{\alpha}{p})}\frac{\Gamma(2j+2)\Gamma(N+1)}{\Gamma(j+1-i\frac{\alpha}{p})\Gamma(j+2+i\frac{\alpha}{p} +N)}
   \left[2i\lambda p\, \Omega^{-1} \right]^{-j-1-i\frac{\alpha}{p}} \Omega^N\\
   & & \times \, _2F_1 \left(j+1+i\frac{\alpha}{p} ,\,\,\,-j+i\frac{\alpha}{p};\,\,\,\frac{r}{\lambda} +j+2+i\frac{\alpha}{p};\,\,\,-\frac{1}{2i\lambda p}\, \Omega \right) \\
    & &  \\
      & +& e^{i\pi(-j-1+i\frac{\alpha}{p})}\frac{\Gamma(2j+2)\Gamma(N+1)}{\Gamma(j+1+i\frac{\alpha}{p})\Gamma(j+2-i\frac{\alpha}{p} +N)}
   \left[-2i\lambda p\, \Omega \right]^{-j-1+i\frac{\alpha}{p}} \Omega^N\\
   & & \times \, _2F_1 \left(j+1-i\frac{\alpha}{p} ,\,\,\,-j-i\frac{\alpha}{p};\,\,\,\frac{r}{\lambda} +j+2-i\frac{\alpha}{p};\,\,\,\frac{1}{2i\lambda p}\, \Omega ^{-1}\right) . \\
   \end{array}
\ee
Keep in mind that our endeavour is to compare the NCQM result with the QM one. It is therefore suitable  to rewrite the prefactors in front of the hypergeometric functions as exponentials. This way we obtain, besides other factors, also logarithms of $\Gamma$-functions, which can be rewritten using the following formula (see \cite{Bat}):
\be\label{lngamma}
\ln \left(\Gamma (z+a) \right)=\left(z+a-\frac{1}{2}\right)\ln(z) - z + \frac{1}{2} \ln(2\pi)+ \sum^{\infty}_{n=1} \frac{B_{n+1}(a)}{n(n+1)}z^{-n} . \ee
Here $B_{n}(a)$ is a Bernoulli polynomial. After certain rearrangements we finally acquire the result (\ref{rozklad}).
\subsection*{Appendix D}
\noindent This part of the Appendix contains various calculations related to the velocity operator. For the computations it is useful to define two sets of auxiliary operators, which act as follows:
\begin{eqnarray} \label{auxAB}
\nonumber \hat{a}_\alpha \Psi = a_\alpha \Psi , & \hat{a}^+_\alpha \Psi = a^+_\alpha \Psi , \\ 
\hat{b}_\alpha \Psi = \Psi a_\alpha , & \hat{b}^+_\alpha \Psi = \Psi a^+_\alpha .
\end{eqnarray}
\\
\textbf{Derivation of operator $\hat{V}^j$}\\
Here we will derive the velocity operator in the form \eqref{V2.0}. Note that for states with equal number of c/a operators it is true that
\begin{equation} \label{Hzeta}
\hat{H}_0 = \frac{1}{2\lambda \hat{r}} \left(\frac{2 \hat{r}}{\lambda} - \hat{a}^+_\alpha \hat{b}_\alpha -\hat{b}^+_\alpha \hat{a}_\alpha \right) ,
\end{equation}
insterting \eqref{Hzeta} into \eqref{V} we obtain (recall that $[x^i, r]=0$)
\begin{equation}
\hat{V}^i = - \frac{i}{2\lambda \hat{r}}\left[\hat{X}^j , \frac{2 \hat{r}}{\lambda} - \hat{a}^+_\alpha \hat{b}_\alpha -\hat{b}^+_\alpha \hat{a}_\alpha \right] = \frac{i}{2\lambda \hat{r}}[\hat{X}^j , \hat{a}^+_\alpha \hat{b}_\alpha +\hat{b}^+_\alpha \hat{a}_\alpha].
\end{equation}
Using \eqref{auxAB} we may write the coordinate operator as $\hat{X}^j =\frac{\lambda}{2} \sigma^j_{\alpha \beta} (\hat{a}^+_\alpha \hat{a}_\beta + \hat{b}^+_\alpha \hat{b}_\beta)$. While 'a' and 'b' operators obviously commutate, for the 'b' operators we have $[\hat{b}_\alpha , \hat{b}^+_\beta]=-\delta_{\alpha \beta}$. Using those we easily obtain the final result
\begin{equation}
\hat{V}^i = \frac{i}{2r} \sigma^i_{\alpha \beta} (\hat{a}^+_\alpha \hat{b}_\beta - \hat{a}_\beta \hat{b}^+_\alpha) = \frac{i}{2r} \sigma^i_{\alpha \beta} \hat{w}_{\alpha \beta}.
\end{equation}
\\
\textbf{Evaluation of the correction terms in the uncertainty relation}\\
We need to evaluate two correction terms
\begin{eqnarray}
 \nonumber {\cal K}^i (x^j,\Psi) &=& -\frac{i}{2r}\sigma ^i _{\alpha \beta} ([a^+_\alpha , x^j][a_\beta , \Psi]- [a_\beta , x^j][a^+_\alpha , \Psi]),\\
{\cal K}^i (\Psi,x^j) &=& -\frac{i}{2r}\sigma ^i _{\alpha \beta} ([a^+_\alpha ,\Psi][a_\beta , x^j]- [a_\beta , \Psi][a^+_\alpha ,  x^j]) ,
\end{eqnarray}
using the commutation relations for c/a operators, definition of $x^j$ and properties of Pauli matrices we get
\begin{eqnarray}
{\cal K}^i (x^j,\Psi) &=& -\frac{i}{2r}\sigma ^i _{\alpha \beta} (\overbrace{[a^+_\alpha , x^j]}^{-\lambda \sigma^j_{\gamma \delta} a^+_\gamma}[a_\beta , \Psi]- \overbrace{[a_\beta , x^j]}^{\lambda\sigma^j_{\gamma \delta} a_\delta}[a^+_\alpha , \Psi])) \\ \nonumber
&=&\frac{i \lambda}{2 r}(\underbrace{\sigma^i_{\alpha \beta}\sigma^j_{\gamma \alpha}}_{\delta^{ij}\delta_{\gamma \beta} + i \varepsilon^{jik}\sigma^k_{\gamma \beta}}  a^+_\gamma (a_\beta \Psi - \Psi a_\beta) +\underbrace{\sigma^i_{\alpha \beta} \sigma^j_{\beta \delta}}_{\delta^{ij}\delta_{\alpha \delta} + i \varepsilon^{ijk}\sigma^k_{\alpha \delta}} a_\delta(a^+_\alpha \Psi - \Psi a^+_\alpha))
\end{eqnarray}
and similarly for ${\cal K}^i (\Psi,x^j)$
\begin{eqnarray}
\nonumber {\cal K}^i (\Psi,x^j) &=&-\frac{i\lambda}{2r}((a^+_\alpha \Psi - \Psi a^+_\alpha) \sigma^i_{\alpha \beta} \sigma^j _{\beta \gamma} a_\delta + (a_\beta \Psi - \Psi a_\beta) \sigma^i_{\alpha \beta} \sigma^j_{\gamma \delta} a^+_\gamma). \\
& &
\end{eqnarray}
Combing those two results we have
\begin{eqnarray}
\nonumber {\cal K}^i (x^j,\Psi) + {\cal K}^i (\Psi,x^j) &=&(\delta^{ij}((a^+_\alpha a_\alpha \Psi - a^+_\alpha \Psi a_\alpha + a_\alpha a^+_\alpha \Psi - a_\alpha \Psi a^+_\alpha)\\ 
&+&(-a_\alpha \Psi a^+_\alpha + \Psi a^+_\alpha a_\alpha - a^+_\alpha \Psi a_\alpha + \Psi a^+_\alpha a_\alpha)) \\ \nonumber
&+&i\varepsilon^{ijk} \sigma^k_{\alpha \beta}((-a^+_\alpha a_\beta \Psi+ a^+_\alpha \Psi a_\beta + a_\beta a^+_\alpha \Psi - a_\beta \Psi a^+_\alpha )\\ \nonumber
 &-&(a^+_\alpha \Psi a_\beta - \Psi a^+_\alpha a_\beta - a_\beta \Psi a^+_\alpha + \Psi a_\beta a^+_\alpha))),
\end{eqnarray}
using the commutation relations for c/a operators and the the tracelessness of $\sigma$ it is evident that terms in third and fourth line add up to zero. On the other hand, by looking at the equation \eqref{Hzeta} it is obvious that
\begin{equation}
\frac{1}{2}({\cal K}^i (x^j,\Psi) + {\cal K}^i (\Psi,x^j) )= i\delta^{ij}\lambda^2 H_0 \Psi .
\end{equation}
\\
\\
\textbf{Commutator $[\hat{V}^i, \hat{V}^j]$}\\
Since this calculation includes many steps, we will just outline it here. Thanks to $[\hat{V}^i, \hat{V}^j]$ being antisymmetric,  we can calculate $\varepsilon^{ijk}[\hat{V}^i, \hat{V}^j]$ instead; we will do so in fact, because of the vector Fierz identity $\varepsilon ^{ijk} \sigma ^ i _{\alpha \beta} \sigma ^ j _{\gamma \delta} = i (\sigma ^k _{\alpha \delta} \delta_{\gamma \beta} - \sigma ^k _{\gamma \beta} \delta _{\alpha \delta})$ which we want to use. Using the notation \eqref{V2.0} we have
\begin{equation}
\varepsilon^{ijk}[\hat{V}^i, \hat{V}^j] = \left(\frac{i}{2}\right)^2\varepsilon^{ijk} \sigma^i_{\alpha \beta} \sigma^j _{\gamma \delta} \left[\frac{1}{\hat{r}} \hat{w}_{\alpha \beta} , \frac{1}{\hat{r}}\hat{w}_{\gamma \delta} \right]
\end{equation}
\begin{equation*}
= \left(\frac{i}{2}\right)^2 i (\sigma ^k _{\alpha \delta} \delta_{\gamma \beta} - \sigma ^k _{\gamma \beta} \delta _{\alpha \delta})\left(\frac{1}{\hat{r}^2}[\hat{w}_{\alpha \beta} , \hat{w}_{\gamma \delta}]+ \frac{1}{\hat{r}}\left[\hat{w}_{\alpha \beta}, \frac{1}{\hat{r}}\right]\frac{1}{\hat{r}}+\frac{1}{\hat{r}}\left[\frac{1}{\hat{r}},\hat{w}_{\gamma \delta}\right]\hat{w}_{\alpha \beta}\right) .
\end{equation*}
The first term is easily evaluated using \eqref{V2.0} and \eqref{auxAB}, yielding
\begin{equation} \label{VVv1}
\varepsilon^{ijk} \sigma^i_{\alpha \beta} \sigma^j_{\gamma \delta} \frac{1}{\hat{r}^2}[\hat{w}_{\alpha \beta} , \hat{w}_{\gamma \delta}] = \frac{8i}{\hat{r}^2} \hat{L}^k .
\end{equation}
The other two terms are a bit more demanding. Let us begin with (following from $[a_\alpha , r] =\lambda a_\alpha$, $[a^+_\alpha , r] = -\lambda a^+_\alpha$)
\begin{eqnarray}
 \label{arN} a_\alpha r^N &=&  a_\alpha \underbrace{r...r}_N = r(a_\alpha + \lambda)\underbrace{r...r}_{N-1} = ... = (r+\lambda)^N a_\alpha \, , \\ \nn
 a^+_\alpha r^N &=& (r-\lambda)^N a^+_\alpha .
\end{eqnarray}
Using those relations we get
\begin{equation} \label{VVv2}
\varepsilon^{ijk}\sigma^i_{\alpha \beta} \sigma^j _{\gamma \delta}\left(\frac{1}{\hat{r}}\left[\hat{w}_{\alpha \beta}, \frac{1}{\hat{r}}\right]\frac{1}{\hat{r}}+\frac{1}{\hat{r}}\left[\frac{1}{\hat{r}},\hat{w}_{\gamma \delta} \right]\hat{w}_{\alpha \beta}\right)=...=-\frac{8i}{\hat{r}^2}\hat{L}^k ,
\end{equation}
adding together results \eqref{VVv1} and \eqref{VVv2} implies $\varepsilon^{ijk}[\hat{V}^i, \hat{V}^j]=0$ and therefore $[\hat{V}^i, \hat{V}^j]=0$.
\\
\\
\textbf{Relation between the velocity operator and the free Hamiltonian}\\
Finding this relation consist of a huge number of trivial steps and a tricky one. We will omit the trivial ones here, showing only the important part. Writing down the velocity operator(s) using \eqref{V2.0} we have
\begin{equation} \label{Vdva}
\hat{V}^2 =  \frac{i}{2\hat{r}}\sigma^i_{\alpha \beta} (\hat{a}^+_\alpha \hat{b}_\beta - \hat{a}_\beta \hat{b}^+_\alpha)\sigma^j_{\gamma \delta}\frac{i}{2\hat{r}}(\hat{a}^+_\gamma \hat{b}_\delta - \hat{a}_\delta \hat{b}^+_\gamma)
\end{equation}
\begin{equation*}
=... \mbox{a lot of trivial modifications}...=
\end{equation*}
\begin{equation*}
= \frac{1}{\lambda^2} - \frac{1}{4\hat{r}(\hat{r}-\lambda)}((\hat{a}^+\hat{b})^2+(\hat{a}^+\hat{b})(\hat{a}\hat{b}^+)))-\frac{1}{4\hat{r}(\hat{r}+\lambda)}((\hat{a}\hat{b}^+)^2+(\hat{a}\hat{b}^+)(\hat{a}^+\hat{b})) ,
\end{equation*}
where $(\hat{a}^+\hat{b})=\hat{a}^+_\alpha \hat{b}_\alpha$ and similarly for other combinations. From \eqref{Hzeta} it is evident that
\begin{equation} \label{Hdva}
\hat{H}_0 - \frac{1}{\lambda^2} = - \frac{1}{2\lambda \hat{r}}((\hat{a}^+\hat{b})+(\hat{b}^+\hat{a})) .
\end{equation}
While in \eqref{Vdva} there are four c/a operators in each term, there are only two of them in \eqref{Hdva}, so we need to square it:
\begin{equation*}
\left(\frac{1}{\lambda ^2} - \hat{H}_0 \right)^2 = \frac{1}{\lambda^2 \hat{r}}( (\hat{a}^+\hat{b}) + (\hat{b}^+\hat{a}))\frac{1}{\hat{r}} ((\hat{a}^+\hat{b}) + (\hat{b}^+\hat{a}))
\end{equation*}
\begin{equation} \label{V^2b}
=  \frac{1}{4\lambda ^2\hat{r}(\hat{r}-\lambda)}((\hat{a}^+\hat{b})^2+(\hat{a}^+\hat{b})(\hat{a}\hat{b}^+))
\end{equation}
\begin{equation*}
+\frac{1}{4\lambda ^2\hat{r} (\hat{r}+\lambda)}((\hat{a}\hat{b}^+)^2 + (\hat{a}\hat{b}^+)(\hat{a}^+\hat{b})).
\end{equation*}
In the last step we have used \eqref{arN}. Now comparing this result with \eqref{Vdva} it is obvious that
\begin{equation}
\left(\frac{1}{\lambda^2}-\hat{H}_0 \right)^2 = \frac{1}{\lambda ^2} \left(\frac{1}{\lambda^2}-\hat{V}^2 \right).
\end{equation}
\\
\textbf{The Ehrenfest theorem}\\
The idea of this calculation is to evaluate the correction term \eqref{kor} for \eqref{ehren}, which with the use of \eqref{arN} turns to
\begin{equation*}
{\cal K}^i ( U(r), \Psi) = -\frac{i}{2r}\sigma^i_{\alpha \beta} ([a^+_\alpha ,U(r)][a_\beta,\Psi]-[a_\beta, U(r)][a^+_\alpha , \Psi])=
\end{equation*}
\begin{equation}
=-\frac{i}{2r}\sigma^i_{\alpha \beta} ((U(r-\lambda)-U(r))\underbrace{a^+_\alpha[a_\beta,\Psi]}_{\hat{A}_{\alpha \beta} \Psi}-(U(r+\lambda)-U(r))\underbrace{a_\beta[a^+_\alpha , \Psi])}_{\hat{B}_{\alpha \beta} \Psi}
\end{equation}
Now we need to evaluate the under-braced terms. As one can easily check, they are equal to
\begin{eqnarray}
\nonumber \hat{A}_{\alpha \beta}&=&\frac{1}{2}(\hat{W}_{\alpha \beta} + {\cal L}_{\alpha \beta} - \hat{w}_{\alpha \beta}), \\ 
\hat{B}_{\alpha \beta}&=&\frac{1}{2}(\hat{W}_{\alpha \beta} + {\cal L}_{\alpha \beta} + \hat{w}_{\alpha \beta}),
\end{eqnarray}
where $\hat{W}_{\alpha \beta} = \hat{a}^+_\alpha \hat{a}_\beta + \hat{b}^+_\alpha \hat{b}_\alpha - \hat{a}^+_\alpha \hat{b}_\beta - \hat{a}_\beta \hat{b}^+_\alpha$, $\frac{\sigma^i_{\alpha \beta}}{2} {\cal L}_{\alpha \beta}=\hat{L}^i $ and $\hat{w}_{\alpha \beta}$ is defined by \eqref{V2.0}. Such decomposition might seem a bit artificial, but will become more transparent in chapter 6. By noting the  definition of $\hat{W}^i$ in \eqref{W_k} we obtain the result
\begin{equation}
-i{\cal K}^i(U(r) , \Psi) = \left(\hat{U}_\lambda'(\hat{r}) \left(\frac{\lambda}{\hat{r}}\hat{L}^i + \lambda^2 \hat{W}^i\right)
+ \frac{\lambda^2}{2} \hat{U}_\lambda''(\hat{r})\hat{V}^i \right) \Psi ,
\end{equation}
where we have denoted $\hat{U}''_\lambda(\hat{r})=\frac{1}{\lambda^2}(\hat{U}(\hat{r}+\lambda) - 2\hat{U}(\hat{r}) + \hat{U}(\hat{r}-\lambda))$.
\subsection*{Appendix E}
\noindent In the chapter about the LRL vector many detailed calculations have been skipped. Here we provide at least some sketches thereof.
We are not about to write down every single step; if any interest in tiny details arises, we suppose it will be easier for the reader to calculate something on their own now and then, than to try to keep track of all the signs and many times renamed indices in this paper. As in Appendix D we provide the main tricks which make the calculations more manageable, and then we write down the sketches of the particular derivations.\\
\\
We can almost say that everything one needs to do is to calculate commutators of various strings of creation and annihilation operators acting from the left or from the right, that is, using the commutation relations for $\hat{a}^+_\alpha ,\,\hat{a}_\beta, \,\hat{b}^+_\alpha ,\,\hat{b}_\beta $ in a suitable way. As we know already, the $b$- operators are simply the $a$-ones acting from the right; the fact that those acting from the opposite sides commute saves us a good deal of work. \\
So the following should be  borne in mind before digging into the calculations:
$$
\begin{array}{lllclll}
 [\hat{a}_\alpha, \hat{a}^+_\beta] &=& \delta_{\alpha \beta}&,\,\,\,\,\, &[\hat{b}_\alpha, \hat{b}^+_\beta] &=&-\delta_{\alpha \beta}, \\
\end{array}
$$
the other commutators are zero.\\
Below we shall use frequently the identities for Pauli matrices a the identities for NC coordinates that follow directly, let us sum them up:
$$
\begin{array}{lllclll}
[\sigma^i,\sigma^j] &=&2i\varepsilon_{ijk} \sigma^k,  &\,\,\,\, & \varepsilon_{ijk} \sigma^i_{\alpha\beta}\sigma^j_{\gamma\delta}  &=&i(\sigma^k_{\alpha\delta}\delta_{\gamma\beta}-\sigma^k_{\gamma\beta}\delta_{\alpha\delta}), \\
 \{\sigma^i,\sigma^j\} &=&2\delta_{ij}\mathbf{1} ,                                &\,\,\,\, &  \sigma^i_{\alpha\beta}\sigma^i_{\gamma\delta}  &=&2\delta_{\alpha\delta}\delta_{\gamma\beta}- \delta_{\alpha\beta}\delta_{\gamma\delta},\\
\sigma^i\sigma^j &=&\delta_{ij}\mathbf{1} +i\varepsilon_{ijk} \sigma^k,             & \,\,\,\,&  Tr\sigma^i &=&\sigma^i_{\alpha\alpha} = 0 ,  \\

 &&&& \\

 [\hat{X}_i,\hat{r}] &=& 0,  &\,\,\,\, & [\hat{X}_i^L, \hat{X}_j^R] &=&0, \\

[\hat{X}_i^L, \hat{X}_j^L] &=& 2i\lambda\varepsilon_{ijk}  \hat{X}_k^L ,     &\,\,\,\, &   [\hat{X}_i^R, \hat{X}_j^R] &=& -\,2i\lambda\varepsilon_{ijk}  \hat{X}_k^R,\\

\end{array}
$$
where
$$
\begin{array}{lcl}
\hat{X}_i^L\, \Psi\, =\, x_i\, \Psi\,,  & \hat{X}_i^R\, \Psi\, =\,  \Psi\, x_i\,, & \hat{X}_i\, =\, \frac{1}{2}\,(\hat{X}_i^L\, +\, \hat{X}_i^R)\,, \\
\ \hat{r}^L\, \Psi\, =\, r\, \Psi\,,  & \hat{r}^R\, \Psi\, =\,  \Psi\, r\,, & \  \hat{r} \,=\, \frac{1}{2}\, (r^L\, +\, \hat{r}^R)\,. \\
\end{array}
$$
Make sure  you have noticed the minus sign in the commutator $[\hat{X}_i^R, \hat{X}_j^R]$. One should be in general  careful about the ordering when dealing with operators acting from the right.\\
\\
The reader may be tired by now of getting so much advice - so to give it a chance to be appreciated, let us start. Calculations related to various forms in which $A_k$ can be written (the derivation of the equation (\ref{A_k})) involve:
\begin{eqnarray}
\hat{A}_k &= &\frac{1}{2}\varepsilon_{ijk} (\hat{L}_i\hat{V}_j + \hat{V}_j\hat{L}_i) + q \frac{\hat{X}_k}{\hat{r}} \,\, = \,\, \hat{A}_k^0 + q \frac{\hat{X}_k}{\hat{r}}, 
\end{eqnarray}
\begin{eqnarray}
\hat{A}^0_k &=& \varepsilon_{ijk} \frac{i}{8\hat{r}} \sigma^i_{\alpha \beta} \sigma^j_{\gamma \delta}  \nonumber \\
&& \times ((\hat{a}^+_\alpha \hat{a}_\beta - \hat{b}_\beta \hat{b}^+_\alpha)(\hat{a}^+_\gamma \hat{b}_\delta - \hat{a}_\delta \hat{b}^+_\gamma) +(\hat{a}^+_\gamma \hat{b}_\delta-\hat{a}_\delta \hat{b}^+_\gamma)(\hat{a}^+_\alpha \hat{a}_\beta - \hat{b}_\beta \hat{b}^+_\alpha))\nonumber\\
&& \nonumber \\
 &=& -\frac{1}{8\hat{r}} (\sigma^k_{\alpha \delta} \delta_{\gamma\beta} - \sigma^k_{\gamma \beta} \delta_{\alpha \delta}) \nonumber \\
  && \times ((\hat{a}^+_\alpha \hat{a}_\beta - \hat{b}_\beta \hat{b}^+_\alpha)(\hat{a}^+_\gamma \hat{b}_\delta - \hat{a}_\delta \hat{b}^+_\gamma)+(\hat{a}^+_\gamma \hat{b}_\delta-\hat{a}_\delta \hat{b}^+_\gamma)(\hat{a}^+_\alpha \hat{a}_\beta - \hat{b}_\beta \hat{b}^+_\alpha))\nonumber\\
&& \nonumber \\
 &=& -\frac{\sigma^k_{\alpha \delta}}{8\hat{r}} ((\hat{a}^+_\alpha \hat{a}_\beta - \hat{b}_\beta \hat{b}^+_\alpha)(\hat{a}^+_\beta \hat{b}_\delta - \hat{a}_\delta \hat{b}^+_\beta)+ (\hat{a}^+_\beta \hat{b}_\delta-\hat{a}_\delta \hat{b}^+_\beta)(\hat{a}^+_\alpha \hat{a}_\beta - \hat{b}_\beta \hat{b}^+_\alpha)) \nonumber\\
 && +\frac{ \sigma^k_{\gamma \beta}}{8\hat{r}}   ((\hat{a}^+_\alpha \hat{a}_\beta - \hat{b}_\beta \hat{b}^+_\alpha)(\hat{a}^+_\gamma \hat{b}_\alpha - \hat{a}_\alpha \hat{b}^+_\gamma)+(\hat{a}^+_\gamma \hat{b}_\alpha-\hat{a}_\alpha \hat{b}^+_\gamma)(\hat{a}^+_\alpha \hat{a}_\beta - \hat{b}_\beta \hat{b}^+_\alpha))\nonumber\\
&& \nonumber \\
&=& -\frac{\sigma^k_{\alpha \delta}}{8\hat{r}} (\hat{a}_\beta \hat{a}^+_\beta \hat{a}^+_\alpha \hat{b}_\delta  - \hat{a}^+_\alpha \hat{b}_\delta -\hat{a}^+_\alpha \hat{a}_\delta  \hat{a}_\beta \hat{b}^+_\beta -  \hat{b}^+_\alpha \hat{b}_\delta \hat{b}_\beta \hat{a}^+_\beta + \hat{a}^+_\alpha \hat{b}_\delta + \hat{b}_\beta \hat{b}^+_\beta \hat{b}^+_\alpha \hat{a}_\delta   )  \nonumber \\
&& -\frac{\sigma^k_{\alpha \delta}}{8\hat{r}}(  \hat{a}^+_\beta \hat{a}_\beta \hat{b}_\delta \hat{a}^+_\alpha - \hat{a}^+_\alpha \hat{b}_\delta  - \hat{b}_\delta \hat{b}^+_\alpha \hat{a}^+_\beta \hat{b}_\beta + \hat{a}^+_\alpha \hat{b}_\delta - \hat{a}_\delta \hat{a}^+_\alpha \hat{b}^+_\beta \hat{a}_\beta + \hat{b}^+_\beta \hat{b}_\beta \hat{a}_\delta \hat{b}^+_\alpha)  \nonumber \\
&& + \frac{ \sigma^k_{\gamma \beta}}{8\hat{r}}(\hat{a}_\beta \hat{a}^+_\gamma \hat{a}^+_\alpha \hat{b}_\alpha - \hat{a}^+_\gamma \hat{b}_\beta  - \hat{a}^+_\alpha \hat{a}_\alpha \hat{a}_\beta \hat{b}^+_\gamma - \hat{b}^+_\alpha \hat{b}_\alpha \hat{b}_\beta \hat{a}^+_\gamma + \hat{a}^+_\gamma \hat{b}_\beta + \hat{b}_\beta \hat{b}^+_\gamma \hat{b}^+_\alpha \hat{a}_\alpha  )   \nonumber \\
&& + \frac{ \sigma^k_{\gamma \beta}}{8\hat{r}}( \hat{a}^+_\gamma \hat{a}_\beta  \hat{b}_\alpha \hat{a}^+_\alpha     - \hat{a}^+_\gamma \hat{b}_\beta  -\hat{b}_\alpha \hat{b}^+_\alpha \hat{a}^+_\gamma \hat{b}_\beta  + \hat{a}^+_\gamma \hat{b}_\beta - \hat{a}_\alpha \hat{a}^+_\alpha \hat{b}^+_\gamma \hat{a}_\beta + \hat{b}^+_\gamma \hat{b}_\beta \hat{a}_\alpha \hat{b}^+_\alpha )   \nonumber \\
&& \nonumber \\
&=& -\frac{\sigma^k_{\alpha \delta}}{8\lambda \hat{r}}( (2\hat{r}^L + 2\hat{r}^R)\hat{a}^+_\alpha \hat{b}_\delta + (2\hat{r}^R + 2\hat{r}^L) \hat{b}^+_\alpha \hat{a}_\delta )  \nonumber \\
&& + \frac{1}{8\lambda \hat{r}} ( (2\hat{X}_k^L + 2\hat{X}_k^R) \hat{a}_\beta \hat{b}^+_\beta  + (2\hat{X}_k^R + 2\hat{X}_k^L) \hat{a}^+_\beta \hat{b}_\beta )              \nonumber \\
&& \nonumber \\
&=& -\frac{1}{2\lambda \hat{r}}\hat{r} \sigma^k_{\alpha \delta}(\hat{a}^+_\alpha \hat{b}_\delta +  \hat{b}^+_\alpha \hat{a}_\delta) + \frac{1}{2\lambda \hat{r}}\hat{X}_k (\hat{a}_\beta \hat{b}^+_\beta  +  \hat{a}^+_\beta \hat{b}_\beta )  \nonumber \\ 
\label{Acalc} &=&-\frac{1}{2\lambda \hat{r}}(\hat{r}\hat{\zeta}_k - \hat{X}_k \hat{\zeta}) .
\end{eqnarray}
Now we are going to derive a few interesting identities which will come in handy when dealing with calculations related to the vanishing commutator $[\hat{W}', \hat{W}'_k]$ in  (\ref{dotA}) (recall  that for states with equal number of c/a operators $\hat{r}=\hat{r}^L=\hat{r}^R$).
\begin{eqnarray}
[\hat{\zeta}_k , \hat{r}^L] &=&     \lambda \sigma^k_{\alpha \beta} [ \hat{a}^+_\alpha \hat{b}_\beta + \hat{a}_\beta \hat{b}^+_\alpha , \hat{a}^+_\delta \hat{a}_\delta +1]
= \lambda \sigma^k_{\alpha \beta} (\hat{a}^+_\delta \hat{b}_\beta (-\delta_{\alpha\delta}) +\delta_{\beta\delta} \, \hat{a}_\delta \hat{b}^+_\alpha)     \nonumber  \\
&=&- \lambda \sigma^k_{\alpha \beta} (\hat{a}^+_\alpha \hat{b}_\beta - \hat{a}_\beta \hat{b}^+_\alpha)=  -\lambda \hat{w}_k =i 2\lambda \hat{r} \hat{V}_k.   \nonumber
\end{eqnarray}
For the sake of completion and next calculations let us calculate as well
\begin{eqnarray}
[\hat{\zeta} , \hat{X}_k] &=& \frac{\lambda}{2}\sigma^k_{\alpha \beta} [\hat{a}^+_\delta \hat{b}_\delta + \hat{a}_\delta \hat{b}^+_\delta, \hat{a}^+_\alpha \hat{a}_\beta + \hat{b}_\beta \hat{b}^+_\alpha ]\nonumber  \\
&=&\frac{\lambda}{2} \sigma^k_{\alpha \beta}(\hat{a}^+_\alpha (- \delta_{\delta \beta}) \hat{b}_\delta + \hat{a}^+_\delta (-\delta_{\delta\alpha})\hat{b}_\beta + \hat{b}^+_\delta(\delta_{\delta \alpha}) \hat{a}_\beta + \hat{a}_\delta \hat{b}^+_\alpha ( \delta_{\delta \beta}) )\nonumber\\
&=&-\lambda \sigma^k_{\alpha \beta} (\hat{a}^+_\alpha \hat{b}_\beta - \hat{a}_\beta \hat{b}^+_\alpha)=-\lambda \hat{w}_k= i 2\lambda \hat{r} \hat{V}_k ,\nonumber
\end{eqnarray}
\be
[\hat{\zeta} , \hat{X}_k]= [\hat{\zeta} _k , \hat{r}]=i 2\lambda \hat{r} \hat{V}_k ,
\ee
\begin{eqnarray}
[\hat{\zeta} , \hat{\zeta} _{\alpha \beta}]&=& [\hat{a}^+_\gamma \hat{b}_\gamma + \hat{a}_\gamma \hat{b}^+_\gamma\,,\,\hat{a}^+_\alpha \hat{b}_\beta + \hat{a}_\beta \hat{b}^+_\alpha ]\nonumber\\
&=&-\delta_{\gamma\beta}\,\hat{b}_\gamma \hat{b}^+_\alpha -\delta_{\gamma\alpha}\,\hat{a}_\beta \hat{a}^+_\gamma + \delta_{\gamma\alpha} \, \hat{b}^+_\gamma \hat{b}_\beta + \delta_{\gamma\beta}\,\hat{a}^+_\alpha \hat{a}_\gamma  \nonumber \\
&=& -\hat{b}_\beta \hat{b}^+_\alpha -\hat{a}_\beta \hat{a}^+_\alpha + \hat{b}^+_\alpha \hat{b}_\beta + \hat{a}^+_\alpha \hat{a}_\beta =\delta_{\alpha\beta} - \delta_{\alpha\beta} =0,\nonumber
\end{eqnarray}
\be
 [\hat{\zeta}, \hat{\zeta} _k]=0 .
\ee
At this point we have all that is needed to prove (\ref{dotA}):
\begin{eqnarray}
[\hat{W}', \hat{W}'_k] &=& [\hat{W} +\omega \hat{r}, \hat{W}_k + \omega \hat{X}_k]\nonumber \\
&=& \left[2\hat{r}\lambda^{-1} - \hat{\zeta} + \omega \hat{r}, 2\hat{X}_k\lambda^{-1} - \hat{\zeta} _k + \omega \hat{X}_k\right]\nonumber \\
&=& \eta^2 [ \hat{r}, \hat{X}_k] -  \eta [\hat{r},\hat{\zeta}_k] -\eta [\hat{\zeta} , \hat{X}_k]+[\hat{\zeta} ,\hat{\zeta}_k]  \\
&=&  0+ \eta 2i\hat{r}\lambda \hat{V}_k- \eta 2i\hat{r}\lambda  \hat{V}_k +0 = 0. \nonumber
\end{eqnarray}
When checking the $so(4)$ algebra relations in (\ref{AiAj}), the knowledge of the commutator $[\hat{W}'_i, \hat{W}'_j]$ is necessary. In order to work it out, we need to derive $[\hat{\zeta}_i,\hat{\zeta}_j],\, [\hat{X}_i, \hat{\zeta} _j ],\,[\hat{X}_i,\hat{X}_j]$ as well. As to the following few calculations, recall that $\hat{X}_i$ is composed of $x_i$ acting from the left and the right side: $\hat{X}_i=(\hat{X}_i^L + \hat{X}_i^R)/2$. 
\begin{eqnarray}
[\hat{\zeta}_i,\hat{\zeta}_j] &=& \sigma^i_{\alpha \beta}\sigma^j _{\gamma \delta}  [\hat{\zeta}_{\alpha \beta} , \hat{\zeta} _{\gamma \delta}]=\sigma^i_{\alpha \beta}\sigma^j_{\gamma \delta} [\hat{a}^+_\alpha \hat{b}_\beta + \hat{a}_\beta \hat{b}^+_\alpha \,,\,\hat{a}^+_\gamma \hat{b}_\delta + \hat{a}_\delta \hat{b}^+_\gamma ]\nonumber\\
&=&-(\sigma^j\sigma^i)_{\gamma\beta} \hat{b}_\beta \hat{b}^+_\gamma - (\sigma^i\sigma^j)_{\alpha\delta } \hat{a}_\delta \hat{a}^+_\alpha + (\sigma^i\sigma^j)_{\alpha\delta} \hat{b}^+_\alpha \hat{b}_\delta + (\sigma^j\sigma^i)_{\gamma\beta} \hat{a}^+_\gamma \hat{a}_\beta
\nonumber\\
&=& \delta_{ij}(\hat{b}^+_\alpha \hat{b}_\alpha -  \hat{b}_\alpha \hat{b}^+_\alpha +    \hat{a}^+_\alpha \hat{a}_\alpha -  \hat{a}_\alpha \hat{a}^+_\alpha ) \nonumber\\
 && + i \varepsilon_{ijk}\sigma^k_{\alpha\beta}(\hat{b}_\beta \hat{b}^+_\alpha + \hat{b}^+_\alpha \hat{b}_\beta - \hat{a}_\beta \hat{a}^+_\alpha - \hat{a}^+_\alpha \hat{a}_\beta) \\
&=& \delta_{ij}(\delta_{\alpha\alpha}-\delta_{\alpha\alpha})+2i\varepsilon_{ijk}\sigma^k_{\alpha\beta}(\hat{b}_\beta \hat{b}^+_\alpha  - \hat{a}^+_\alpha \hat{a}_\beta)\nonumber \\
& =& 2i\lambda^{-1}\varepsilon_{ijk}(\hat{X}_k^R  - \hat{X}_k^L) = -4i\varepsilon_{ijk} \hat{L}_k \nonumber .
\end{eqnarray}
\begin{eqnarray}
[\hat{X}_i^L, \hat{\zeta}_j] &=& \lambda \sigma^i _{\alpha \beta} \sigma^j _{\gamma \delta} [ \hat{a}^+_\alpha \hat{a}_\beta , \hat{a}^+_\gamma \hat{b}_\delta + \hat{a}_\delta \hat{b}^+_\gamma] \nonumber \\
&=&\lambda  \sigma ^i _{\alpha \beta} \sigma^j _{\gamma \delta}  (\hat{a}^+_\alpha \hat{b}_\delta \delta_{\beta \gamma} - \hat{a}_\beta \hat{b}^+_\gamma \delta_{\alpha \delta} \nonumber \\
& = &\lambda( (\sigma ^i \sigma ^j)_{\alpha \beta}\hat{a}^+_\alpha \hat{b}_\beta -(\sigma^j \sigma^i)_{\alpha \beta} \hat{a}_\beta \hat{b}^+_\alpha)\nonumber ,
\end{eqnarray}
\begin{eqnarray}
[\hat{X}_i^R, \hat{\zeta} _j] &=& \lambda \sigma ^i _{\alpha \beta} \sigma^j _{\gamma \delta} [ \hat{b}_\beta \hat{b}^+_\alpha  , \hat{a}^+_\gamma \hat{b}_\delta + \hat{a}_\delta \hat{b}^+_\gamma] \nonumber \\
&=& \lambda  \sigma ^i _{\alpha \beta} \sigma^j _{\gamma \delta}  (\hat{b}_\beta \hat{a}^+_\gamma \delta_{\alpha \delta} - \hat{a}_\delta \hat{b}^+_\alpha \delta_{\beta \gamma} )\nonumber \\
&=& \lambda (-(\sigma ^i \sigma ^j)_{\alpha \beta}\hat{a}_\beta \hat{b}^+_\alpha + (\sigma^j \sigma^i)_{\alpha \beta}\hat{a}^+_\alpha \hat{b}_\beta) \nonumber ,
\end{eqnarray}
\begin{eqnarray}
[\hat{X}_i, \hat{\zeta} _j ] &=& \frac{\lambda}{2}(\sigma ^i \sigma ^j+\sigma^j \sigma^i)_{\alpha \beta}(\hat{a}^+_\alpha \hat{b}_\beta - \hat{a}_\beta \hat{b}^+_\alpha) \nonumber \\
&=& \frac{\lambda}{2}\{\sigma^i, \sigma^j\} _{\alpha \beta} \hat{w}_{\alpha \beta} = \lambda\delta_{ij} \hat{w} .
\end{eqnarray}
\noindent The following calculation takes use of the fact that $\hat{X}_i^L$ commutes with $\hat{X}_i^R$. It is also useful to have in mind
$[\hat{X}_i^L ,\hat{X}_j^L]= 2i\lambda\varepsilon_{ijk} \hat{X}_k^L$ and $[\hat{X}_i^R ,\hat{X}_j^R]=  -2i\lambda\varepsilon_{ijk} \hat{X}_k^R$
\begin{eqnarray}
[\hat{X}_i,\hat{X}_j] &=& \frac{1}{4}[\hat{X}_i^L + \hat{X}_i^R, \hat{X}_j^L+ \hat{X}_j^R] =\frac{1}{4}([\hat{X}_i^L , \hat{X}_j^L] +[ \hat{X}_i^R,  \hat{X}_j^R] ) \nonumber \\
&=&     \frac{1}{4}(2i\varepsilon_{ijk}\lambda \hat{X}_k^L + 2i\varepsilon_{jik}\lambda \hat{X}_k^R)=  \frac{i}{2}\varepsilon_{ijk}(\hat{X}_k^L - \hat{X}_k^R)=  i \lambda^2 \varepsilon_{ijk}\hat{L}_k.
\end{eqnarray}

So for the commutators involved in (\ref{AiAj}) we can write:
\begin{eqnarray}
 [\hat{W}'_i, \hat{W}'_j] &=& [\eta \hat{X}_i - \hat{\zeta} _i , \eta \hat{X}_j - \hat{\zeta} _j]\nonumber \\
&=& \eta^2 [\hat{X}_i, \hat{X}_j] - \eta ([\hat{X}_i , \hat{\zeta} _j] - [\hat{X}_j, \hat{\zeta} _i]) +[\hat{\zeta}_i , \hat{\zeta} _j]\nonumber\\
&=&\eta^2 i \varepsilon_{ijk} \lambda ^2 \hat{L}_k - \eta((\delta_{ij} -\delta_{ji})\lambda w) - i4 \varepsilon_{ijk} \hat{L}_k \label{tildWW} \\
&=& 4i\lambda\omega\left(1+ \left(\frac{\lambda\omega}{4}\right)\right)\varepsilon_{ijk}\hat{L}_k.  \nonumber
\end{eqnarray}
\noindent To compute the commutator $[\hat{L}_i,\hat{A}_j]$  acting on the vectors from $\mathcal{H_{\lambda}^E}$, we actually need  to find $[\hat{L}_i, \hat{W}'_j]$ and multiply it with certain constants. To manage this, let us begin with
\begin{eqnarray}
[\hat{L}_i ,\hat{X}_j]&=& \frac{1}{4\lambda}[\hat{X}_i^L- \hat{X}_i^R, \hat{X}_j^L+ \hat{X}_j^R] =\frac{1}{4\lambda}([\hat{X}_i^L, \hat{X}_j^L]- [ \hat{X}_i^R,  \hat{X}_j^R] ) \nonumber\\
&=& \frac{1}{4\lambda}(2i\varepsilon_{ijk}\lambda \hat{X}_k^L - 2i\varepsilon_{jik}\lambda \hat{X}_k^R)=  \frac{i}{2}\varepsilon_{ijk}(\hat{X}_k^L +\hat{X}_k^R)= i \varepsilon_{ijk}\hat{X}_k. \label{L_ix_j}
\end{eqnarray}
Somewhere above we have already computed $[\hat{X}_i^L, \hat{\zeta}_j]$ and  $[\hat{X}_i^R, \hat{\zeta}_j]$, we can use it also now:
\begin{eqnarray}
[\hat{L}_i , \hat{\zeta}_j]&=& \frac{1}{2\lambda}[\hat{X}_i^L- \hat{X}_i^R , \hat{\zeta}_j]=\frac{1}{2\lambda}([\hat{X}_i^L , \hat{\zeta}_j] -[ \hat{X}_i^R , \hat{\zeta}_j])\nonumber \\
&=& \frac{1}{2\lambda}\lambda((\sigma^i\sigma^j)_{\alpha\beta}(\hat{a}^+_\alpha \hat{b}_\beta + \hat{a}_\beta \hat{b}^+_\alpha )-(\sigma^j\sigma^i)_{\alpha\beta}(\hat{a}^+_\alpha \hat{b}_\beta + \hat{a}_\beta \hat{b}^+_\alpha )) \label{L_izeta_j} \\
&=&\frac{1}{2}[\sigma^i,\,\sigma^j]_{\alpha\beta}(\hat{a}^+_\alpha \hat{b}_\beta + \hat{a}_\beta \hat{b}^+_\alpha )=i\varepsilon_{ijk} \sigma^k_{\alpha\beta}(\hat{a}^+_\alpha \hat{b}_\beta + \hat{a}_\beta \hat{b}^+_\alpha ) = i\varepsilon_{ijk}\hat{\zeta}_k, \nonumber
\end{eqnarray}
\begin{eqnarray}\label{liwj}
[\hat{L}_i , \hat{W}'_j]&=& [\hat{L}_i , \eta \hat{X}_j - \hat{\zeta}_j]=- i \varepsilon_{ijk} (\eta \hat{X}_k - \hat{\zeta}_k )\nonumber \\
&=&   i \varepsilon_{ijk} \hat{W}'_k .
\end{eqnarray}
At this point to obtain the commutator $[\hat{L}_i , \hat{A}_j]$ in $(\ref{LA})$ we just need to recall that $\frac{\hat{W}'_j}{2\lambda}= \hat{A}_j$, and therefore
\be
 [\hat{L}_i , \hat{A}_j]=i \varepsilon_{ijk} \hat{A}_k .
\ee
\vspace*{0.3cm}
\noindent We shall now provide some details related to the first Casimir operator in (\ref{CAS}). We are going to prove that  $\hat{L}_j \hat{A}_j \propto \hat{L}_j (\eta \hat{X}_k - \hat{\zeta}_j)=0 $ when restricted to the subspace $\mathcal{H}_{\lambda}^E$. In the very next calculation a lot of renaming of the dummy indices appears and the reader will probably appreciate having a piece of paper and a pencil at hand in order to keep track of the procedure. The final zero in the last line is obtained after realizing that there is no difference between $\hat{r}^L$ and  $\hat{r}^R$ if they act on anything which has the same number of creation and annihilation  operators. Our considered operator wave functions $\Psi$ have this property and the same holds for $\hat{a}^+_\alpha \hat{b}_\alpha \Psi$ and $\hat{a}_\alpha \hat{b}^+_\alpha \Psi$.
\begin{eqnarray}
2 \hat{L}_j \hat{\zeta} _j &=& \sigma ^j _{\alpha \beta} \sigma^j_{\gamma \delta}  (\hat{a}^+_\alpha \hat{a}_\beta - \hat{b}_\beta \hat{b}^+_\alpha )(\hat{a}^+_\gamma \hat{b}_\delta + \hat{a}_\delta \hat{b}^+_\gamma) \nonumber \\
 &=& (2 \delta_{\alpha\delta}\delta_{\beta\gamma}-\delta_{\alpha\beta}\delta_{\gamma\delta})  (\hat{a}^+_\alpha \hat{a}_\beta - \hat{b}_\beta \hat{b}^+_\alpha )(\hat{a}^+_\gamma \hat{b}_\delta + \hat{a}_\delta \hat{b}^+_\gamma) \nonumber \\
&=& 2 ( \hat{a}^+_\alpha \hat{a}_\beta \hat{a}^+_\beta \hat{b}_\alpha + \hat{a}^+_\alpha \hat{a}_\beta \hat{a}_\alpha \hat{b}^+_\beta -  \hat{b}_\beta \hat{b}^+_\alpha \hat{a}^+_\beta \hat{b}_\alpha - \hat{b}_\beta \hat{b}^+_\alpha \hat{a}_\alpha \hat{b}^+_\beta) \nonumber \\
&& - ( \hat{a}^+_\alpha \hat{a}_\alpha \hat{a}^+_\beta \hat{b}_\beta + \hat{a}^+_\alpha \hat{a}_\alpha \hat{a}_\beta \hat{b}^+_\beta -  \hat{b}_\alpha \hat{b}^+_\alpha \hat{a}^+_\beta \hat{b}_\beta -  \hat{b}_\alpha \hat{b}^+_\alpha \hat{a}_\beta \hat{b}^+_\beta) \nonumber \\
&=& 2(\hat{a}_\beta \hat{a}^+_\beta \hat{a}^+_\alpha \hat{b}_\alpha - \hat{a}^+_\alpha \hat{b}_\alpha + \hat{a}^+_\alpha \hat{a}_\alpha \hat{a}_\beta \hat{b}^+_\beta -\hat{b}^+_\alpha \hat{b}_\alpha \hat{b}_\beta \hat{a}^+_\beta + \hat{a}^+_\alpha \hat{b}_\alpha -\hat{b}_\beta \hat{b}^+_\beta \hat{b}^+_\alpha \hat{a}_\alpha )  \nonumber \\
&& -\hat{a}_\beta \hat{a}^+_\beta \hat{a}^+_\alpha \hat{b}_\alpha +2 \hat{a}^+_\alpha \hat{b}_\alpha - \hat{a}^+_\alpha \hat{a}_\alpha \hat{a}_\beta \hat{b}^+_\beta + \hat{b}^+_\alpha \hat{b}_\alpha \hat{b}_\beta \hat{a}^+_\beta -2\hat{a}^+_\alpha \hat{b}_\alpha \nonumber \\
&\,& + \hat{b}_\beta \hat{b}^+_\beta \hat{b}^+_\alpha \hat{a}_\alpha  \nonumber \\
&=& \hat{a}_\beta \hat{a}^+_\beta \hat{a}^+_\alpha \hat{b}_\alpha + \hat{a}^+_\beta \hat{a}_\beta \hat{a}_\alpha \hat{b}^+_\alpha - \hat{b}^+_\beta \hat{b}_\beta \hat{a}^+_\alpha \hat{b}_\alpha - \hat{b}_\beta \hat{b}^+_\beta \hat{a}_\alpha \hat{b}^+_\alpha \nonumber \\
&=& \lambda^{-1}\left( (\hat{r}^L + \lambda)-(\hat{r}^R + \lambda) \right) \hat{a}^+_\alpha \hat{b}_\alpha + \lambda^{-1}\left( (\hat{r}^L - \lambda)-(\hat{r}^R - \lambda) \right) \hat{a}_\alpha \hat{b}^+_\alpha \nonumber \\ &=& \,\,0, \label{zeroCas1}
\end{eqnarray}
\begin{eqnarray}
\hat{L}_j \hat{X} _j &=& \frac{1}{4\lambda} (\hat{X}_j^L - \hat{X}_j^R)(\hat{X}_j^L + \hat{X}_j^R) = \frac{1}{4\lambda}(\hat{X}_j^L \hat{X}_j^L + \hat{X}_j^L  \hat{X}_j^R - \hat{X}_j^R  \hat{X}_j^L - \hat{X}_j^R  \hat{X}_j^R )  \nonumber \\
&=& \frac{1}{4\lambda}(\hat{X}_j^L \hat{X}_j^L  - \hat{X}_j^R  \hat{X}_j^R ) =  \frac{1}{2\lambda}(((\hat{r}^L)^2-\lambda^2) - ((\hat{r}^2)^R - \lambda^2) )= 0 .\label{zeroCas2}
\end{eqnarray}
\noindent Putting the previous two calculations together we can claim
\be
\hat{L}_j \hat{A}_j = \frac{\hat{L}_j \hat{W}'_j}{2 \lambda}  = \frac{\hat{L}_j (\eta \hat{X}_j -\hat{\zeta}_j)}{2 \lambda} = 0.
\ee
\vspace*{0.3cm}
\noindent Now let us move to the second Casimir operator in (\ref{CAS}). We have mentioned already that the relevant calculation is based on the fact which we are about to prove now:
\be \label{id}
 \hat{W}'_i\hat{W}'_i +(\eta^2\lambda^2-4)(\hat{L}_i\hat{L}_i+1) = (\hat{W}')^2 .
\ee
Undoubtedly it is a useful identity, but the reader may be slightly uncertain about where did the very idea come from.
Perhaps the following few words  will fail to  give  a satisfactory motivation, but it will have to suffice:
There are not too many equations which relate operators and c-numbers, so we naturally try to use them whenever possible. If we encounter a term which seems someway related to the Schr\"{o}dinger equation,  we are only glad to work towards rewriting that expression in terms of $\hat{W}'$. (Just to refresh the reader's memory, one form of the Schr\"{o}dinger equation reads $(\hat{W}' - 2 \lambda q) \Psi_E= 0$.)\\
So let us start, calculations of many identities follow. The  motivation for some of them may not be  straightforward, but all should be clear by the end of the appendix at the latest.
\begin{eqnarray}
 (\hat{W}')^2 &=&  (\eta \hat{r} - \hat{\zeta})^2 = \eta^2 \hat{r}^2 -\eta \{\hat{r}, \hat{\zeta} \} + (\hat{\zeta} )^2, \nonumber \\
 & &    \nonumber \\
 \hat{W}'_i \hat{W}'_i&=&  (\eta \hat{X}_i - \hat{\zeta}_i)^2 = \eta^2 \hat{X}_i \hat{X}_i -\eta \{\hat{X}_i, \hat{\zeta}_i\} + \hat{\zeta}_i \hat{\zeta}_i ,
\end{eqnarray}
\begin{eqnarray}
 \hat{L}_i \hat{L}_i &=&  \frac{1}{4\lambda^2}( \hat{X}^L_i - \hat{X}^R_i)^2 = \frac{1}{4\lambda^2}( (r^L)^2 + (\hat{r}^R)^2 - 2\hat{X}^L_i\hat{X}^R_i - 2\lambda^2), 
 \end{eqnarray}
\begin{eqnarray}
 \hat{X}_i \hat{X}_i &=&  \frac{1}{4}( \hat{X}^L_i + \hat{X}^R_i)^2 = \frac{1}{4}( (\hat{r}^L)^2 + (\hat{r}^R)^2 + 2\hat{X}^L_i\hat{X}^R_i - 2\lambda^2), 
 \end{eqnarray}
\begin{eqnarray}
 \lambda^{-2}\hat{X}^L_i \hat{X}^R_i &=& \sigma^i_{\alpha\beta}\sigma^i_{\gamma\delta}(\hat{a}^+_\alpha \hat{a}_\beta \hat{b}_\delta \hat{b}^+_\gamma)= (2\delta_{\alpha\delta}\delta_{\gamma\beta}-\delta_{\alpha\beta}\delta_{\gamma\delta})(\hat{a}^+_\alpha \hat{a}_\beta \hat{b}_\delta \hat{b}^+_\gamma) \\
 &=&   2 \hat{a}^+_\alpha \hat{a}_\beta \hat{b}_\alpha \hat{b}^+_\beta -     \hat{a}^+_\alpha \hat{a}_\alpha \hat{b}_\beta \hat{b}^+_\beta =    2 \hat{a}^+_\alpha \hat{a}_\beta \hat{b}_\alpha \hat{b}^+_\beta -     \lambda^{-2}(\hat{r}^L-\lambda)(\hat{r}^R-\lambda) . \nonumber
 \end{eqnarray}
The last result is to be used to rewrite the $2\hat{a}^+_\alpha \hat{a}_\beta \hat{b}_\alpha \hat{b}^+_\beta $ term in the following calculation:
\begin{eqnarray}
( \hat{\zeta} )^2 &=& (\hat{a}^+_\alpha \hat{b}_\alpha + \hat{a}_\alpha \hat{b}^+_\alpha )(\hat{a}^+_\beta \hat{b}_\beta + \hat{a}_\beta \hat{b}^+_\beta )\nonumber \\
 &=& \hat{a}^+_\alpha \hat{a}^+_\beta \hat{b}_\alpha \hat{b}_\beta + \hat{a}_\alpha \hat{a}_\beta \hat{b}^+_\alpha \hat{b}^+_\beta +       \hat{a}^+_\alpha \hat{a}_\beta \hat{b}_\alpha \hat{b}^+_\beta  + (\hat{a}^+_\beta \hat{a}_\alpha +\delta_{\alpha\beta} )(\hat{b}_\beta \hat{b}^+_\alpha +\delta_{\alpha\beta})                         \nonumber \\
&=&  \hat{a}^+_\alpha \hat{a}^+_\beta \hat{b}_\alpha \hat{b}_\beta + \hat{a}_\alpha \hat{a}_\beta \hat{b}^+_\alpha \hat{b}^+_\beta + 2 \hat{a}^+_\alpha \hat{a}_\beta \hat{b}_\alpha \hat{b}^+_\beta + 2\lambda^{-1}\hat{r}  \\
&=&  \hat{a}^+_\alpha \hat{a}^+_\beta \hat{b}_\alpha \hat{b}_\beta + \hat{a}_\alpha \hat{a}_\beta \hat{b}^+_\alpha \hat{b}^+_\beta + \lambda^{-2}(\hat{X}^L_i \hat{X}^R_i + \hat{r}^2 + \lambda^2) . \nonumber
 \end{eqnarray}
The above equation will be used to rewrite the terms $\hat{a}^+_\alpha \hat{a}^+_\beta \hat{b}_\alpha \hat{b}_\beta + \hat{a}_\alpha \hat{a}_\beta \hat{b}^+_\alpha \hat{b}^+_\beta$ in the following one:
\begin{eqnarray}
 \hat{\zeta}^{i}  \hat{\zeta}^{i} &=&   \sigma^i_{\alpha\beta}\sigma^i_{\gamma\delta}\hat{\zeta}_{\alpha\beta}\hat{\zeta}_{\gamma\delta}= (2\delta_{\alpha\delta}\delta_{\gamma\beta}-\delta_{\alpha\beta}\delta_{\gamma\delta})(\hat{a}^+_\alpha \hat{b}_\beta + \hat{a}_\beta \hat{b}^+_\alpha)(\hat{a}^+_\gamma \hat{b}_\delta + \hat{a}_\delta \hat{b}^+_\gamma)  \nonumber \\
 &=&  2 (\hat{a}^+_\alpha \hat{a}^+_\beta \hat{b}_\alpha \hat{b}_\beta + \hat{a}^+_\alpha \hat{a}_\alpha \hat{b}_\beta \hat{b}^+_\beta + \hat{a}_\beta \hat{a}^+_\beta \hat{b}^+_\alpha \hat{b}_\alpha + \hat{a}_\alpha \hat{a}_\beta \hat{b}^+_\alpha \hat{b}^+_\beta ) -    ( \hat{\zeta}^{0})^2                  \\
&=&  2 (\hat{a}^+_\alpha \hat{a}^+_\beta \hat{b}_\alpha \hat{b}_\beta +  \hat{a}_\alpha \hat{a}_\beta \hat{b}^+_\alpha \hat{b}^+_\beta ) +4\lambda^{-2}(\hat{r}^2 + \lambda^2) -    ( \hat{\zeta} )^2                         \nonumber \\
&=&  ( \hat{\zeta} )^2  + 2\lambda^{-2}(\hat{r}^2 - \hat{X}_i^L \hat{X}_i^R +\lambda^2), \nonumber
 \end{eqnarray}
\begin{eqnarray}
[\hat{r}, \hat{\zeta} ] &=& \frac{\lambda}{2}[\hat{a}^+_\alpha \hat{a}_\alpha + \hat{b}_\alpha \hat{b}^+_\alpha + 2,\, \hat{a}^+_\beta \hat{b}_\beta + \hat{a}_\beta \hat{b}^+_\beta]   \nonumber \\
 &=& \lambda(\hat{a}^+_\alpha \hat{b}_\alpha - \hat{a}_\alpha \hat{b}^+_\alpha)  = \lambda \hat{w},  \\
 & & \nonumber \\
 \{\hat{r}, \hat{\zeta} \} &=& [\hat{r}, \hat{\zeta} ] + 2 \hat{\zeta}  \hat{r} =  \lambda \hat{w} + 2 \hat{\zeta}  \hat{r}, 
 \end{eqnarray}
\begin{eqnarray}
\hat{\zeta}_i \hat{X}_i &=&  \frac{\lambda}{2}\sigma^i_{\alpha\beta}\sigma^i_{\gamma\delta}( \hat{a}^+_\alpha \hat{b}_\beta + \hat{a}_\beta \hat{b}^+_\alpha) ( \hat{a}^+_\gamma \hat{a}_\delta + \hat{b}_\delta \hat{b}^+_\gamma) \nonumber \\
&=& \frac{\lambda}{2}(2\delta_{\alpha\delta}\delta_{\gamma\beta}-\delta_{\alpha\beta}\delta_{\gamma\delta}) ( \hat{a}^+_\alpha \hat{b}_\beta + \hat{a}_\beta \hat{b}^+_\alpha) ( \hat{a}^+_\gamma \hat{a}_\delta + \hat{b}_\delta \hat{b}^+_\gamma) \nonumber \\
 &=&  \lambda (\hat{a}^+_\alpha \hat{a}_\alpha \hat{a}^+_\beta \hat{b}_\beta -\hat{a}^+_\alpha \hat{b}_\alpha  + \hat{b}_\beta \hat{b}^+_\beta \hat{a}^+_\alpha \hat{b}_\alpha -\hat{a}^+_\alpha \hat{b}_\alpha + \hat{a}_\beta \hat{a}^+_\beta \hat{a}_\alpha \hat{b}^+_\alpha + \hat{b}^+_\alpha \hat{b}_\alpha \hat{a}_\beta \hat{b}^+_\beta  )  \nonumber \\
 & & - \frac{1}{2}\hat{\zeta} (\hat{r}^L + \hat{r}^R - 2\lambda)\nonumber \\
 & = &  (\hat{r}^L + \hat{r}^R - 2\lambda) \hat{a}^+_\alpha \hat{b}_\alpha +  (\hat{r}^L + \hat{r}^R + 2\lambda) \hat{a}_\alpha \hat{b}^+_\alpha -2\lambda \hat{a}^+_\alpha \hat{b}_\alpha - \hat{\zeta}  \hat{r} + \lambda \hat{\zeta}        \nonumber \\
  & =& \hat{r}\hat{\zeta}  + [\hat{r},\,\hat{\zeta} ] - 3\lambda \hat{w} \ =\ \hat{r}\hat{\zeta}  - 2\lambda \hat{w} , \nonumber
 \end{eqnarray}
 \begin{eqnarray}
\hat{X}_i\hat{\zeta}_i  &=&  \frac{\lambda}{2}\sigma^i_{\alpha\beta}\sigma^i_{\gamma\delta}( \hat{a}^+_\alpha \hat{a}_\beta + \hat{b}_\beta \hat{b}^+_\alpha) ( \hat{a}^+_\gamma \hat{b}_\delta + \hat{a}_\delta \hat{b}^+_\gamma) \nonumber \\
&=&  \frac{\lambda}{2}(2\delta_{\alpha\delta}\delta_{\gamma\beta}-\delta_{\alpha\beta}\delta_{\gamma\delta})( \hat{a}^+_\alpha \hat{a}_\beta + \hat{b}_\beta \hat{b}^+_\alpha) ( \hat{a}^+_\gamma \hat{b}_\delta + \hat{a}_\delta \hat{b}^+_\gamma) \nonumber \\
 &=&  \lambda (\hat{a}_\beta \hat{a}^+_\beta \hat{a}^+_\alpha \hat{b}_\alpha -\hat{a}^+_\alpha \hat{b}_\alpha  + \hat{a}^+_\alpha \hat{a}_\alpha \hat{a}_\beta \hat{b}^+_\beta + \hat{b}^+_\alpha \hat{b}_\alpha \hat{a}^+_\beta \hat{b}_\beta   -\hat{a}^+_\alpha \hat{b}_\alpha  + \hat{b}_\beta \hat{b}^+_\beta \hat{a}_\alpha \hat{b}^+_\alpha  )  \nonumber \\
 & & - \frac{1}{2}(\hat{r}^L + \hat{r}^R - 2\lambda)\hat{\zeta}  \nonumber \\
 & = &  (\hat{r}^L + \hat{r}^R + 2\lambda) \hat{a}^+_\alpha \hat{b}_\alpha +  (\hat{r}^L + \hat{r}^R - 2\lambda) \hat{a}_\alpha \hat{b}^+_\alpha -2\lambda \hat{a}^+_\alpha \hat{b}_\alpha - \hat{r}\hat{\zeta}  + \lambda \hat{\zeta}       \nonumber \\
  & =& \hat{r}\hat{\zeta}  +\lambda \hat{w} ,  \nonumber
 \end{eqnarray}
  \be
  [\hat{X}_i,\, \hat{\zeta}_i]  = 3\lambda \hat{w} ,   \ \ \ \ \ \ \ \    \{  \hat{X}_i,\, \hat{\zeta}_i \}  =  2 \hat{r}\hat{\zeta}  - \lambda \hat{w}  .
\ee
Now we can use all the derived stuff to prove (\ref{id}):
 \begin{eqnarray}
 (\hat{W}' )^2 &=&  \eta^2 \hat{r}^2 -\eta \{\hat{r}, \hat{\zeta} \} + (\hat{\zeta} )^2 =   \eta^2 \hat{r}^2 - \eta (\lambda \hat{w} + 2\hat{\zeta}  \hat{r}) + (\hat{\zeta} )^2, \nonumber \\
  (\hat{W}'_i)^2 &+& (\eta^2\lambda^2 -4 )((\hat{L}_i)^2 + 1) = \nonumber \\
 &=&  \frac{\eta^2}{2}(\hat{r}^2 + \hat{X}_i^L \hat{X}_i^R - \lambda^2) -\eta (2\hat{r}\hat{\zeta} -\lambda w) + (\hat{\zeta} )^2 + \frac{2}{\lambda^2}(\hat{r}^2 - \hat{X}_i^L \hat{X}_i^R + \lambda^2)   \nonumber \\
  & & + \frac{\eta^2}{2}(\hat{r}^2 - x_i^L x_i^R - \lambda^2) - \frac{2}{\lambda^2}(r^2 - x_i^L x_i^R - \lambda^2) + \eta^2\lambda^2 -4 \nonumber \\
 & = & \eta^2 \hat{r}^2 - \eta(2\hat{r}\hat{\zeta}  - \lambda \hat{w}) + (\hat{\zeta} )^2  \nonumber \\
 &=& (\hat{W}')^2 .\nonumber
  \end{eqnarray}
In the last line we have used $[\hat{r}, \hat{\zeta} ]=\lambda \hat{w}$ and consequently
\be
 \lambda \hat{w} + 2\hat{\zeta}  \hat{r} =  2\hat{r}\hat{\zeta}  - \lambda \hat{w} .
  \ee
The proof of    (\ref{id}) is finished, and  (\ref{CAS2}) follows immediately.\\
\\
\\
Although  the attention and patience of the reader may have been taxed just enough by this Appendix, hopefully they have kept sufficiently positive attitude to at least appreciate the absence of further details.



%


\bibliography{aipsamp}%

%
\end{document}